\documentclass[a4paper,11pt]{article}
\pdfoutput=1 

\usepackage{jheppub} 
\usepackage[vcentermath]{youngtab}
\usepackage{url,amsmath,amssymb,amsfonts,amsxtra, mathrsfs,graphics,graphicx,amsthm,epsfig, youngtab,bm,longtable,float,tikz,caption,MnSymbol}
\usepackage[T1]{fontenc} 
\usepackage[margin=0.5cm]{caption}
\usepackage{jheppub}
\usepackage{float}
\restylefloat{table}
\usepackage{comment}
\usepackage{subfig}
\usepackage{mdframed}
\usepackage{verbatim}

\addtocontents{toc}{\setcounter{tocdepth}{2}}

\makeatletter
\def\@fpheader{\relax}
\makeatother

\title{Exact quantization conditions for the elliptic Ruijsenaars-Schneider model}

\author[a]{Yasuyuki Hatsuda,}
\author[b]{Antonio Sciarappa}
\author[c]{and Szabolcs Zakany}

\affiliation[a]{Department of Physics, Rikkyo University \\
Toshima, Tokyo 171-8501, Japan}

\emailAdd{yhatsuda@rikkyo.ac.jp}

\affiliation[b]{School of Physics, Korea Institute for Advanced Study \\
85 Hoegiro, Dongdaemun-gu, Seoul 130-722, Republic of Korea}

\emailAdd{asciara@kias.re.kr}

\affiliation[c]{D\'{e}partement de Physique Th\'{e}orique et Section de Math\'{e}matiques \\
Universit\'{e} de Gen\'{e}ve, Gen\'{e}ve, CH-1211 Switzerland}

\emailAdd{szabolcs.zakany@unige.ch}

\abstract{We propose and test exact quantization conditions for the $N$-particle quantum elliptic Ruijsenaars-Schneider integrable system, as well as its Calogero-Moser limit, based on the conjectural correspondence to the five-dimensional $\mathcal{N} = 1^*$ $SU(N)$ gauge theory in the Nekrasov-Shatashvili limit. 
We discuss two natural sets of quantization conditions, related by the electro-magnetic duality, and the importance of non-perturbative corrections in the Planck constant. We also comment on the eigenfunction problem, by reinterpreting the Separation of Variables approach in gauge theory terms.}

 \preprint{KIAS-P18082, RUP-18-25}

\begin{document}

\maketitle


\section{Introduction and summary}

Among all quantum integrable systems, 
elliptic ones are the hardest to analyse due to the doubly-periodic nature of their potential. They are however also the most fundamental ones, since  many other integrable systems can be obtained from them by taking opportune limits. 
The most famous example is the $N$-particle elliptic Calogero-Moser system, which in various special limits reduces to its trigonometric or hyperbolic version, or even to the $N$-particle closed Toda chain \cite{Olshanetsky:1983wh}: although much progress has been made in the last few years, its analytic solution is not known if not for very special cases. 
The situation is even worse if we consider its relativistic generalization, also known as $N$-particle elliptic Ruijsenaars-Schneider system \cite{RUIJSENAARS1986370,ruijsenaars1987,Ruijsenaars:1992ua,Ruijsenaars1999}: in that case not only the analytic solution is not known, but also determining its Hilbert integrability (i.e. its domain of self-adjointness) is a rather non-trivial problem since the Hamiltonian is not a differential but a finite-difference operator  \cite{0305-4470-32-9-018,Ruijsenaars1999,2015SIGMA..11..004R}. It would therefore be quite helpful to develop new techniques to approach these elliptic problems.

As it became clear recently, supersymmetric gauge theories and topological string theory can provide an alternative, non-conventional framework to study quantum integrable systems. This was first pointed out in \cite{2010maph.conf..265N}, where it was conjectured that all-orders WKB quantization conditions and energy spectrum for a particular class of integrable systems (such as closed Toda chains, elliptic Calogero-Moser and their relativistic versions) can be obtained by considering certain observables in $\Omega$-background deformed four or five-dimensional gauge theories with at least eight supercharges and unitary gauge group, in a special limit known as Nekrasov-Shatashvili (NS) limit. In this limit only one (say, $\epsilon_1$) of the two $\Omega$-background parameters $\epsilon_1$ and $\epsilon_2$ is kept, and plays the role of the Planck constant $\hbar$ of the quantum integrable system; other gauge theory parameters can also be identified with parameters of the quantum Hamiltonians. 
The most important advantage of the gauge theory approach is that it naturally gives exact WKB quantization conditions and energy spectrum expressed not as an asymptotic series in $\hbar$, but as a convergent power series in an auxiliary parameter (the instanton counting parameter $Q$, associated to some parameter in the Hamiltonians) whose coefficients are exact in $\hbar$.


The conjecture by \cite{2010maph.conf..265N} was tested in \cite{Mironov:2009uv,Mironov:2009dv,Kozlowski:2010tv} for the case of the $N$-particle closed Toda chain, associated to four-dimensional $\mathcal{N} = 2$ pure $SU(N)$ super Yang-Mills theory in the NS limit, and was shown to reproduce the known Toda solution studied in \cite{GUTZWILLER1980347,GUTZWILLER1981304,0305-4470-25-20-007,Kharchev:1999bh,Kharchev:2000ug,Kharchev:2000yj,An2009}. 
More recently other tests were performed in \cite{2015arXiv151102860H} for the case of the $N$-particle relativistic closed Toda chain \cite{ruijsenaars1990}, associated to the NS limit of five-dimensional $\mathcal{N} = 1$ pure $SU(N)$ super Yang-Mills theory compactified on a circle, and in \cite{Franco:2015rnr} for the case of other relativistic quantum integrable systems of cluster type \cite{2011arXiv1107.5588G}, which in general can only be associated to a topological string theory on a local toric Calabi-Yau 3-fold (or $(p,q)$-brane web) rather than a five-dimensional gauge theory. Quite interestingly, the conjecture by \cite{2010maph.conf..265N} turns out to be inconsistent for all the relativistic systems considered in \cite{2015arXiv151102860H,Franco:2015rnr}: this is because even if the exact WKB quantization conditions obtained from gauge theory/topological string theory are exact in $\hbar$ and convergent as a series in $Q$, they present an infinitely dense set of poles as functions of $\hbar$, 
which is inconsistent with the fact that the quantum integrable systems are perfectly well-defined for any $\hbar \in \mathbb{R}_+$.

This is however not too surprising: in principle in any quantum mechanical system
complex instantons may generate important non-perturbative contributions in $\hbar$, but these will be missed if we only consider the WKB solution. One can therefore imagine that inconsistencies of the conjecture by \cite{2010maph.conf..265N} are due to the fact that complex instantons are present in all relativistic integrable systems, but the gauge theory approach does not take them into account. We thus expect the poles in $\hbar$ should disappear when non-perturbative effects are considered. This is indeed what seems to happen: as conjectured and tested in \cite{2015arXiv151102860H,Franco:2015rnr}, if we add the relevant non-perturbative contributions to the exact WKB quantization conditions we obtain pole-free quantization conditions for relativistic Toda chains and relativistic cluster integrable systems, and the resulting energy spectrum matches the one we can compute numerically by standard quantum mechanical arguments. 

Of course, the biggest problem is to precisely determine the non-perturbative contributions. In \cite{2015arXiv151102860H,Franco:2015rnr} (based on previous works \cite{Hatsuda:2012dt,2013arXiv1308.6485K,Huang:2014eha,Grassi:2014zfa,Wang:2015wdy,Gu:2015pda,Codesido:2015dia,2015arXiv150704799H}) it is suggested that these are completely fixed by requiring the full (WKB + non-perturbative) exact quantization conditions to be invariant under the exchange $\hbar \leftrightarrow 4 \pi^2 / \hbar$. Although the effectiveness of such a simple proposal may seem surprising at first, this is not completely unexpected, being in line with the quantum group and modular double structure underlying the relativistic Toda chain \cite{Kharchev:2001rs} and, more in general, any relativistic cluster integrable system \cite{2011arXiv1107.5588G}. \\

The purpose of the present work is to further test the conjecture by \cite{2010maph.conf..265N}, or its non-perturbative refinement \`{a} la \cite{2015arXiv151102860H,Franco:2015rnr}, in the case of $N$-particle elliptic quantum integrable systems of Calogero-Moser and Ruijsenaars-Schneider type, related to four-dimensional $\mathcal{N} = 2^*$ $SU(N)$ and five-dimensional $\mathcal{N} = 1^*$ $SU(N)$ gauge theories respectively, mostly focussing on the 2-particle case for simplicity. 
In order to test these conjectures against numerical results, we will restrict our attention to the region in parameter space for which the elliptic quantum integrable systems admit a discrete set of energy levels.

As we will see, the situation is pretty much similar to what happens for the closed Toda chain, relativistic and not: the exact WKB quantization conditions conjectured by \cite{2010maph.conf..265N} seem to be sufficient to reproduce the Calogero-Moser spectrum, while in the Ruijsenaars-Schneider case we also need to include non-perturbative corrections, fixed by the requirement of invariance under $\hbar \leftrightarrow 4\pi^2 / \hbar$ along the lines of \cite{2015arXiv151102860H,Franco:2015rnr}.
It is important to remark that although the elliptic Ruijsenaars-Schneider system is not of cluster type, i.e. it is not related to a toric Calabi-Yau 3-fold, the $\hbar \leftrightarrow 4\pi^2/\hbar$ symmetry is still expected due to its underlying modular double structure  \cite{0305-4470-32-9-018,Ruijsenaars1999,2015SIGMA..11..004R}.\footnote{Because of different conventions for the parameters, in this work the $\hbar \leftrightarrow 4\pi^2 / \hbar$ invariance will be replaced by invariance under $\hbar \leftrightarrow 2 \omega $ with $2\omega$ real period of the torus on which the integrable system is defined.}

There is however a new ingredient to the story, which is special to elliptic system. When the potential is doubly-periodic with real and imaginary periods $(2\omega, 2\omega')$, we can either consider the  $L^2([0,2\omega]^{N-1})$ quantum mechanical problem relative to the real period or the $L^2([0,2\omega']^{N-1})$ one relative to the imaginary period. Clearly the two problems are the same since they are simply related by the exchange $\omega \leftrightarrow \omega'$, but as already pointed out in \cite{2010maph.conf..265N} the gauge theory approach treats them differently: quantization conditions for the first and second case correspond to quantize the $\epsilon_1$-deformed Seiberg-Witten periods $a_j$ or the $\epsilon_1$-deformed dual periods $a_{D,j}$ respectively. This is not unexpected: since the complex structure $\tau = \omega'/\omega$ of the torus is identified with the complexified gauge coupling, exchanging $\omega \leftrightarrow \omega'$ corresponds to electro-magnetic duality, which leads to the exchange $a \leftrightarrow a_D$.
When $\epsilon_1 = 0$ this was already made clear in \cite{Seiberg:1994aj} using the fact that $a, a_D$ are related by the Seiberg-Witten prepotential; we will see that this remains true even at $\epsilon_1 \neq 0$ if we replace the Seiberg-Witten prepotential by its appropriate $\epsilon_1$-deformed version (usually called twisted effective superpotential), which however in the five-dimensional case needs to include contributions which are non-perturbative in $\epsilon_1 \sim \hbar$ as we already discussed: we therefore conclude that away from the Seiberg-Witten limit, realizing electro-magnetic duality in five dimensions can only be achieved by also considering non-perturbative effects in the Omega background parameters. \\

The plan of the paper is as follows. Section \ref{CMsection} will be dedicated to studying the 2-particle elliptic Calogero-Moser system. We first introduce the system as well as its trigonometric and hyperbolic limits. After review what is known about its analytic solution, both in these special limits and in general, we explain how to study the problem numerically; we then move to discuss how gauge theory can be used to determine the energy spectrum analytically. Finally, we comment on what gauge theory can tell us about the problem of constructing eigenfunctions. Section \ref{RSsection} will follow the same steps, but for the 2-particle elliptic Ruijsenaars-Schneider system; Section \ref{NRSsection} instead is devoted to the study of the $N$-particle case. Useful formulae are collected in Appendices \ref{appA}, \ref{appgauge}.

\section{2-particle systems of Calogero-Moser type} \label{CMsection}

Let us consider a complex coordinate $z = x + i y$ ($x, y \in \mathbb{R}$) on a rectangular torus of half-periods $(\omega, \omega')$ with $\omega \in \mathbb{R}_+$, $\omega' \in i\mathbb{R}_+$; this means that
\begin{equation}
x \sim x + 2 \omega , \;\;\;\; y \sim y + 2 \vert \omega' \vert.
\end{equation}
With gauge theory applications in mind, we will sometimes find it more convenient to reparametrize the periods as 
\begin{equation}
(2\omega, 2 \omega') \,=\,  ( 2\omega, -i \frac{\omega}{\pi} \ln Q_{\text{4d}} ),
\end{equation}
where
\begin{equation}
Q_{\text{4d}} = e^{2\pi i \tau}, \;\;\;\;\;\; \tau = \frac{\omega'}{\omega}.
\end{equation}
The 2-particle elliptic Calogero-Moser system (2-eCM) Hamiltonian is given by the second-order ordinary differential equation (see Appendix \ref{appell} for our conventions on elliptic functions)
\begin{equation}
\left[- \partial_z^2 + (\frac{g^2}{\hbar^2} - \frac{1}{4}) \wp(z\vert \omega, \omega') \right] \psi(z) = \frac{E}{\hbar^2} \psi(z). \label{ellCM}
\end{equation}
Here all parameters $z,\hbar, g, E \in \mathbb{C}$. One possible problem one could study is to look for the two linearly independent solutions to this equation for generic complex values of the parameters. We are however interested in quantum mechanical applications of \eqref{ellCM}, so we should only be considering values of parameters for which \eqref{ellCM} realizes a well-defined quantum mechanical problem with self-adjoint Hamiltonian. Among various other possibilities,\footnote{For example the choice of parameters considered in \cite{He:2011zk,Piatek:2013ifa,2015JHEP...02..160B,Beccaria:2016wop}, which leads to a quantum mechanical problem with continuous energy spectrum and a band/gap structure, will not be studied here.} this happens in the following two cases:

\paragraph{B-model:} Focussing on the real slice $z = x \in \mathbb{R}$ and requiring $\hbar = \hbar_x \in \mathbb{R}_+$, $g = g_x \in \mathbb{R}_+$, $E = E^{(\text{B})} \in \mathbb{R}_+$ the Hamiltonian \eqref{ellCM} reduces to
\begin{equation}
\left[- \partial_x^2 + (\frac{g^2_x}{\hbar^2_x} - \frac{1}{4}) \wp(x \vert \omega, \omega') \right] \psi^{(\text{B})}(x) = \frac{E^{(\text{B})}}{\hbar^2_x} \psi^{(\text{B})}(x), \label{ellCMB}
\end{equation}
or alternatively, by introducing the sometimes more familiar parameter $\alpha_x = \frac{g_x}{\hbar_x} + \frac{1}{2}$,
\begin{equation}
\left[- \partial_x^2 + \alpha_x(\alpha_x - 1) \wp(x \vert \omega, \omega') \right] \psi^{(\text{B})}(x) = \frac{E^{(\text{B})}}{\hbar^2_x} \psi^{(\text{B})}(x). \label{ellCMBbis}
\end{equation}
For $g_x \geqslant \frac{\hbar_x}{2}$ (or $\alpha_x \geqslant 1$) the potential is confining, and \eqref{ellCMB} defines a quantum mechanical problem on $L^2_x([0,2\omega])$ whose energy spectrum is real and discrete; we will refer to \eqref{ellCMB} as the B-type 2-particle elliptic Calogero-Moser quantum integrable system (2-eCM$_{\text{B}}$). 

In the limit $\omega' \rightarrow i\infty$ (i.e. $Q_{\text{4d}} \rightarrow 0$) this reduces (modulo constants) to the 2-particle trigonometric Calogero-Moser quantum integrable system (2-tCM) 
\begin{equation}
\left[- \partial_x^2 + (\frac{g^2_x}{\hbar^2_x} - \frac{1}{4}) \dfrac{\pi^2}{\omega^2} \dfrac{1}{4 \sin^2(\frac{\pi x}{2 \omega})} \right] \psi^{(\text{T})}(x) = \frac{E^{(\text{T})}}{\hbar^2_x} \psi^{(\text{T})}(x) ; \label{tCM}
\end{equation}
this is still a quantum mechanical problem on $L^2_x([0,2\omega])$ with a discrete set of energy levels for $g_x \geqslant \frac{\hbar_x}{2}$, and its solution is known analytically.

\paragraph{A-model:} Focussing instead on the slice $z = i y \in i \mathbb{R}$ and requiring $\hbar = i \hbar_y \in i\mathbb{R}_+$, $g = i g_y \in i\mathbb{R}_+$, $E = E^{(\text{A})} \in \mathbb{R}_+$ the Hamiltonian \eqref{ellCM} reduces to
\begin{equation}
\left[- \partial_y^2 - (\frac{g^2_y}{\hbar^2_y} - \frac{1}{4}) \wp(i y \vert \omega, \omega' ) \right] \psi^{(\text{A})}(y) = \frac{E^{(\text{A})}}{\hbar^2_y} \psi^{(\text{A})}(y), \label{ellCMA}
\end{equation}
or alternatively, by introducing the parameter $\alpha_y = \frac{g_y}{\hbar_y} + \frac{1}{2}$,
\begin{equation}
\left[- \partial_y^2 - \alpha_y(\alpha_y - 1) \wp(i y \vert \omega, \omega') \right] \psi^{(\text{A})}(y) = \frac{E^{(\text{A})}}{\hbar^2_y} \psi^{(\text{A})}(y). \label{ellCMAbis}
\end{equation}
For $g_y \geqslant \frac{\hbar_y}{2}$ (or $\alpha_y \geqslant 1$) the potential is confining, and \eqref{ellCMA} defines a quantum mechanical problem on $L^2_y([0,2 \vert \omega' \vert])$ with energy spectrum real and discrete; we will refer to \eqref{ellCMA} as the A-type 2-particle elliptic Calogero-Moser quantum integrable system (2-eCM$_{\text{A}}$). 

In the limit $\omega' \rightarrow i \infty$ (i.e. $Q_{\text{4d}} \rightarrow 0$) this reduces (modulo constants) to the 2-particle hyperbolic Calogero-Moser quantum integrable system (2-hCM) 
\begin{equation}
\left[- \partial_y^2 + (\frac{g^2_y}{\hbar^2_y} - \frac{1}{4}) \dfrac{\pi^2}{\omega^2} \dfrac{1}{4 \sinh^2(\frac{\pi y}{2 \omega})} \right] \psi^{(\text{H})}(y) = \frac{E^{(\text{H})}}{\hbar^2_y} \psi^{(\text{H})}(y) ; \label{hCM}
\end{equation}
this is still a well-defined quantum mechanical problem for $y \in (0,\infty)$ and $g_y \geqslant \frac{\hbar_y}{2}$ whose solution is known analytically, however its energy spectrum is continuous.

\noindent Although we introduced them as different, it is important to remark that the two problems 2-eCM$_{\text{B}}$ and 2-eCM$_{\text{A}}$ are 
actually the same: in fact since \cite{lawden1989elliptic}
\begin{equation}
\wp(i y\vert \omega, \omega') = - \wp(y \vert -i \omega, -i \omega') = 
- \wp(y \vert -i \omega', i \omega),
\end{equation}
the 2-eCM$_{\text{A}}$ problem with half-periods $(\omega, \omega')$ is equivalent to the 2-eCM$_{\text{B}}$ problem with half-periods $(-i \omega', i \omega)$ if we also identify $\hbar_y$, $g_y$ with $\hbar_x$, $g_x$, that is
\begin{equation}
\underbrace{ - \partial_y^2 - (\frac{g^2_y}{\hbar^2_y} - \frac{1}{4}) \wp(i y \vert \omega, \omega') }_\textrm{A-problem}
\;=\; \underbrace{ - \partial_y^2 + (\frac{g^2_y}{\hbar^2_y} - \frac{1}{4}) \wp(y \vert -i \omega', i\omega) }_\textrm{B-problem (inverted periods)}. \label{relationeCM}
\end{equation}
Despite this fact, we prefer to keep distinguishing them in order to facilitate later comparison with gauge theory. In fact, we will see that the gauge theory approach treats B- and A-model differently; verifying that these different treatments respect \eqref{relationeCM} will therefore be a non-trivial consistency check for the validity of the gauge theory approach.


\subsection{Trigonometric and hyperbolic cases} \label{tCMsection}

Before discussing the 2-eCM$_{\text{B}}$ and 2-eCM$_{\text{A}}$ systems, we will first review the known analytic solution to their trigonometric (2-tCM) and hyperbolic (2-hCM) limits. This will be of help later in Section \ref{seceCMB} to understand how to numerically evaluate the spectrum of the elliptic system.

\subsection*{Trigonometric case} 

Let us start by considering the 2-tCM system. For generic half-period $\omega$, the 2-tCM problem reads
\begin{equation}
\widehat{H}^{(\text{T})} \psi^{(\text{T})}(x) = \left[ -\hbar_x^2 \partial_x^2 + \hbar_x^2 (\frac{g_x^2}{\hbar_x^2} - \frac{1}{4}) \frac{\pi^2}{\omega^2} \dfrac{1}{4 \sin^2 (\frac{\pi x}{2 \omega})} \right] \psi^{(\text{T})}(x) = 
\dfrac{\pi^2}{\omega^2} a^2 \psi^{(\text{T})}(x), \label{tCMbis}
\end{equation}
where for later convenience we reparametrized the energy $E^{(\text{T})}$ in terms of an auxiliary variable $a \in \mathbb{R}$ as
\begin{equation}
E^{(\text{T})}(a) = \frac{\pi^2}{\omega^2} a^2. \label{entCM}
\end{equation}
Equation \eqref{tCMbis} will in general have two linearly independent solutions; however, since we will ultimately be interested in $L^2([0,2\omega])$ eigenfunctions, we start by considering only a particular linear combination of 
them which vanishes at $x = 0$, given by (modulo normalization factors) 
\begin{equation}
\psi^{(\text{T})}_+(x) \propto 
\left[ \sin \frac{\pi x}{2 \omega} \right]^{\frac{1}{2} + \frac{g_x}{\hbar_x}}
{}_2 F_1 \left(\frac{1}{4}+\frac{g_x}{2\hbar_x}+\frac{a}{\hbar_x},\frac{1}{4}+\frac{g_x}{2\hbar_x}-\frac{a}{\hbar_x},1+\frac{g_x}{\hbar_x},\sin^2 \frac{\pi x}{2 \omega}\right) , \;\; x \in (0, \omega) \label{tCMsol}
\end{equation}
for $a,g_x,\hbar_x \in \mathbb{R}_+$. Expression \eqref{tCMsol} is actually a solution only in the interval $x \in [0,\omega]$; from the symmetry of the problem, the solution in the remaining interval $x \in [\omega, 2\omega]$ should be of the same form of \eqref{tCMsol}, possibly up to an overall sign. However, the $x \in [0, \omega]$ and $x \in [\omega, 2\omega]$ solutions are not connected smoothly at $x = \omega$ for generic values of $a$ (or, equivalently, $E^{(\text{T})}$). In order for them to be smoothly connected, i.e. for the total solution to be in $L^2([0,2\omega])$, we have to tune $a$ (i.e. the energy $E^{(\text{T})}$) carefully; this is why the energy spectrum of the 2-tCM system is quantized. More concretely, the smoothness conditions we must impose are
\begin{equation}
\partial_x \psi^{(\text{T})}(x)\big\vert_{x = \omega} = 0 \;\;\;\; \text{or} \;\;\;\; \psi^{(\text{T})}(x = \omega) = 0; 
\end{equation}
these conditions determine the only 
allowed values $a_n$ and $E^{(\text{T})}_n = E^{(\text{T})}(a_n)$ $(n = 0, 1, \ldots)$, which are simply
\begin{equation}
a_n = \pm \,\dfrac{g_x + (n+\frac{1}{2})\hbar_x}{2},\;\;\;\;\;\; E_n^{(\text{T})} = \dfrac{\pi^2}{\omega^2} a_n^2,\;\;\;\;\;\; n \in \mathbb{N}. \label{quanttCM}
\end{equation}
Quantization conditions \eqref{quanttCM} give rise to parity-even (with respect to $x = \omega$) or parity-odd $L^2([0,2\omega])$ eigenfunctions $\psi^{(\text{T})}_n(x)$ for $n$ even or odd respectively:
\begin{equation}
\begin{split}
\psi^{(\text{T})}_{2m}(x) &\propto \left\{
\begin{array}{rl}
\psi^{(\text{T})}_+(x) & \;\text{if } \; x \in [0,\omega],\\
\psi^{(\text{T})}_+(x) & \;\text{if } \; x \in [\omega,2\omega],
\end{array} \right. , \;\;\;\;\;\; n = 2m; \\
\psi^{(\text{T})}_{2m + 1}(x) &\propto \left\{
\begin{array}{rl}
\psi^{(\text{T})}_+(x) & \;\text{if } \; x \in [0,\omega],\\
-\psi^{(\text{T})}_+(x) & \;\text{if } \; x \in [\omega,2\omega],
\end{array} \right. , \;\;\; n = 2m+1.
\end{split}
\end{equation}
By using the identity
\begin{equation}
{}_2 F_1 \left( a,b,a+b+\frac{1}{2},z \right) = {}_2 F_1 \left( 2a,2b,a+b+\frac{1}{2}, \frac{1 - \sqrt{1-z}}{2} \right) ,
\end{equation}
these (unnormalized) eigenfunctions can be rewritten in the more compact form
\begin{equation}
\begin{split}
\psi_n^{(\text{T})}(x) & \propto  
\left[ \sin \frac{\pi x}{2 \omega} \right]^{\frac{1}{2} + \frac{g_x}{\hbar_x}}
{}_2 F_1 \left(-n,\frac{2g_x}{\hbar_x}+1+n,\frac{g_x}{\hbar_x}+1, 
\frac{1 - \sqrt{\cos^2 \frac{\pi x}{2 \omega}}}{2}\right) \\
& \propto \left[ \sin \frac{\pi x}{2 \omega} \right]^{\frac{1}{2} + \frac{g_x}{\hbar_x}}
C_n^{\frac{1}{2} + \frac{g_x}{\hbar_x}}(\cos \frac{\pi x}{2 \omega}),  
\;\;\;\;\;\; x \in [0, 2\omega],
\end{split}
\end{equation}
where $C^{\alpha}_n (z)$ are the Gegenbauer polynomials
\begin{equation}
C^{\alpha}_n (z) = \dfrac{\Gamma(n + 2\alpha)}{n! \,\Gamma(2 \alpha)}
{}_2 F_1 \left( -n, 2\alpha + n, \alpha + \frac{1}{2}, \frac{1 - z}{2} \right).
\end{equation}
Thanks to the orthogonality properties of the Gegenbauer polynomials we can then construct a normalized basis of 2-tCM eigenfunctions
\begin{equation} 
\psi_n^{(\text{T})}(x) = \left[ \dfrac{2^{\frac{2g_x}{\hbar_x}}(n+\frac{1}{2}+\frac{g_x}{\hbar_x})n!\Gamma^2(\frac{1}{2}+\frac{g_x}{\hbar_x})}{2 \omega \, \Gamma(n+\frac{2g_x}{\hbar_x}+1)} \right]^{1/2}
\left[ \sin \frac{\pi x}{2 \omega} \right]^{\frac{1}{2} + \frac{g_x}{\hbar_x}}
C_n^{\frac{1}{2} + \frac{g_x}{\hbar_x}}(\cos \frac{\pi x}{2 \omega}), \label{tCMsolnorm}
\end{equation}
which satisfy the orthonormality condition
\begin{equation}
\int_0^{2 \omega} dx\, \psi_m^{(\text{T})}(x) \psi_n^{(\text{T})}(x) = \delta_{mn}.
\end{equation}
We may also replace $n$ by $a_n$ via \eqref{quanttCM} and write
\begin{equation}
\begin{split}
\psi_n^{(\text{T})}(x) = & \left[ \dfrac{2^{\frac{2g_x}{\hbar_x}}}{2\omega} \dfrac{2a_n}{\hbar_x} 
\dfrac{\Gamma(\frac{1}{2}-\frac{g_x}{\hbar_x}+\frac{2a_n}{\hbar_x})\Gamma^2(\frac{1}{2}+\frac{g_x}{\hbar_x})}{\Gamma(\frac{1}{2}+\frac{g_x}{\hbar_x}+\frac{2a_n}{\hbar_x})} \right]^{1/2}
\left[ \sin \frac{\pi x}{2 \omega} \right]^{\frac{1}{2} + \frac{g_x}{\hbar_x}} \\
& \dfrac{\Gamma(\frac{1}{2} + \frac{g_x}{\hbar_x} + \frac{2a_n}{\hbar_x})}{\Gamma(\frac{1}{2} - \frac{g_x}{\hbar_x} + \frac{2a_n}{\hbar_x}) \Gamma(1 + \frac{2g_x}{\hbar_x})}
{}_2 F_1 \left( \frac{1}{4} + \frac{g_x}{2\hbar_x} + \frac{a_n}{\hbar_x}, \frac{1}{4} + \frac{g_x}{2\hbar_x} - \frac{a_n}{\hbar_x}, 1 + \frac{g_x}{\hbar_x}, \sin^2 \frac{\pi x}{2 \omega} \right).
\end{split}
\end{equation}
We have thus completely determined the normalized $L^2([0, 2\omega])$ 2-tCM eigenfunctions \eqref{tCMsolnorm} and its discrete energy levels $E^{(\text{T})}_n = E^{(\text{T})}(a_n)$ \eqref{quanttCM}.

\subsection*{Hyperbolic case} \label{hCMsection}

Let us now briefly discuss the 2-hCM problem
\begin{equation}
\widehat{H}^{(\text{H})} \psi^{(\text{H})}(y) = \left[ -\hbar_y^2 \partial_y^2 + \hbar_y^2 (\frac{g_y^2}{\hbar_y^2} - \frac{1}{4}) \dfrac{\pi^2}{\omega^2} \dfrac{1}{4 \sinh^2 (\frac{\pi y}{2\omega})} \right] \psi^{(\text{H})}(y) = 
\dfrac{\pi^2}{\omega^2} a^2 \psi^{(\text{H})}(y), \label{hCMbis}
\end{equation}
where again we reparametrized the energy $E^{(\text{H})}$ in terms of $a \in \mathbb{R}$ as
\begin{equation}
E^{(\text{H})}(a) = \dfrac{\pi^2}{\omega^2} a^2. \label{enhCM}
\end{equation}
Clearly, the 2-hCM problem can be obtained from the 2-tCM one \eqref{tCMbis} by sending $x \rightarrow i y$, $g_x \rightarrow i g_y$, $\hbar_x \rightarrow i \hbar_y$; therefore, a formal solution to this problem can be recovered from \eqref{tCMsol} by performing the same substitutions, i.e.
\begin{equation}
\psi^{(\text{H})}(y) \propto \left[ \sinh \frac{\pi y}{2 \omega} \right]^{\frac{1}{2} + \frac{g_y}{\hbar_y}}
{}_2 F_1 \left(\frac{1}{4}+\frac{g_y}{2\hbar_y}+\frac{i a}{\hbar_y},\frac{1}{4}+\frac{g_y}{2\hbar_y}-\frac{i a}{\hbar_y},1+\frac{g_y}{\hbar_y},-\sinh^2 \frac{\pi y}{2 \omega}\right) , \;\; y \in (0, \infty). \label{hCMsol}
\end{equation}
Let us stress again that in principle there are two linearly independent solutions to \eqref{hCMbis}, but since we are interested in solutions vanishing at $y=0$ (due to the divergence of the potential) we only consider their particular linear combination \eqref{hCMsol}. Since this combination also vanishes fast enough at $y=+\infty$, \eqref{hCMsol} is actually the complete $y \in (0, \infty)$ solution to the 2-hCM system; given that there is no need to impose further conditions on the solution, the energy \eqref{enhCM} is continuous and positive.

\subsection{Elliptic case: B-model} \label{seceCMB}

We are now ready to consider the 2-eCM$_{\text{B}}$ quantum mechanical problem, defined as
\begin{equation}
\widehat{H}_{\text{B}}\psi^{(\text{B})}(x, Q_{\text{4d}}) = 
\left[- \hbar^2_x \partial_x^2 + \hbar^2_x (\frac{g^2_x}{\hbar^2_x} - \frac{1}{4}) \wp(x \vert \omega, \omega') \right] \psi^{(\text{B})}(x, Q_{\text{4d}}) = E^{(\text{B})} \psi^{(\text{B})}(x, Q_{\text{4d}}), \label{ellCMBter}
\end{equation}
where for convenience we made the explicit dependence on $Q_{\text{4d}}$, that is on $(\omega, \omega')$, of the eigenfunction $\psi^{(\text{B})}(x, Q_{\text{4d}})$ (which of course also depends on all the other parameters of the problem). This problem can be studied either analytically as a perturbation series in $Q_{\text{4d}}$ or numerically, as we are going to illustrate now.

\subsubsection*{Analytical study}

In order to solve the 2-eCM$_{\text{B}}$ problem analytically on $L^2([0,2\omega])$ we can use perturbation theory around $Q_{\text{4d}} = 0$ (or $\omega' \rightarrow i \infty$), that is around the trigonometric limit.\footnote{This is the same approach followed in \cite{Langmann:2004sj}.} We will therefore consider the $Q_{\text{4d}} $-series expansion of the Hamiltonian
\begin{equation}
\begin{split}
\widehat{H}_{\text{B}} &\,=\, - \hbar^2_x \partial_x^2 + \hbar^2_x (\frac{g^2_x}{\hbar^2_x} - \frac{1}{4}) \wp(x \vert \omega, \omega') \\[6 pt]
& \,=\, \widehat{H}_{\text{B}}^{(0)} + Q_{\text{4d}} \widehat{H}_{\text{B}}^{(1)} + Q_{\text{4d}}^2 \widehat{H}_{\text{B}}^{(2)} + Q_{\text{4d}}^3 \widehat{H}_{\text{B}}^{(3)} + O(Q_{\text{4d}}^4),
\end{split}
\end{equation}
where
\begin{equation}
\begin{split}
\widehat{H}_{\text{B}}^{(0)} &= -\hbar_x^2 \partial_x^2 +  (g_x^2 - \frac{\hbar_x^2}{4}) \left[ \dfrac{\pi^2}{\omega^2} \dfrac{1}{4 \sin^2 (\frac{\pi x}{2 \omega})} - \dfrac{\pi^2}{12 \omega^2} \right] \\[4 pt]
& = \widehat{H}^{(\text{T})} - \dfrac{\pi^2}{12 \omega^2}  (g_x^2 - \frac{\hbar_x^2}{4}) , \\[4 pt]
\widehat{H}_{\text{B}}^{(1)} &= (g^2 - \frac{\hbar^2}{4}) \dfrac{4 \pi^2}{\omega^2} \sin^2 \frac{\pi x}{2 \omega} , \\[4 pt]
\widehat{H}_{\text{B}}^{(2)} &= (g^2 - \frac{\hbar^2}{4}) \dfrac{4 \pi^2}{\omega^2} \left[ \sin^2 \frac{\pi x}{2 \omega} + 2\sin^2 \frac{\pi x}{\omega} \right] , \\[4 pt]
\widehat{H}_{\text{B}}^{(3)} &= \ldots .
\end{split}
\end{equation}
The discrete energy levels and eigenfunctions will admit similar expansions:
\begin{equation}
\begin{split}
E^{(\text{B})}_n & \,=\, E^{(0)}_n + Q_{\text{4d}} E^{(1)}_n + Q_{\text{4d}}^2 E^{(2)}_n + Q_{\text{4d}}^3 E^{(3)}_n + O(Q_{\text{4d}}^4) , \\[6 pt]
\psi_n^{(\text{B})}(x, Q_{\text{4d}}) & \,=\, \psi_n^{(0)}(x) + Q_{\text{4d}} \psi_n^{(1)}(x) + Q_{\text{4d}}^2 \psi_n^{(2)}(x) + Q_{\text{4d}}^3 \psi_n^{(3)}(x) + O(Q_{\text{4d}}^4),
\end{split}
\end{equation}
where clearly
\begin{equation}
E^{(0)}_n \,=\, E^{(\text{T})}_n - \dfrac{\pi^2}{12 \omega^2}  (g_x^2 - \frac{\hbar_x^2}{4}) \,=\, 
\frac{\pi^2}{\omega^2} a_n^2 - \dfrac{\pi^2}{12 \omega^2}  (g_x^2 - \frac{\hbar_x^2}{4}) 
\end{equation}
with $a_n$ as in \eqref{quanttCM}, while $\psi_n^{(0)}(x)$ will be the normalized 2-tCM eigenfunction \eqref{tCMsolnorm}
\begin{equation}
\psi_n^{(0)}(x) = \psi_n^{(\text{T})}(x) \equiv \vert n \rangle =\,\eqref{tCMsolnorm}.
\end{equation}
With these considerations, we can now look for a solution to the 2-eCM$_{\text{B}}$ problem
\begin{equation}
\widehat{H}_{\text{B}}\psi^{(\text{B})}(x, Q_{\text{4d}}) =  E^{(\text{B})} \psi^{(\text{B})}(x, Q_{\text{4d}})
\end{equation}
order by order in the $Q_{\text{4d}}$ expansion. This can be done if we further consider expanding all coefficients $ \psi^{(l)}_n(x) \equiv \vert \psi^{(l)}_n \rangle$, $l \geqslant 1$ in terms of the $L^2([0,2\omega])$ 2-tCM orthonormal basis $\psi_n^{(\text{T})}(x) = \psi_n^{(0)}(x) \equiv \vert n \rangle $, i.e.
\begin{equation}
\vert \psi^{(l)}_n \rangle = \sum_{m = 0}^{\infty} c^{(l)}_{nm} \vert m \rangle,
\end{equation}
where the coefficients $c_{nn}^{(l)}$ should be fixed by the normalization condition $\langle \psi^{(l)}_n \vert \psi^{(l)}_n \rangle = 1$.
By doing this we find for example
\begin{equation}
\begin{split}
E^{(0)}_n & = \dfrac{\pi^2}{\omega^2} \left( \frac{(n+\frac{1}{2})\hbar_x + g_x}{2} \right)^2  -\frac{\pi^2}{12 \omega^2} (g_x^2 - \frac{\hbar_x^2}{4}), \\
E^{(1)}_n & = \langle n \vert \widehat{H}^{(1)}_{\text{B}} \vert n \rangle = 
\dfrac{2\pi^2}{\omega^2}(g_x^2 - \frac{\hbar_x^2}{4}) \left[ 1 + 
\dfrac{\frac{g_x^2}{\hbar_x^2} - \frac{1}{4}}{(n+ \frac{g_x}{\hbar_x} + \frac{1}{2} + 1)(n+ \frac{g_x}{\hbar_x} + \frac{1}{2} - 1)} \right], \\
E^{(2)}_n & = \ldots, \label{enCMBanalytic}
\end{split}
\end{equation}
as well as
\begin{equation}
\begin{split}
\psi_n^{(0)}(x) & = \vert n \rangle = \left[ \dfrac{2^{\frac{2g_x}{\hbar_x}}(n+\frac{1}{2}+\frac{g_x}{\hbar_x})n!\Gamma^2(\frac{1}{2}+\frac{g_x}{\hbar_x})}{2 \omega \, \Gamma(n+\frac{2g_x}{\hbar_x}+1)} \right]^{1/2}
\left[ \sin \frac{\pi x}{2 \omega} \right]^{\frac{1}{2} + \frac{g_x}{\hbar_x}}
C_n^{\frac{1}{2} + \frac{g_x}{\hbar_x}}(\cos \frac{\pi x}{2 \omega}) , \\[5 pt]
\psi_n^{(1)}(x) & = \sum_{m \geqslant 0} c^{(1)}_{n m} \vert m \rangle , \\[5 pt]
\psi_n^{(2)}(x) & = \ldots ,
\end{split}
\end{equation}
where $c^{(1)}_{n n} = 0$ for $m = n$, 
while for $m \neq n$
\begin{equation}
c^{(1)}_{n m} = \dfrac{\langle m \vert \widehat{H}^{(1)}_{\text{B}} \vert n \rangle}{E^{(0)}_n - E^{(0)}_m},
\end{equation}
given more explicitly by
\begin{equation}
\begin{split}
c^{(1)}_{n m} \,=\; & (\frac{g_x^2}{\hbar_x^2} - \frac{1}{4})
\dfrac{1}{(n + \frac{g_x}{\hbar_x} + \frac{1}{2} + 1)^2}\sqrt{ \dfrac{(n+1)(n+2)(n+ \frac{2g_x}{\hbar_x} + 1)(n+ \frac{2g_x}{\hbar_x} + 2)}{(n + \frac{g_x}{\hbar_x} + \frac{1}{2})(n + \frac{g_x}{\hbar_x} + \frac{1}{2} + 2)} } \delta_{m, n+2} \\[5 pt]
& -(\frac{g_x^2}{\hbar_x^2} - \frac{1}{4})
\dfrac{1}{(n + \frac{g_x}{\hbar_x} + \frac{1}{2} - 1)^2}\sqrt{ \dfrac{n(n-1)(n+ \frac{2g_x}{\hbar_x})(n+ \frac{2g_x}{\hbar_x} - 1)}{(n + \frac{g_x}{\hbar_x} + \frac{1}{2})(n + \frac{g_x}{\hbar_x} + \frac{1}{2} - 2)} } \delta_{m, n-2}.
\end{split}
\end{equation}
Similar results can be obtained for higher orders in the $Q_{\text{4d}}$ expansion. 

To compute the higher order corrections more efficiently, one can use the idea in \cite{PhysRev.184.1231}.
The perturbative expansion of the wave function takes the following form:
\begin{equation}
\psi_n^{(\text{B})}(x,Q_\text{4d})=\mathcal{N}_n \left[ \sin \frac{\pi x}{2\omega} \right]^{\frac{1}{2}+\frac{g_x}{\hbar_x}} 
\sum_{\ell=0}^{\infty} Q_\text{4d}^\ell P_n^{(\ell)}(\cos \frac{\pi x}{2\omega}),
\end{equation}
where $\mathcal{N}_n$ is an irrelevant normalization constant. The important point is that the functions $P_n^{(\ell)}(z)$ are polynomials in $z$. Of course, we have $P_n^{(0)}(z)=C_n^{\frac{1}{2}+\frac{g_x}{\hbar_x}}(z)$. One can fix unknown coefficients in $P_n^{(\ell)}(z)$ as well as $E^{(\text{B})}$ in order to satisfy the eigenvalue equation.


\subsubsection*{Numerical study}

The 2-eCM$_{\text{B}}$ problem \eqref{ellCMBter} can also be studied numerically. For doing so, we simply have to numerically diagonalize the matrix constructed out of the matrix elements
\begin{equation}
\langle m \vert \widehat{H}_{\text{B}} \vert n \rangle =
\int_{0}^{2\omega} dx \, \psi_m^{(\text{T})}(x) \left[- \hbar^2_x \partial_x^2 + \hbar^2_x (\frac{g^2_x}{\hbar^2_x} - \frac{1}{4}) \wp(x \vert \omega, \omega') \right] \psi_n^{(\text{T})}(x),
\end{equation}
at fixed $g_x > \frac{\hbar_x}{2}$ and $Q_{\text{4d}}$, where we are using the $L^2([0,2\omega])$ 2-tCM orthonormal basis $\psi_n^{(\text{T})}(x) = \vert n \rangle$ \eqref{tCMsolnorm} to perform diagonalization. Clearly, this procedure would in principle require diagonalizing an $\infty \times \infty$ matrix; at the practical level what we do instead is to compute the eigenvalues of a truncated version of this matrix, and then check convergence of the eigenvalues as we increase the matrix size. In this way we can obtain both the numerical spectrum and the numerical eigenfunctions of the 2-eCM$_{\text{B}}$ system.

\subsection{Elliptic case: A-model} \label{seceCMA}

Let us finally comment on the 2-eCM$_{\text{A}}$ quantum mechanical problem
\begin{equation}
\widehat{H}_{\text{A}}\psi^{(\text{A})}(y, Q_{\text{4d}}) = 
\left[- \hbar^2_y \partial_y^2 - \hbar^2_y (\frac{g^2_x}{\hbar^2_y} - \frac{1}{4}) \wp(i y \vert \omega, \omega') \right] \psi^{(\text{A})}(y, Q_{\text{4d}}) = E^{(\text{A})} \psi^{(\text{A})}(y, Q_{\text{4d}}). \label{ellCMAter}
\end{equation}
Trying to solve this problem analytically on $L^2([0, 2\vert\omega'\vert]) = L^2([0, - \frac{\omega}{\pi}\ln Q_{\text{4d}}])$ in terms of perturbation theory around $Q_{\text{4d}} = 0$ (or $\omega' \rightarrow i \infty$) appears to be more complicated than in the 2-eCM$_{\text{B}}$ case: this is because in the $Q_{\text{4d}} = 0$ limit we reduce to the 2-hCM system, which has continuous spectrum and eigenfunctions in $y \in (0, \infty)$ rather than $L^2([0, 2\vert\omega'\vert])$. It should however be possible to do so, by using the quantum Separation of Variables approach which involves the construction of entire solutions to the associated Baxter equation.\footnote{A similar quantum Separation of Variables approach was followed in \cite{Gerasimov:2002cf} to construct the eigenfunction for a special case of the 2-particle hyperbolic Calogero-Moser system.}
We will outline this procedure, reinterpreted in gauge theory language, in Section \ref{eCMeigenfunctions}. \\

For our purposes, it will however be sufficient to study the solution to \eqref{ellCMAter} numerically. As for the case of the B-model, we just have to diagonalize the matrix constructed out of the matrix elements
\begin{equation}
\langle m \vert \widehat{H}_{\text{A}} \vert n \rangle =
\int_{0}^{2\vert \omega' \vert} dy \, \psi_m^{(\text{T})}(y) \left[- \hbar^2_y \partial_y^2 - \hbar^2_y (\frac{g^2_y}{\hbar^2_y} - \frac{1}{4}) \wp(i y \vert \omega, \omega') \right] \psi_n^{(\text{T})}(y),
\end{equation}
at fixed $g_y > \frac{\hbar_y}{2}$ and $Q_{\text{4d}}$, where this time diagonalization is performed by using the $L^2([0,2\vert \omega' \vert])$ orthonormal basis $\psi_n^{(\text{T})}(y) = \vert n \rangle$ with half-period $\vert \omega' \vert = - \frac{\pi}{2\omega} \ln Q_{\text{4d}}$ rather than $\omega$. Clearly, because of \eqref{relationeCM} diagonalizing $\widehat{H}_{\text{A}}$ with half-periods $(\omega, \omega')$ is equivalent to diagonalizing $\widehat{H}_{\text{B}}$ with half-periods $(-i \omega', i\omega)$ and parameters $g_x = g_y$, $\hbar_x = \hbar_y$, that is
\begin{equation}
E^{(\text{A})}_n(g_y, \hbar_y \vert \omega, \omega') =
E^{(\text{B})}_n(g_x, \hbar_x \vert -i\omega', i\omega) \;\;\;\;
\text{for} \;\;\;\; g_x = g_y, \; \hbar_x = \hbar_y. \label{Sdual}
\end{equation}
In fact, as we already mentioned the 2-eCM$_{\text{A}}$ and 2-eCM$_{\text{B}}$ problems are related by S-duality (in $\tau$), i.e. by the exchange of half-periods $\omega, \omega'$, which also implies the exchange
\begin{equation}
Q_{\text{4d}} = e^{2\pi i \frac{\omega'}{\omega}} = e^{2\pi i \tau} \;\;\;\; \longleftrightarrow 
\;\;\;\; Q_{\text{4d}}^{(D)} = e^{2\pi i \frac{i \omega}{-i\omega'}} = e^{2\pi i (-1/\tau)}.
\end{equation}
An alternative option to compute the 2-eCM$_{\text{A}}$ spectrum would therefore be to study the 2-eCM$_{\text{A}}$ problem in terms of perturbation theory around $Q_{\text{4d}}^{(D)} = 0$ (or $\omega \rightarrow \infty$), i.e. around the 2-tCM system of period $2\vert \omega' \vert$, along the lines of what we did in Section \ref{seceCMB}.

\subsection*{A special case with analytic solution} \label{special}

We should also mention that there are special values of the parameter $g$ for which the solution to the 2-eCM$_{\text{B}}$ and 2-eCM$_{\text{A}}$ problems is known analytically. This happens for example when $\frac{g}{\hbar} = \frac{3}{2}$, in which case the generic elliptic problem \eqref{ellCM} reduces to
\begin{equation}
\left[- \partial_z^2 + 2\, \wp(z \vert \omega, \omega') \right] \psi(z, Q_{\text{4d}}) = \frac{E}{\hbar^2} \psi(z, Q_{\text{4d}}). \label{red}
\end{equation}
The solution in this case can be found for example Section 1 of \cite{Olshanetsky:1983wh}: if we define $b$, $k$ as
\begin{equation}
\frac{E}{\hbar^2} = -\wp(b \vert \omega, \omega') \,, \;\;\;\; \xi(b \vert \omega, \omega') = k ,
\end{equation}
then the linear combination
\begin{equation}
\psi(z, Q_{\text{4d}}) = e^{k z} \dfrac{\sigma(z-b \vert \omega, \omega')}{\sigma(z \vert \omega, \omega')} + e^{-k z} \dfrac{\sigma(z+b \vert \omega, \omega')}{\sigma(z \vert \omega, \omega')}
\end{equation}
is a formal solution to \eqref{red} vanishing at $z = 0$. Requiring this to also vanish at $z = 2 \omega$ or $z = 2 \omega'$, i.e. requiring the solution to be in $L^2([0,2\omega])$ or $L^2([0,2\vert\omega\vert'])$, imposes restrictions on the values of the energy, which in turn determine the discrete spectrum of the 2-eCM$_{\text{B}}$ and 2-eCM$_{\text{A}}$ system respectively. More in general, analytic solutions exist for $\frac{g}{\hbar} = \frac{1}{2} + n$ with $n \in \mathbb{N}$, see \cite{Olshanetsky:1983wh,whittaker1996course}.

\subsection{Calogero-Moser energy spectrum from gauge theory} \label{seceCMenergygauge}

\renewcommand{\arraystretch}{1.5}
\begin{table}
\begin{center}
\begin{tabular}{|c|c|c|c|}
\hline 
Gauge theory (NS) & 2-eCM model (complex) & 2-eCM$_{\text{B}}$ model & 2-eCM$_{\text{A}}$ model \\ 
\hline 
$\epsilon_1$ & $\hbar$ & $\hbar_x$ & $i \hbar_y$ \\ 
\hline 
$m$ & $g$ & $g_x$ & $i g_y$ \\ 
\hline 
$Q_{\text{4d}}$ & $e^{2\pi i \frac{\omega'}{\omega}}$ & $e^{2\pi i \frac{\omega'}{\omega}}$ & $e^{2\pi i \frac{\omega'}{\omega}}$ \\ 
\hline 
$ - \frac{i}{2} \frac{\omega}{\pi} \ln Q_{\text{4d}}$ &  $\omega'$ & $\omega'$ & $\omega'$ \\ 
\hline 
\end{tabular} \caption{Map of 4d $SU(2)$ $\mathcal{N} = 2^*$ gauge theory and 2-eCM$_{\text{B}}$, 2-eCM$_{\text{A}}$ parameters.} \label{mapCM}
\end{center}
\end{table}
\renewcommand{\arraystretch}{1}

Up until now we have been discussing how to solve (analytically or numerically) the 2-eCM$_{\text{B}}$ and 2-eCM$_{\text{A}}$ problems, as well as their trigonometric and hyperbolic limit, only relying on standard quantum mechanics techniques. We now want to discuss how the same results can be obtained from gauge theory considerations, at least for what the energy spectrum is concerned.

The idea, first suggested in \cite{2010maph.conf..265N}, is that it should possible to recover the discrete spectrum of both 2-eCM$_{\text{B}}$ and 2-eCM$_{\text{A}}$ (as well as their eigenfunctions) by computing the vacuum expectation value of certain observables in the $\mathcal{N} = 2^*$ $SU(2)$ supersymmetric gauge theory on $\mathbb{R}^4_{\epsilon_1, \epsilon_2}$, in the particular limit (usually referred to as NS limit) in which $\epsilon_2 \rightarrow 0$. 
Here $\epsilon_1$ and $\epsilon_2$ are the Omega background parameters used to regularize the infinite volume of $\mathbb{R}^4$; additional gauge theory parameters are the complex mass $m$ of the adjoint $\mathcal{N} = 2$ hypermultiplet, the complex VEV $a$ of the scalar field in the 
$\mathcal{N} = 2$ vector multiplet, and the instanton counting parameter $Q_{\text{4d}} = e^{2\pi i \tau}$ with $\tau$ complexified gauge coupling. 
The dictionary between gauge theory parameters and 2-eCM$_{\text{B}}$, 2-eCM$_{\text{A}}$ parameters is provided in Table \ref{mapCM}: here $\omega$ should be thought as an overall scale,
while as we will see $a$ is just another way of parametrizing the energy (similarly to \eqref{entCM} and \eqref{enhCM}) which appears to be more natural from the gauge theory point of view.


Let us illustrate the proposal by \cite{2010maph.conf..265N} (and subsequent works) in some more detail. The idea is to proceed in two steps:

\subsection*{Step I: general analysis}


First, we would need to construct two (generically) linearly independent formal solutions 
 $\psi^{(i)}(z, g, \hbar, a, \omega, \omega')$, $i = 1,2$ to the \textit{complex} 2-particle elliptic Calogero-Moser system \eqref{ellCM}, such that
\begin{equation}
\left[- \partial_z^2 + (\frac{g^2}{\hbar^2} - \frac{1}{4}) \wp(z\vert \omega, \omega') \right] \psi^{(i)}(z, a, \hbar, g, \omega, \omega') 
= \frac{E(a, \hbar, g,\omega, \omega')}{\hbar^2} \psi^{(i)}(z, a, \hbar, g, \omega, \omega') \label{eCMcomplex}
\end{equation}
for any generic $z, a, \hbar, g \in \mathbb{C}$. Gauge theory provides explicit, convergent $Q_{\text{4d}}$-series expressions both for the energy and the formal solutions as functions of $a$; these in fact coincide with the NS limit of VEVs of certain codimension four and two defects, if we identify $\hbar = \epsilon_1$ and $g = m$ as in Table \ref{mapCM}:
\begin{equation}
\begin{array}{ccc}
E(a, \hbar, g, \omega, \omega') &\;\;\;\; \underset{\hbar = \epsilon_1, \; g = m}{\Longleftrightarrow} \;\;\;\; & \text{codim. 4 defect} \; \langle \text{Tr}\, \phi^2(a, \epsilon_1, m, Q_{\text{4d}}) \rangle_{\text{NS}}, \\[6 pt] 
\psi^{(i)}(z, a, \hbar, g, \omega, \omega') & \;\;\;\; \underset{\hbar = \epsilon_1, \; g = m}{\Longleftrightarrow} \;\;\;\; & \text{codim. 2 defect (monodromy, NS)}. \label{generalCM}
\end{array} 
\end{equation}
The codimension two defects of interest are monodromy (or Gukov-Witten) defects \cite{2006hep.th...12073G,Alday:2010vg,Kanno:2011fw}, whose interpretation as formal solutions to \eqref{eCMcomplex} has been discussed in a number of papers \cite{Alday:2010vg,Maruyoshi:2010iu}; since we will not need them for our purposes, we refer to these works for further details.\footnote{Although most works focus on checking that these monodromy defects satisfy \eqref{eCMcomplex} order by order in $Q_{\text{4d}}$, this can actually be proven at all orders via qq-characters technology \cite{Nekrasov:2015wsu,Nekrasov:2016qym,Nekrasov:2016ydq,Nekrasov:2017rqy,Nekrasov:2017gzb}.} 
The codimension four defect corresponding to the energy is instead a local observable which basically coincides with $\langle \text{Tr}\phi^2 \rangle$, where $\phi$ is the scalar field in the $\mathcal{N} = 2$ vector multiplet; explicitly, the gauge theory expression for the energy reads
\begin{equation}
\begin{split}
E(a, \hbar, g, \omega, \omega') \,=\, \frac{\pi^2}{\omega^2} \Big[ & a^2 - \dfrac{1}{12} (m^2 - \frac{\epsilon_1^2}{4})  
+ (m^2 - \frac{\epsilon_1^2}{4}) \dfrac{1 - E_2(Q_{\text{4d}})}{6} \\[5 pt]
& + Q_{\text{4d}} \partial_{Q_{\text{4d}}} \mathcal{W}^{SU(2)}_{\text{4d,\,inst}} (a,\epsilon_1,m,Q_{\text{4d}}) \Big] \Big\vert_{\substack{\epsilon_1 = \hbar \\ m = g}} \,, \label{enCMgauge}
\end{split}
\end{equation}
where $\mathcal{W}^{SU(2)}_{\text{4d,\,inst}}$ is the instanton part of the twisted effective superpotential for the $SU(2)$ $\mathcal{N} = 2^*$ theory (with $a_1 = -a_2 = a$) defined as
\begin{equation}
\mathcal{W}^{SU(2)}_{\text{4d,\,inst}}(a,\epsilon_1,m,Q_{\text{4d}}) 
= \lim_{\epsilon_2 \to 0} \left[ -\epsilon_1 \epsilon_2 \ln Z^{SU(2)}_{\text{4d,\,inst}}(a,\epsilon_1,\epsilon_2,m,Q_{\text{4d}})  \right],
\end{equation}
with $Z^{SU(2)}_{\text{4d,\,inst}}$ as in Appendix \ref{appgauge}. 
The function $E_2(Q)$ is the quasi-modular form, defined in \eqref{eq:E2}. For the reader's convenience, we report here the first few terms in the $Q_{\text{4d}}$ expansion: 
\begin{equation}
\begin{split}
& \mathcal{W}^{SU(2)}_{\text{4d,\,inst}}(a,\epsilon_1,m,Q_{\text{4d}})  = - Q_{\text{4d}} (m^2 - \frac{\epsilon_1^2}{4})\dfrac{16a^2 - 4m^2 - 3\epsilon_1^2}{2(4a^2 - \epsilon_1^2)}  
- Q_{\text{4d}}^2 \dfrac{(m^2 - \frac{\epsilon_1^2}{4})}{256(a^2 - \epsilon_1^2)(4a^2 - \epsilon_1^2)^3} \times \\
& \Big[ 631 \epsilon _1^8 + 300 m^2 \epsilon _1^6 + 1104 m^4 \epsilon _1^4 - 448 m^6 \epsilon _1^2 - a^2 \left(1280 m^6 + 5184 m^4 \epsilon _1^2 + 4848 m^2 \epsilon _1^4 + 8428 \epsilon _1^6\right) \\
& + a^4 \left(12288 m^4+18432 m^2 \epsilon _1^2+40704 \epsilon _1^4\right)
- a^6 \left(24576 m^2 + 79872 \epsilon _1^2\right) + 49152 a^8 \Big] + O(Q_{\text{4d}}^3).
\end{split}
\end{equation}

\renewcommand{\arraystretch}{1.5}
\begin{table}
\begin{center}
\begin{tabular}{|c|c|c|}
\hline 
$\omega = \pi$, $\;\omega' = - \frac{2\pi^2 i}{\ln(0.0025)}$ & $E^{(\text{B})}_{0}/\hbar^2_x$ & $E^{(\text{B})}_{1}/\hbar^2_x$ \\ 
\hline 
Gauge theory - $O(Q_{\text{4d}}^2)$ & \underline{0.8425}472339439172\ldots & \underline{2.0902}551277519722\ldots \\ 
\hline 
Gauge theory - $O(Q_{\text{4d}}^4)$ & \underline{0.842525630}9667935\ldots & \underline{2.090240423}9427707\ldots \\ 
\hline 
Gauge theory - $O(Q_{\text{4d}}^6)$ & \underline{0.84252563084654}72\ldots & \underline{2.090240423863342}8\ldots
\\ 
\hline 
Analytic & \underline{0.8425256308465468}\ldots & \underline{2.0902404238633425}\ldots \\ 
\hline 
Numerical & \underline{0.8425256308465468}\ldots & \underline{2.0902404238633425}\ldots \\ 
\hline 
\end{tabular} \caption{Energies for the 2-eCM$_{\text{B}}$ model at $g_x = \frac{3}{2} \hbar_x$.} \label{ex1tab}
\end{center}
\end{table}
\renewcommand{\arraystretch}{1}

\renewcommand{\arraystretch}{1.5}
\begin{table}
\begin{center}
\begin{tabular}{|c|c|c|}
\hline 
$\omega = \frac{5}{2}$, $\; \omega' = i \pi$ & $E^{(\text{B})}_{0}/\hbar^2_x$ & $E^{(\text{B})}_{1}/\hbar^2_x$ \\ 
\hline 
Gauge theory - $O(Q_{\text{4d}})$ & \underline{2.3}4149047707890\ldots & \underline{4.89}658733731540\ldots
\\ 
\hline 
Gauge theory - $O(Q_{\text{4d}}^2)$ & \underline{2.3399}9535457005\ldots & \underline{4.89304}903548720\ldots
\\ 
\hline  
Gauge theory - $O(Q_{\text{4d}}^3)$ & \underline{2.33998900}362183\ldots & \underline{4.89304471}502009\ldots \\ 
\hline
Gauge theory - $O(Q_{\text{4d}}^4)$ & \underline{2.33998900200}670\ldots & \underline{4.8930447117}4149\ldots
\\ 
\hline 
Numerical & \underline{2.33998900200391}\ldots & \underline{4.89304471173978}\ldots \\ 
\hline 
\end{tabular} \caption{Energies for the 2-eCM$_{\text{B}}$ model at $g_x = \sqrt{5} \hbar_x$.} \label{ex2tab}
\end{center}
\end{table}
\renewcommand{\arraystretch}{1}

\subsection*{Step II: Hilbert integrability}

Once we have determined the formal solution to the generic complex 2-eCM system \eqref{eCMcomplex} from gauge theory, we can study Hilbert integrability: that is, we want to understand if this formal eigenfunction can belong to the Hilbert space of a quantum mechanical system with self-adjoint Hamiltonian, for example 2-eCM$_{\text{B}}$ or 2-eCM$_{\text{A}}$. Usually, for generic values of $a$ (or, equivalently, the energy) this is not the case; however, the conjecture by \cite{2010maph.conf..265N} states that Hilbert integrability should emerge in the following two special situations:


\paragraph{2-eCM$_{\text{B}}$ model:}
Fixing $\epsilon_1 = \hbar_x \in \mathbb{R}_+$ and $m = g_x \in \mathbb{R}_+$, if $a$ satisfies the B-type quantization conditions
\begin{equation}
a = \dfrac{g_x + (n + \frac{1}{2}) \hbar_x}{2} , \;\;\;\; n \in \mathbb{N}, \label{quantCMB}
\end{equation}
our gauge theory expression \eqref{enCMgauge} is conjecturally expected to produce the discrete energy levels for the 2-eCM$_{\text{B}}$ model with Hilbert space $L^2([0,2\omega])$. This can be checked in various ways. First of all, it indeed reproduces the $Q_{\text{4d}}$-series analytic expression for the energy we computed earlier in \eqref{enCMBanalytic}. Moreover, it is also in agreement with the energy spectrum computed numerically (or exactly for $g_x = \frac{3}{2} \hbar_x$):
 in fact, as we can see from Tables \ref{ex1tab} and \ref{ex2tab} the gauge theory results approach the numerical/analytical ones better and better when considering more and more instanton corrections. 
Finally, it correctly reduces to the known 2-tCM energy spectrum in the trigonometric limit $Q_{\text{4d}} \rightarrow 0$.
It is also interesting to notice that the B-type quantization conditions for 2-eCM$_{\text{B}}$ \eqref{quantCMB} actually coincide with the quantization conditions for 2-tCM \eqref{quanttCM}, i.e. they do not receive corrections in $Q_{\text{4d}}$. 

\renewcommand{\arraystretch}{1.5}
\begin{table}
\begin{center}
\begin{tabular}{|c|c|c|}
\hline 
$\omega = - \frac{2\pi^2}{\ln(0.0025)}$, $\; \omega' = i \pi$ & $E^{(\text{A})}_{0}/\hbar^2_y$ & $E^{(\text{A})}_{1}/\hbar^2_y$ \\ 
\hline 
Gauge theory - $O(Q_{\text{4d}}^3)$ & \underline{0.84252}45641197144\ldots & \underline{2.0902}399533002937\ldots \\ 
\hline 
Gauge theory - $O(Q_{\text{4d}}^6)$ & \underline{0.842525630846}4403\ldots & \underline{2.090240423863}1993\ldots
\\ 
\hline 
Gauge theory - $O(Q_{\text{4d}}^9)$ & \underline{0.842525630846546}9\ldots & \underline{2.09024042386334}19\ldots \\ 
\hline 
Analytic & \underline{0.8425256308465468}\ldots & \underline{2.0902404238633425}\ldots \\ 
\hline 
Numerical & \underline{0.8425256308465468}\ldots & \underline{2.0902404238633425}\ldots \\ 
\hline 
\end{tabular} \caption{Energies for the 2-eCM$_{\text{A}}$ model at $g_y = \frac{3}{2} \hbar_y$. The spectrum is the same as the one in Table~\ref{ex1tab}.} \label{ex3tab}
\end{center}
\end{table}
\renewcommand{\arraystretch}{1}

\renewcommand{\arraystretch}{1.5}
\begin{table}
\begin{center}
\begin{tabular}{|c|c|c|}
\hline 
$\omega = \pi$, $\; \omega' = \frac{5 i}{2}$ & $E^{(\text{A})}_{0}/\hbar^2_y$ & $E^{(\text{A})}_{1}/\hbar^2_y$ \\ 
\hline 
Gauge theory - $O(Q_{\text{4d}}^3)$ & \underline{2.339}89242414480\ldots & \underline{4.8930}1216734932\ldots \\ 
\hline 
Gauge theory - $O(Q_{\text{4d}}^6)$ & \underline{2.33998}899746720\ldots & \underline{4.893044711}08014\ldots
\\ 
\hline 
Gauge theory - $O(Q_{\text{4d}}^9)$ & \underline{2.33998900200}500\ldots & \underline{4.8930447117397}6\ldots \\ 
\hline 
Numerical & \underline{2.33998900200391}\ldots & \underline{4.89304471173978}\ldots \\ 
\hline 
\end{tabular} \caption{Energies for the 2-eCM$_{\text{A}}$ model at $g_y = \sqrt{5} \hbar_y$. The spectrum is the same as the one in Table~\ref{ex2tab}.} \label{ex4tab}
\end{center}
\end{table}
\renewcommand{\arraystretch}{1}

\paragraph{2-eCM$_{\text{A}}$ model:} Fixing $\epsilon_1 = i \hbar_y$ and $m = i g_y$ with $\hbar_y, g_y \in \mathbb{R}_+$, if $a$ satisfies the A-type quantization conditions
\begin{equation}
\begin{split}
\dfrac{a_D}{\hbar_y} \,=\, & -\dfrac{2a}{\hbar_y} \ln Q_{\text{4d}} 
-2i \ln \dfrac{\Gamma(\frac{2 i a}{\hbar_y})\Gamma(-\frac{2ia}{\hbar_y} + \frac{g_y}{\hbar_y} + \frac{1}{2})}{\Gamma(-\frac{2 i a}{\hbar_y})\Gamma(\frac{2ia}{\hbar_y} + \frac{g_y}{\hbar_y} + \frac{1}{2}) } \\[5 pt]
& -\dfrac{1}{\hbar_y} \partial_a \mathcal{W}^{SU(2)}_{\text{4d,\,inst}}(a, i \hbar_y, i g_y, Q_{\text{4d}}) = 2\pi n \, , \;\;\;\; n \in \mathbb{N}, \label{quantCMA}
\end{split}
\end{equation}
our gauge theory expression \eqref{enCMgauge} is conjecturally expected to produce the discrete energy levels for the 2-eCM$_{\text{A}}$ model with Hilbert space $L^2([0,2\vert\omega'\vert])$. In fact, as we can see from the examples in Tables \ref{ex3tab} and \ref{ex4tab} the gauge theory results seem to be in agreement with the energy spectrum computed numerically (or exactly for $g_y = \frac{3}{2} \hbar_y$). A-type quantization conditions \eqref{quantCMA} can be thought as an S-dual (electro-magnetic dual) version of the B-type conditions \eqref{quantCMB}, where we quantize $a_D$ rather than $a$. 
These two variables are related by
\begin{equation}
a_D = - \partial_a \mathcal{W}^{SU(2)}_{\text{4d}}(a, i \hbar_y, i g_y, Q_{\text{4d}}) 
, \label{finale}
\end{equation}
where $\mathcal{W}^{SU(2)}_{\text{4d}}$ is the full twisted effective superpotential containing both the instanton and the classical + 1-loop part. This is nothing else but the $\epsilon_1$-deformed version of the Seiberg-Witten relation between $a$ and $a_D$, in which the Seiberg-Witten prepotential is replaced by $\mathcal{W}^{SU(2)}_{\text{4d}}$. We will later see in Section \ref{eCMeigenfunctions} how A-type quantization conditions may arise from the quantum Separation of Variables approach to the 2-eCM$_{\text{A}}$ problem.

To sum up, the conjecture by \cite{2010maph.conf..265N} seems to work nicely in the 2-eCM case:
the discrete energy levels for the two quantum mechanical systems 2-eCM$_{\text{B}}$ and 2-eCM$_{\text{A}}$ can indeed be obtained from the \textit{same} expression for the energy \eqref{enCMgauge} by quantizing $a$ in two different ways, related by S-duality/electro-magnetic duality $\tau \leftrightarrow - \frac{1}{\tau}$. S-duality is also reflected in the relations \eqref{relationeCM} and \eqref{Sdual}, whose realization in gauge theory is evident from comparing Tables \ref{ex1tab}, \ref{ex3tab} or Tables \ref{ex2tab}, \ref{ex4tab}.


\subsection{Comments on Baxter equation and quantum Separation of Variables} 
\label{eCMeigenfunctions}

As we mentioned in Section \ref{seceCMenergygauge}, formal solutions to the complex 2-eCM problem can be computed in gauge theory by considering codimension two defects of monodromy type in four-dimensional $\mathcal{N} = 2^*$ $SU(2)$ theory. Conjecturally, these formal solutions should reduce to true eigenfunctions of the 2-eCM$_{\text{B}}$ and 2-eCM$_{\text{A}}$ systems when imposing B-type \eqref{quantCMB} or A-type \eqref{quantCMA} quantization conditions, and the resulting B-type and A-type eigenfunctions should be related by S-duality $\tau \leftrightarrow - \frac{1}{\tau}$.

In this Section we would like to elaborate, without any pretence of rigorousness, about a different way of constructing eigenfunctions for the 2-eCM$_{\text{A}}$ system, based on a gauge theory reinterpretation of the quantum Separation of Variables approach. As we remarked in Section \ref{seceCMA}, it is hard to study 2-eCM$_{\text{A}}$ eigenfunctions analytically in terms of perturbation theory around $Q_{\text{4d}} = 0$, i.e. around the hyperbolic limit, since the 2-eCM$_{\text{A}}$ and 2-hCM problems have different Hilbert spaces and in addition 2-hCM eigenfunctions are not bound states. Quantum Separation of Variables was developed as a general framework to overcome these and other similar difficulties, and it was successfully applied to the $N$-particle Toda chain (open and closed) \cite{Kharchev:1999bh,Kharchev:2000ug,Kharchev:2000yj}, the $N$-particle hyperbolic Calogero-Moser system \cite{Gerasimov:2002cf}, as well as the relativistic version of the open Toda chain \cite{Kharchev:2001rs} and the hyperbolic Ruijsenaars-Schneider system \cite{2012arXiv1206.3787H,2016arXiv160706672H}.\footnote{More recently \cite{Sciarappa:2017hds,2018arXiv180306196B,2018arXiv180401749B} quantum Separation of Variables was also applied to the relativistic closed Toda chain, based on preliminary discussion in \cite{Kharchev:2001rs} and gauge theory intuition.}

Roughly, the idea of the quantum Separation of Variables approach is to construct the $N$-particle elliptic Calogero-Moser (of type A) eigenfunction 
by means of a particular integral transformation which involves the eigenfunction of the $N-1$-particle hyperbolic Calogero-Moser system (with center of mass not decoupled) as well as an appropriate entire, fast-decaying solution to an \textit{auxiliary} 1-dimensional problem known as Baxter equation, which can also be thought of as a quantized version of the spectral curve of the  classical integrable system. Since the 1-particle hCM system is just a free system, its eigenfunction will be a plane wave, and as a consequence the 2-eCM$_{\text{A}}$ eigenfunction will simply be obtained from the solution to the Baxter equation via Fourier transformation.

Let us try to be more quantitative. First of all we need to consider the Baxter equation associated to the 2-eCM$_{\text{A}}$ system: as we said, this arises from quantization of the classical spectral curve, which for the model at hand was constructed in \cite{kri}. From the gauge theory point of view, the 2-eCM$_{\text{A}}$ spectral curve is nothing but the Seiberg-Witten curve for the four-dimensional $\mathcal{N} = 2^*$ $SU(2)$ theory as originally studied in \cite{Donagi:1995cf,DHoker:1997hut}. It will however be more useful for our purposes to consider the novel approach by \cite{Nekrasov:2012xe,Nekrasov:2013xda} to Seiberg-Witten curves and their quantum version. In this approach, the Seiberg-Witten curve for a theory can be constructed from the VEV of a particular observable $\langle \chi_{\text{4d}}(\sigma,\vec{a}, \epsilon_1, \epsilon_2, m, Q_{\text{4d}}) \rangle$, usually known as (fundamental) qq-character \cite{Nekrasov:2015wsu}, in the ``classical'' or Seiberg-Witten limit $\epsilon_{1,2} \rightarrow 0$, while the quantum version of the curve (or Baxter equation in the integrable system language) is constructed out of the NS limit $\epsilon_2 \rightarrow 0$ of the qq-character.
The NS limit $\langle \chi_{\text{4d}}^{SU(2)}(\sigma, a, \epsilon_1, m, Q_{\text{4d}} ) \rangle_{\text{NS}}$ of the $\mathcal{N} = 2^*$ $SU(2)$ qq-character can be explicitly  computed in gauge theory with the formulae given in Appendix \ref{appgauge}: this is a polynomial in an auxiliary variable $\sigma$ of degree $N = 2$ given by (for $a_1 = -a_2 = a$) 
\begin{equation}
\Big\langle \chi_{\text{4d}}^{SU(2)}(\sigma, a, \epsilon_1, m, Q_{\text{4d}} ) \Big\rangle_{\text{NS}}
= \sigma^2 - E'(a, \epsilon_1, m, Q_{\text{4d}}), \label{qqSU24d}
\end{equation}
where\footnote{Let us remark that $Q_{\text{4d}} \partial_{Q_{\text{4d}}} \mathcal{W}_{\text{4d,\,inst}}^{U(1)}(\epsilon_1, m, Q_{\text{4d}}) = -\left( m^2 - \frac{\epsilon_1^2}{4} \right) \frac{1 - E_2(Q_{\text{4d}})}{24}$ for the $\mathcal{N} = 2^*$ $U(1)$ theory.}
\begin{equation}
E'(a, \epsilon_1, m, Q_{\text{4d}}) = a^2 + Q_{\text{4d}} \partial_{Q_{\text{4d}}} \mathcal{W}_{\text{4d,\,inst}}^{SU(2)}(a, \epsilon_1, m, Q_{\text{4d}}) 
+ (m^2 - \frac{\epsilon_1^2}{4} ) \frac{1 - E_2(Q_{\text{4d}})}{12}. \label{E2}
\end{equation}
As we can see, $E'(a, \epsilon_1, m, Q_{\text{4d}})$ is closely related to the gauge theory expression for the complex 2-eCM system \eqref{enCMgauge}: 
\begin{equation}
E(a, \hbar, g, \omega, \omega') = \frac{\pi^2}{\omega^2} \bigg[ E'(a, \epsilon_1, m, Q_{\text{4d}}) -\frac{1}{12} (m^2 - \frac{\epsilon_1^2}{4}) E_2(Q_{\text{4d}}) \bigg] 
\bigg\vert_{\substack{ \epsilon_1 = \hbar \\ m = g }}.
\end{equation}
From \eqref{qqSU24d}, by following \cite{Nekrasov:2013xda}, specializing to $\epsilon_1 = i \hbar_y$, $m = i g_y$ as from Table \ref{mapCM} and restricting to $\sigma \in \mathbb{R}$, we obtain the 2-eCM$_{\text{A}}$ Baxter equation or 4d $\mathcal{N} = 2^*$ $SU(2)$ quantum Seiberg-Witten curve 
\begin{equation}
\sum_{n \in \mathbb{Z}} (-1)^n Q_{\text{4d}}^{\frac{n(n-1)}{2}} \Big\langle \chi_{\text{4d}}^{SU(2)}\left(\sigma + i n (g_y + \frac{\hbar_y}{2}), a, i \hbar_y, i g_y, Q_{\text{4d}}\right) \Big\rangle_{\text{NS}} e^{i n \hbar_y \partial_{\sigma}} = 0. \label{bax}
\end{equation}
As we can see, our Baxter equation is a rather complicated infinite-order finite-difference operator on $\mathbb{R}$. Two (generically) linearly independent functions $\mathcal{Q}^{(f)}(\sigma, a, \hbar_y, g_y, Q_{\text{4d}})$, $\mathcal{Q}^{(af)}(\sigma, a, \hbar_y, g_y, Q_{\text{4d}})$ formally satisfying 
\eqref{bax} can be obtained by considering the NS limit of the VEVs of other codimension two defects, different from the monodromy one considered in \eqref{generalCM}: these correspond to coupling the four-dimensional theory to two free ``fundamental'' $(f)$ or ``antifundamental'' $(af)$ two-dimensional $\mathcal{N} = (2,2)^*$ hypermultiplets,\footnote{These are hypermultiplet in the two-dimensional $\mathcal{N} = (4,4)$ sense; ``fundamental'' or ``antifundamental'' is related to their charge under a $U(1)_{\sigma}$ flavour symmetry which is not gauged.} whose $SU(2)$ part of the flavour symmetry is gauged and coincides with the $SU(2)$ gauge symmetry of the four-dimensional theory under exam \cite{Gaiotto:2014ina,Bullimore:2014awa,Gaiotto:2015una}.
Schematically, these two functions can be written as
\begin{equation}
\begin{split}
\mathcal{Q}^{(f)}(\sigma, a, \hbar_y, g_y, Q_{\text{4d}}) & \;=\; 
\mathcal{Q}^{(f)}_{0}(\sigma, a, \hbar_y, g_y) 
\big\langle \mathcal{Q}^{(f)}_{\text{4d,inst}}(\sigma, a, i \hbar_y, i g_y, Q_{\text{4d}}) \big\rangle_{\text{NS}} \;, \\
\mathcal{Q}^{(af)}(\sigma, a, \hbar_y, g_y, Q_{\text{4d}}) & \;=\; 
Q_{\text{4d}}^{-\frac{i \sigma}{\hbar}}\mathcal{Q}^{(af)}_{0}(\sigma, a, \hbar_y, g_y) 
\big\langle \mathcal{Q}^{(af)}_{\text{4d,inst}}(\sigma, a, i \hbar_y, i g_y, Q_{\text{4d}}) \big\rangle_{\text{NS}} \;. \label{bbb}
\end{split}
\end{equation}
Here 
\begin{equation}
\begin{split}
& \mathcal{Q}^{(f)}_{0}(\sigma, a, \hbar_y, g_y)  = \dfrac{\Gamma(-i\frac{\sigma - a}{\hbar_y}) 
\Gamma(-i\frac{\sigma + a}{\hbar_y})}{
\Gamma(-i\frac{\sigma - a + i g_y + i \hbar_y/2}{\hbar_y})
\Gamma(-i\frac{\sigma + a + i g_y + i \hbar_y/2}{\hbar_y})}, \\
& \mathcal{Q}^{(af)}_{0}(\sigma, a, \hbar_y, g_y)  = \dfrac{\Gamma(i\frac{\sigma - a}{\hbar_y}) 
\Gamma(i\frac{\sigma + a}{\hbar_y})}{
\Gamma(i\frac{\sigma - a - i g_y - i \hbar_y/2}{\hbar_y})
\Gamma(i\frac{\sigma + a - i g_y - i \hbar_y/2}{\hbar_y})}, \label{q0}
\end{split}
\end{equation}
while $\langle \mathcal{Q}^{(f)}_{\text{4d,inst}}(\sigma, a, \hbar_y, g_y, Q_{\text{4d}}) \rangle_{\text{NS}}$, 
$\langle \mathcal{Q}^{(af)}_{\text{4d,inst}}(\sigma, a, \hbar_y, g_y, Q_{\text{4d}}) \rangle_{\text{NS}}$ contain the $Q_{\text{4d}}$ instanton corrections: these can be computed with the formulae collected in Appendix \ref{appgauge}, however already at one instanton the result is too lengthy to be reported here and we will therefore just mention that
\begin{equation}
\big \langle \mathcal{Q}^{(af)}_{\text{4d,inst}}(\sigma, a, \hbar_y, g_y, Q_{\text{4d}}) \big\rangle_{\text{NS}} = \big\langle \mathcal{Q}^{(f)}_{\text{4d,inst}}(\sigma, a, -\hbar_y, -g_y, Q_{\text{4d}}) \big\rangle_{\text{NS}} \,.
\end{equation}
Clearly, chosen any constant $\xi \in \mathbb{C}$ the linear combination
\begin{equation}
\mathcal{Q}(\sigma, a, \hbar_y, g_y, Q_{\text{4d}}) \; \propto \;
\mathcal{Q}^{(f)}(\sigma, a, \hbar_y, g_y, Q_{\text{4d}}) 
- \xi^{-1} \mathcal{Q}^{(af)}(\sigma, a, \hbar_y, g_y, Q_{\text{4d}}) \label{linear}
\end{equation}
formally satisfies \eqref{bax}, that is
\begin{equation}
\left[ \sum_{n \in \mathbb{Z}} (-1)^n Q_{\text{4d}}^{\frac{n(n-1)}{2}} \big\langle \chi_{\text{4d}}^{SU(2)} \left(\sigma + i n (g_y + \frac{\hbar_y}{2}), a, i \hbar_y, i g_y, Q_{\text{4d}}\right) \big\rangle e^{i n \hbar_y \partial_{\sigma}} \right] \mathcal{Q}(\sigma, a, \hbar_y, g_y, Q_{\text{4d}}) = 0. \label{bax2}
\end{equation}
It is important at this moment to point out that both functions $\mathcal{Q}^{(f)}(\sigma, a, \hbar_y, g_y, Q_{\text{4d}})$, $\mathcal{Q}^{(af)}(\sigma, a, \hbar_y, g_y, Q_{\text{4d}})$ are not entire in $\sigma$, but seem to have two infinite sets of simple poles at $\sigma = \pm a + i n \hbar_y$, $n \in \mathbb{Z}$; for example, $\mathcal{Q}^{(f)}(\sigma, a, \hbar_y, g_y Q_{\text{4d}})$ has poles at $\sigma = \pm a - i n \hbar_y$, $n \geqslant 0$ coming from the Gamma functions in \eqref{q0} and poles at $\sigma = \pm a + i n \hbar_y$, $n \geqslant 1$ coming from the instanton corrections,\footnote{We do not have a proof of this fact: it just comes as an observation from explicit computation of the first few instanton corrections. In addition, instanton corrections for, say, $\mathcal{Q}^{(f)}(\sigma, a, \hbar_y, g_y, Q_{\text{4d}})$ seem to have additional simple poles at $\sigma = \pm a - i g_y - i \frac{\hbar_y}{2} - i n \hbar_y$, $n \geqslant 0$; these are however cancelled by zeroes coming from the Gamma functions in \eqref{q0}.} and viceversa for $\mathcal{Q}^{(af)}(\sigma, a, \hbar_y, g_y, Q_{\text{4d}})$. 

What is interesting is that all these simple poles seem to cancel each other in the linear combination \eqref{linear} at those very special values of $\xi$ and $a$ satisfying the A-type quantization conditions \eqref{quantCMA}. More precisely, if $\xi$ and $a$ are such that
\begin{equation}
\xi = \dfrac{
\underset{\sigma = a}{\text{Res}} \left[\mathcal{Q}^{(af)}(\sigma, a, \hbar_y, g_y, Q_{\text{4d}}) \right]}{\underset{\sigma = a}{\text{Res}} 
\left[ \mathcal{Q}^{(f)}(\sigma, a, \hbar_y, g_y, Q_{\text{4d}}) \right]} 
= \dfrac{\underset{\sigma = - a}{\text{Res}} 
\left[\mathcal{Q}^{(af)}(\sigma, a, \hbar_y, g_y, Q_{\text{4d}}) \right]}{\underset{\sigma = - a}{\text{Res}} 
\left[ \mathcal{Q}^{(f)}(\sigma, a, \hbar_y, g_y, Q_{\text{4d}}) \right]} , \label{quant4d}
\end{equation}
which appears to be equivalent to the A-type quantization conditions
\begin{equation}
-\dfrac{2a}{\hbar_y} \ln Q_{\text{4d}} 
-2i \ln \dfrac{\Gamma(\frac{2 i a}{\hbar_y})\Gamma(-i\frac{2a + i g_y + i \hbar_y/2}{\hbar_y})}{\Gamma(-\frac{2 i a}{\hbar_y})\Gamma(i\frac{2a - i g_y - i \hbar_y/2}{\hbar_y}) } -\dfrac{1}{\hbar_y} \partial_a \mathcal{W}^{SU(2)}_{\text{4d,\,inst}}(a, i \hbar_y, i g_y, Q_{\text{4d}}) = 2\pi n \, , \;\; n \in \mathbb{N}
\end{equation}
due to the fact that
\begin{equation}
-2i \ln \left[ \dfrac{\big\langle\mathcal{Q}^{(af)}_{\text{4d,inst}}(\sigma = \pm a, a, i \hbar_y, i g_y, Q_{\text{4d}})\big\rangle_{\text{NS}}}{\big\langle\mathcal{Q}^{(f)}_{\text{4d,inst}}(\sigma = \pm a, a, i\hbar_y, i g_y, Q_{\text{4d}})\big\rangle_{\text{NS}}} \right] = - \dfrac{1}{\hbar} \partial_a \mathcal{W}^{SU(2)}_{\text{4d,\,inst}}(a, i \hbar_y, i g_y, Q_{\text{4d}}),
\end{equation}
then \eqref{linear} is regular at $\sigma = \pm a$; what is more, this seems to also imply that for any $n \in \mathbb{Z}$
\begin{equation}
\xi = \dfrac{
\underset{\sigma = a + i n \hbar_y}{\text{Res}} \left[\mathcal{Q}^{(af)}(\sigma, a, \hbar_y, g_y, Q_{\text{4d}}) \right]}{\underset{\sigma = a + i n \hbar_y}{\text{Res}} 
\left[ \mathcal{Q}^{(f)}(\sigma, a, \hbar_y, g_y, Q_{\text{4d}}) \right]} 
= \dfrac{\underset{\sigma = - a + i n \hbar_y}{\text{Res}} 
\left[\mathcal{Q}^{(af)}(\sigma, a, \hbar_y, g_y, Q_{\text{4d}}) \right]}{\underset{\sigma = - a + i n \hbar_y}{\text{Res}} 
\left[ \mathcal{Q}^{(f)}(\sigma, a, \hbar_y, g_y, Q_{\text{4d}}) \right]}  .
\end{equation}
This would then imply that the linear combination \eqref{linear} is free of poles and entire in $\sigma$ for these particular values of $\xi$ and $a$ satisfying the A-type quantization conditions \eqref{quant4d} (at the end we would expect $\xi = -(-1)^n$ for some positive integer $n$ because of the symmetry of our problem). 
Although we cannot make any rigorous statement due to the lack of information about the analytic structure of the instanton corrections in \eqref{bbb}, it is still quite remarkable to find that our gauge theory analysis of the 2-eCM$_{\text{A}}$ Baxter equation involves exactly the same requirements and mechanisms at work in the quantum Separation of Variables treatment of the Baxter equation for the similar case of the 2-particle closed Toda chain  \cite{GUTZWILLER1980347,GUTZWILLER1981304,Sklyanin:1984sb,Kharchev:1999bh,Kharchev:2000yj,Kharchev:2000ug,0305-4470-25-20-007,An2009,Kozlowski:2010tv}.

We are therefore encouraged to proceed along the lines of the quantum Separation of Variables approach, reinterpreted in gauge theory language. The next step would then be to consider the Fourier transform of \eqref{linear}; more precisely, let us consider the Fourier transform multiplied by a prefactor depending on $\theta(x,Q_{\text{4d}})$ defined in \eqref{theta}, that is\footnote{The additional theta function factor may be interpreted as dividing by a codimension two defect of monodromy type for an $\mathcal{N} = 2^*$ $U(1)$ theory.}
\begin{equation}
\psi_a(y, \hbar_y, g_y, Q_{\text{4d}}) \,\propto\, \left[\theta( -\frac{i \pi y}{2 \omega} ,Q_{\text{4d}})\right]^{\frac{\hbar_y/2 - g_y}{\hbar_y}}
\int_{\mathbb{R}} d\sigma e^{- \frac{i \pi y\, \sigma }{\omega \hbar_y}} \mathcal{Q}(\sigma, a, \hbar_y, g_y, Q_{\text{4d}}). \label{psi4d}
\end{equation}
If we only focus on the values of $\xi$ and $a = a_n$ for which $\mathcal{Q}(\sigma, a_n, \hbar_y, g_y, Q_{\text{4d}})$ is an entire function, and assuming $\mathcal{Q}(\sigma, a_n, \hbar_y, g_y, Q_{\text{4d}})$ decays fast enough at $\sigma = \pm \infty$, we can bring $y$-derivatives inside the integral and shift the contour of integration; using these facts and \eqref{bax2} one can then show that
\begin{equation}
\begin{split}
& -\hbar_y^2 \partial_y^2 \psi_{a_n}(y, \hbar_y, g_y, Q_{\text{4d}}) 
- (g_y^2 - \frac{\hbar_y^2}{4})
 \left[ \partial_y^2 \ln \theta( -\frac{i \pi y}{2 \omega} ,Q_{\text{4d}}) \right] \psi_{a_n}(y, \hbar_y, g_y, Q_{\text{4d}})  \\
& = \frac{\pi^2}{\omega^2} E'(a_n, i \hbar_y, i g_y, Q_{\text{4d}}) \psi_{a_n}(y, \hbar_y, g_y, Q_{\text{4d}}).
 \end{split}
\end{equation}
Using the identities collected in Appendix \ref{appell}, this equation can be recast in the more familiar form
\begin{equation}
\begin{split}
& \left[ -\hbar_y^2 \partial_y^2 - (g_y^2 - \frac{\hbar_y^2}{4}) \wp(i y \vert \omega, \omega') \right] \psi_{a_n}(y, \hbar_y, g_y, Q_{\text{4d}}) \\
& = E(a_n, i \hbar_y, i g_y, \omega, \omega') \psi_{a_n}(y, \hbar_y, g_y, Q_{\text{4d}}), 
\end{split}
\end{equation}
which is nothing else but the 2-eCM$_{\text{A}}$ quantum mechanical problem. Although once again we cannot make any rigorous statement, the expectation is that \eqref{psi4d} will be the correct $L^2([0,2\vert \omega' \vert])$ 2-eCM$_{\text{A}}$ eigenfunction for $a = a_n$ satisfying A-type quantization conditions \eqref{quant4d}, as it should follow from entirety and decay properties at $\sigma = \pm \infty$ of the solution to the Baxter equation \eqref{linear} at the values $a = a_n$. 
A similar procedure should work for the $N$-particle eCM$_{\text{A}}$ system, although for $N > 2$ the integral transformation involved will be more complicated than a Fourier transform.

\subsection*{Hyperbolic limit}

In the limit $Q_{\text{4d}} \rightarrow 0$ the Baxter solution \eqref{linear} reduces to $\mathcal{Q}^{(f)}_0(\sigma, a, \hbar_y, g_y)$ only, and the Baxter equation \eqref{bax2} becomes
\begin{equation}
\left[ (\sigma - a)(\sigma + a) - (\sigma - a + i (g_y + \frac{\hbar_y}{2}))(\sigma + a + i (g_y + \frac{\hbar_y}{2})) e^{i \hbar_y \partial_{\sigma}} \right] \mathcal{Q}^{(f)}_{0}(\sigma, a, \hbar_y, g_y) = 0. \label{baxhyp}
\end{equation}
It is easy to verify that this equation is indeed satisfied. Assuming $y > 0$,\footnote{Due of the singularity of the 2-hCM potential \eqref{hypcal} at $y=0$ the ordering of particles in the course of motion cannot be changed, so we can only consider $y>0$ or $y<0$.} the Fourier transform (multiplied by a prefactor) $\psi(y, a, \hbar_y, g_y)$ of $\mathcal{Q}^{(f)}_0(\sigma, a, \hbar_y, g_y)$ defined as
\begin{equation}
\psi(y, a, \hbar_y, g_y) \propto (1-e^{-\frac{\pi y}{\omega}})^{\frac{\hbar_y/2 - g_y}{\hbar_y}} \int_{\mathbb{R}} d\sigma e^{-\frac{i \pi y \sigma}{\omega \hbar_y}} \mathcal{Q}^{(f)}_0(\sigma, a, \hbar_y, g_y) \label{hCMsol2}
\end{equation}
satisfies the 2-hCM problem \eqref{hCMbis}
\begin{equation}
\left[ -\hbar_y^2 \partial_y^2 + (g_y^2 - \frac{\hbar_y^2}{4}) \frac{\pi^2}{\omega^2} \dfrac{1}{4 \sinh^2 (\frac{\pi y}{2\omega})} \right] \psi(y, a, \hbar_y, g_y) = \frac{\pi^2}{\omega^2} a^2 \psi(y, a, \hbar_y, g_y). \label{hypcal}
\end{equation}
It is important to notice that in this case $a$, and therefore the 2-hCM energy, is a continuous variable: there is no need to impose quantization conditions, nor to require entirety of the solution to the Baxter equation, since $\mathcal{Q}^{(f)}_0(\sigma, a, \hbar_y, g_y)$ already possesses the correct decay properties at infinity and we are also free to shift the integration contour in the upper half $\sigma$-plane due to the absence of poles. In the special case $g_y = 0$, our solution \eqref{hCMsol2} reduces to the one constructed via the same quantum Separation of Variables approach in \cite{Gerasimov:2002cf}; more in general, for $g_y > \frac{\hbar_y}{2}$ we find (after evaluating the integral)
\begin{equation}
\psi(y, a, \hbar_y, g_y) \propto \left[\sinh \frac{\pi y}{2\omega}\right]^{\frac{1}{2}+\frac{g_y}{\hbar_y}}
{}_2 F_1 \left(\frac{i a}{\hbar_y}+\frac{1}{4}+\frac{g_y}{2\hbar_y},-\frac{i a}{\hbar_y}+\frac{1}{4}+\frac{g_y}{2\hbar_y},1+\frac{g_y}{\hbar_y}, -\sinh^2\frac{\pi y}{2 \omega}\right) ,
\end{equation}
which is nothing but the previously discussed 2-hCM eigenfunction \eqref{hCMsol}.

\section{2-particle systems of Ruijsenaars-Schneider type} \label{RSsection}

Similarly to what is done in Section \ref{CMsection}, let us consider a complex coordinate $z = x + i y$ ($x,y \in \mathbb{R}$) on a rectangular torus of half-periods $(\tilde{\omega},  \tilde{\omega}')$\footnote{We will use tilded half-periods $(\tilde{\omega},\tilde{\omega}')$ to distinguish them from the half-periods $(\omega, \omega')$ of 2-eCM.} with $\tilde{\omega} \in \mathbb{R}_+$, $\tilde{\omega}' \in i \mathbb{R}_+$ so that
\begin{equation}
x \sim x + 2 \tilde{\omega} , \;\;\;\; y \sim y + 2 \vert \tilde{\omega}' \vert.
\end{equation}
Also, for gauge theory applications we will sometimes reparametrize the periods as
\begin{equation}
(2\tilde{\omega}, 2 \tilde{\omega}') \,=\, (2\tilde{\omega}, -i \frac{\tilde{\omega}}{\pi}\ln Q_{\text{5d}}),
\end{equation} 
with 
\begin{equation}
Q_{\text{5d}} = e^{2\pi i \tilde{\tau}}, \;\;\;\; \tilde{\tau} = \frac{\tilde{\omega}'}{\tilde{\omega}}.
\end{equation}
The 2-particle elliptic Ruijsenaars-Schneider system (2-eRS) Hamiltonian is given by the second-order \textit{finite-difference equation}
\begin{equation}
\widehat{H}_{\text{eRS}} \psi(z, Q_{\text{5d}}) = E \psi(z, Q_{\text{5d}}), \label{ellRSp}
\end{equation}
where 
\begin{equation}
\widehat{H}_{\text{eRS}} = \sqrt{\dfrac{\sigma\big(z + i g \vert \tilde{\omega}, \tilde{\omega}' \big)}{\sigma(z \vert \tilde{\omega}, \tilde{\omega}')}} e^{i \hbar \partial_{z}} 
\sqrt{\dfrac{\sigma\big(z - i g \vert \tilde{\omega}, \tilde{\omega}' \big) }{\sigma(z \vert \tilde{\omega}, \tilde{\omega}')}} 
+ \sqrt{\dfrac{\sigma\big(z - i g \vert \tilde{\omega}, \tilde{\omega}' \big)}{\sigma(z \vert \tilde{\omega}, \tilde{\omega}')}} e^{- i \hbar \partial_{z}}
\sqrt{\dfrac{\sigma\big(z + i g \vert \tilde{\omega}, \tilde{\omega}' \big) }{\sigma(z \vert \tilde{\omega}, \tilde{\omega}')}} 
\label{ellRS2p}
\end{equation}
and all parameters $z, \hbar, g, E \in \mathbb{C}$.
The elliptic functions appearing in this section are defined in Appendix~\ref{appell}.
Although this is the standard form of the complex 2-eRS Hamiltonian, we can also work with the slightly different Hamiltonian
\begin{equation}
\begin{split}
& \widehat{H}'_{\text{eRS}} = \sqrt{\dfrac{\theta_1\big(\pi \frac{z + i g}{2\tilde{\omega}} , Q_{\text{5d}}^{1/2} \big)}{\theta_1(\pi \frac{z}{2 \tilde{\omega}}, Q_{\text{5d}}^{1/2})}} e^{i \hbar \partial_{z}} 
\sqrt{\dfrac{\theta_1\big(\pi \frac{z - i g}{2\tilde{\omega}} , Q_{\text{5d}}^{1/2} \big) }{\theta_1(\pi \frac{z}{2\tilde{\omega}} , Q_{\text{5d}}^{1/2})}} 
+ \sqrt{\dfrac{\theta_1\big(\pi \frac{z - i g}{2\tilde{\omega}} , Q_{\text{5d}}^{1/2} \big)}{\theta_1(\pi \frac{z}{2 \tilde{\omega}}, Q_{\text{5d}}^{1/2})}} e^{-i \hbar \partial_{z}} 
\sqrt{\dfrac{\theta_1\big(\pi \frac{z + i g}{2\tilde{\omega}} , Q_{\text{5d}}^{1/2} \big) }{\theta_1(\pi \frac{z}{2\tilde{\omega}} , Q_{\text{5d}}^{1/2})}} ;
\label{ellRS3} 
\end{split}
\end{equation}
this differs from \eqref{ellRS2p} only by an overall multiplicative constant:
\begin{equation}
\widehat{H}_{\text{eRS}} = 
\exp \left( -\frac{ g(g-\hbar)}{24}
\frac{\pi^2}{\tilde{\omega}^2} E_2(Q_{\text{5d}}) \right)
\widehat{H}'_{\text{eRS}}.
\end{equation}
We can therefore alternatively study the problem
\begin{equation}
\widehat{H}'_{\text{eRS}} \psi(z, Q_{\text{5d}}) = E' \psi(z, Q_{\text{5d}}), \label{ellRS}
\end{equation}
with $\widehat{H}'_{\text{eRS}}$ given by \eqref{ellRS3}. As for the Calogero-Moser case, one possible question would be to construct two linearly independent solutions to this finite-difference equation for generic complex values of the parameters. Since we are however interested in quantum mechanical applications, we should only be considering values of parameters for which \eqref{ellRSp}, \eqref{ellRS} realize a well-defined quantum mechanical problem. Among various possibilities, this happens in the following two cases:

\paragraph{B-model:} Restricting to the real slice $z = x \in \mathbb{R}$ and requiring $\hbar = \hbar_x \in \mathbb{R}_+$, $g = g_x \in \mathbb{R}_+$, $E = E^{\text{(B)}} \in \mathbb{R}_+$ problems \eqref{ellRSp}, \eqref{ellRS} reduce respectively to
\begin{equation}
\begin{split}
\widehat{H}_{\text{B}} \psi^{(\text{B})}(x,Q_{\text{5d}}) & = 
\Bigg[ \sqrt{\dfrac{\sigma\big(x + i g_x \vert \tilde{\omega}, \tilde{\omega}' \big)}{\sigma(x \vert \tilde{\omega}, \tilde{\omega}')}} e^{i \hbar_x \partial_{x}} 
\sqrt{\dfrac{\sigma\big(x - i g_x \vert \tilde{\omega}, \tilde{\omega}' \big) }{\sigma(x \vert \tilde{\omega}, \tilde{\omega}')}}  \\[10 pt]
& + \sqrt{\dfrac{\sigma\big(x - i g_x \vert \tilde{\omega}, \tilde{\omega}' \big)}{\sigma(x \vert \tilde{\omega}, \tilde{\omega}')}} e^{- i \hbar_x \partial_{x}}
\sqrt{\dfrac{\sigma\big(x + i g_x \vert \tilde{\omega}, \tilde{\omega}' \big) }{\sigma(x \vert \tilde{\omega}, \tilde{\omega}')}} \Bigg] \psi^{(\text{B})}(x,Q_{\text{5d}}) \\[10 pt]
& = E^{(\text{B})} \psi^{(\text{B})}(x,Q_{\text{5d}}) \label{ellRSBp}
\end{split}
\end{equation}
and
\begin{equation}
\begin{split}
\widehat{H}'_{\text{B}} \psi^{(\text{B})}(x,Q_{\text{5d}}) & = 
\Bigg[ \sqrt{\dfrac{\theta_1\big(\pi \frac{x + i g_x}{2\tilde{\omega}} , Q_{\text{5d}}^{1/2} \big)}{\theta_1(\pi \frac{x}{2 \tilde{\omega}}, Q_{\text{5d}}^{1/2})}} e^{i \hbar_x \partial_{x}} 
\sqrt{\dfrac{\theta_1\big(\pi \frac{x - i g_x}{2\tilde{\omega}} , Q_{\text{5d}}^{1/2} \big) }{\theta_1(\pi \frac{x}{2\tilde{\omega}} , Q_{\text{5d}}^{1/2})}}  \\[10 pt]
& + \sqrt{\dfrac{\theta_1\big(\pi \frac{x - i g_x}{2\tilde{\omega}} , Q_{\text{5d}}^{1/2} \big)}{\theta_1(\pi \frac{x}{2 \tilde{\omega}}, Q_{\text{5d}}^{1/2})}} e^{-i \hbar_x \partial_{x}} 
\sqrt{\dfrac{\theta_1\big(\pi \frac{x + i g_x}{2\tilde{\omega}} , Q_{\text{5d}}^{1/2} \big) }{\theta_1(\pi \frac{x}{2\tilde{\omega}} , Q_{\text{5d}}^{1/2})}} \Bigg] \psi^{(\text{B})}(x,Q_{\text{5d}}) \\[10 pt]
& = E'^{(\text{B})} \psi^{(\text{B})}(x,Q_{\text{5d}}). \label{ellRSB}
\end{split}
\end{equation}
For $0 < g_x < \hbar_x + 2 \vert \tilde{\omega}' \vert$, \eqref{ellRSBp} and \eqref{ellRSB} should define quantum mechanical problems on $L^2_x([0,2\tilde{\omega}])$ whose energy spectrum is real and discrete \cite{0305-4470-32-9-018,Ruijsenaars1999,2015SIGMA..11..004R}; we will refer to \eqref{ellRSBp} (or equivalently \eqref{ellRSB}) as the B-type 2-particle elliptic Ruijsenaars-Schneider quantum integrable system (2-eRS$_{\text{B}}$). 

In the limit $\tilde{\omega}' \rightarrow i\infty$ (i.e. $Q_{\text{5d}} \rightarrow 0$), \eqref{ellRSB} reduces to the 2-particle trigonometric Ruijsenaars-Schneider quantum integrable system (2-tRS) 
\begin{equation}
\begin{split}
\widehat{H}^{(\text{T})}\psi^{(\text{T})}(x) & = 
\Bigg[ \sqrt{\frac{\sin(\pi\frac{x + i g_x}{2\tilde{\omega}}) }{\sin(\pi\frac{x}{2\tilde{\omega}})}} e^{i \hbar_x \partial_x} \sqrt{\frac{\sin(\pi\frac{x - i g_x}{2\tilde{\omega}}) }{\sin(\pi\frac{x}{2\tilde{\omega}})}} \\[10 pt]
& + \sqrt{\frac{\sin(\pi\frac{x - i g_x}{2\tilde{\omega}}) }{\sin(\pi\frac{x}{2\tilde{\omega}})}} e^{-i \hbar_x \partial_x} \sqrt{\frac{\sin(\pi\frac{x + i g_x}{2\tilde{\omega}}) }{\sin(\pi\frac{x}{2\tilde{\omega}})}} \Bigg] \psi^{(\text{T})}(x) \\[10 pt]
& = E^{(\text{T})}\psi^{(\text{T})}(x) ; \label{tRS}
\end{split}
\end{equation}
this is still a quantum mechanical problem on $L^2_x([0,2\tilde{\omega}])$ with discrete spectrum and its solution is known analytically.

\paragraph{A-model:} Restricting instead to the slice $z = i y \in i \mathbb{R}$ and requiring 
$\hbar = i \hbar_y \in i\mathbb{R}_+$, $g = i g_y \in i\mathbb{R}_+$, $E = E^{(\text{A})} \in \mathbb{R}_+$ problems \eqref{ellRSp} and \eqref{ellRS} reduce respectively to
\begin{equation}
\begin{split}
\widehat{H}_{\text{A}} \psi^{(\text{A})}(y,Q_{\text{5d}}) & = 
\Bigg[ \sqrt{\dfrac{\sigma\big(i(y + i g_y) \vert \tilde{\omega}, \tilde{\omega}' \big)}{\sigma(i y \vert \tilde{\omega}, \tilde{\omega}')}} e^{i \hbar_y \partial_{y}} 
\sqrt{\dfrac{\sigma\big(i(y - i g_y) \vert \tilde{\omega}, \tilde{\omega}' \big) }{\sigma(i y \vert \tilde{\omega}, \tilde{\omega}')}}  \\[10 pt]
& + \sqrt{\dfrac{\sigma\big(i(y - i g_y) \vert \tilde{\omega}, \tilde{\omega}' \big)}{\sigma(i y \vert \tilde{\omega}, \tilde{\omega}')}} e^{- i \hbar_y \partial_{y}}
\sqrt{\dfrac{\sigma\big(i(y + i g_y) \vert \tilde{\omega}, \tilde{\omega}' \big) }{\sigma(i y \vert \tilde{\omega}, \tilde{\omega}')}} \Bigg] \psi^{(\text{A})}(y,Q_{\text{5d}}) \\[10 pt]
& = E^{(\text{A})} \psi^{(\text{A})}(y,Q_{\text{5d}}) \label{ellRSAp}
\end{split}
\end{equation}
and
\begin{equation}
\begin{split}
\widehat{H}'_{\text{A}} \psi^{(\text{A})}(y,Q_{\text{5d}}) & = 
\Bigg[ \sqrt{\dfrac{\theta_1\big(i \pi \frac{y + i g_y}{2\tilde{\omega}} , Q_{\text{5d}}^{1/2} \big)}{\theta_1(i \pi \frac{y}{2 \tilde{\omega}}, Q_{\text{5d}}^{1/2})}} e^{i \hbar_y \partial_{y}} 
\sqrt{\dfrac{\theta_1\big(i\pi \frac{y - i g_y}{2\tilde{\omega}} , Q_{\text{5d}}^{1/2} \big) }{\theta_1(i\pi \frac{y}{2\tilde{\omega}} , Q_{\text{5d}}^{1/2})}}  \\[10 pt] 
& + \sqrt{\dfrac{\theta_1\big(i\pi \frac{y - i g_y}{2\tilde{\omega}} , Q_{\text{5d}}^{1/2} \big)}{\theta_1(i\pi \frac{y}{2 \tilde{\omega}}, Q_{\text{5d}}^{1/2})}} e^{-i \hbar_y \partial_{y}} 
\sqrt{\dfrac{\theta_1\big(i\pi \frac{y + i g_y}{2\tilde{\omega}} , Q_{\text{5d}}^{1/2} \big) }{\theta_1(i\pi \frac{y}{2\tilde{\omega}} , Q_{\text{5d}}^{1/2})}} \Bigg] \psi^{(\text{A})}(y,Q_{\text{5d}}) \\[10 pt]
& = E'^{(\text{A})} \psi^{(\text{A})}(y,Q_{\text{5d}}). \label{ellRSA}
\end{split}
\end{equation}
For $0 \leqslant g_y \leqslant \hbar_y + 2\tilde{\omega}$, \eqref{ellRSAp} and \eqref{ellRSA} should define quantum mechanical problems on $L^2_y([0,2\vert \tilde{\omega}'\vert])$ whose energy spectrum is real and discrete; we will refer to \eqref{ellRSAp} (or equivalently \eqref{ellRSA}) as the A-type 2-particle elliptic Ruijsenaars-Schneider quantum integrable system (2-eRS$_{\text{A}}$). 

In the limit $\tilde{\omega}' \rightarrow i \infty$ (i.e. $Q_{\text{5d}} \rightarrow 0$), \eqref{ellRSA} reduces to the 2-particle hyperbolic Ruijsenaars-Schneider quantum integrable system (2-hRS) 
\begin{equation}
\begin{split}
\widehat{H}^{(\text{H})} \psi^{(\text{H})}(y) & = 
\Bigg[ \sqrt{\frac{\sinh(\pi\frac{y + i g_y}{2 \tilde{\omega}}) }{\sinh(\pi\frac{y}{2\tilde{\omega}})}} e^{i \hbar_y \partial_y} \sqrt{\frac{\sinh(\pi\frac{y - i g_y}{2\tilde{\omega}}) }{\sinh(\pi\frac{y}{2\tilde{\omega}})}} \\[10 pt]
& + \sqrt{\frac{\sinh(\pi\frac{y - i g_y}{2\tilde{\omega}}) }{\sinh(\pi\frac{y}{2\tilde{\omega}})}} e^{-i \hbar_y \partial_y} \sqrt{\frac{\sinh(\pi\frac{y + i g_y}{2\tilde{\omega}}) }{\sinh(\pi\frac{y}{2\tilde{\omega}})}} \Bigg] \psi^{(\text{H})}(y) \\[10 pt]
& = E^{(\text{H})}\psi^{(\text{H})}(y) ; \label{hRS}
\end{split}
\end{equation}
this is still a quantum mechanical problem, with continuous spectrum, for $y \in (0,\infty)$ and $0 \leqslant g_y \leqslant \hbar_y + 2\tilde{\omega}$ \cite{rhilbert}, whose analytic solution was studied in \cite{2012arXiv1206.3787H,2016arXiv160706672H}.

\noindent Similarly to what we saw happening for the elliptic Calogero-Moser case, the two problems 2-eRS$_{\text{B}}$ and 2-eRS$_{\text{A}}$ are actually related: in fact, since \cite{lawden1989elliptic}
\begin{equation}
\sigma(i y \vert \tilde{\omega}, \tilde{\omega}') 
= i \sigma(y \vert -i \tilde{\omega}', i \tilde{\omega}) ,
\end{equation}
problem 2-eCM$_{\text{A}}$ \eqref{ellRSAp} with half-periods $(\tilde{\omega}, \tilde{\omega}')$ is equivalent to problem 2-eCM$_{\text{B}}$ \eqref{ellRSBp} with half-periods $(-i \tilde{\omega}', i \tilde{\omega})$ once we also identify $\hbar_y, g_y$ with $\hbar_x, g_x$, that is
\begin{equation}
\underbrace{ \widehat{H}_{\text{A}}(y, \hbar_y, g_y, \tilde{\omega}, \tilde{\omega}') }_\textrm{A-problem \eqref{ellRSAp}}
\;\;\;\;=\; \underbrace{ \widehat{H}_{\text{B}}(y, \hbar_y, g_y, -i\tilde{\omega}', i\tilde{\omega})  }_\textrm{B-problem \eqref{ellRSBp} (inverted periods)}; \label{relationeRS}
\end{equation}
as a consequence,
\begin{equation}
E^{(\text{A})}(\hbar_y, g_y, \tilde{\omega}, \tilde{\omega}')  = 
E^{(\text{B})}(\hbar_y, g_y, -i \tilde{\omega}', i\tilde{\omega}). \label{relationeRSbis}
\end{equation}
We however prefer to keep problems B and A distinguished, since the gauge theory approach treats them differently; verifying that gauge theory results respect \eqref{relationeRSbis} will therefore be a non-trivial consistency check for the validity of the proposed gauge theory solution.

\subsubsection*{Non-relativistic limit}

Before moving to the study of the various models, let us quickly mention that the various Ruijsenaars-Schneider systems can be reduced to their Calogero-Moser analogues by taking the ``non-relativistic'' limit $\tilde{\omega} \rightarrow \infty$, $\tilde{\omega}' \rightarrow i \infty$ while keeping $w = \frac{\omega}{\tilde{\omega}} z$ and $\tilde{\tau}$ constant (so that $\tilde{\tau} \rightarrow \tau$, $Q_{\text{5d}} \rightarrow Q_{\text{4d}}$).
For example, the complex 2-eRS Hamiltonian \eqref{ellRS2p} reduces to the complex 2-eCM one:
\begin{equation}
\widehat{H}_{\text{eRS}} \;\;\;\longrightarrow \;\;\; 2 + \dfrac{\omega^2}{\tilde{\omega}^2} \left[ -\hbar^2 \partial_w^2 + g(g-\hbar) \wp(w \vert \omega, \omega') \right] + O(\tilde{\omega}^{-3}),
\end{equation}
modulo having redefined $g$ by a $\hbar/2$ shift.
Similarly, the 2-tRS Hamiltonian reduces to the 2-tCM one:
\begin{equation}
\begin{split}
& \sqrt{\frac{\sin(\pi\frac{z + i g}{2\tilde{\omega}}) }{\sin(\pi\frac{z}{2\tilde{\omega}})}} e^{i \hbar \partial_z} \sqrt{\frac{\sin(\pi\frac{z - i g}{2\tilde{\omega}}) }{\sin(\pi\frac{z}{2\tilde{\omega}})}} 
+ \sqrt{\frac{\sin(\pi\frac{z - i g}{2\tilde{\omega}}) }{\sin(\pi\frac{z}{2\tilde{\omega}})}} e^{-i \hbar \partial_z} \sqrt{\frac{\sin(\pi\frac{z + i g}{2\tilde{\omega}}) }{\sin(\pi\frac{z}{2\tilde{\omega}})}} \\[10 pt]
& \longrightarrow \;\; 2 + \dfrac{\omega^2}{\tilde{\omega}^2} \left[ -\hbar^2 \partial_w^2 + 
g(g-\hbar) \dfrac{\pi^2}{\omega^2} \dfrac{1}{4 \sin^2(\frac{\pi w}{2 \omega})} \right] 
+ O(\tilde{\omega}^{-3}).
\end{split}
\end{equation}

\vspace*{0.3 cm}

\subsection{Trigonometric and hyperbolic cases} \label{SectRS}

Before discussing the 2-eRS$_{\text{B}}$ and 2-eRS$_{\text{A}}$ systems of interest, let us first collect a few known facts about the analytic solution to their trigonometric (2-tRS) and hyperbolic (2-hRS) limits. This will be of particular importance in Section \ref{eRSBsection}, where we will show how to numerically evaluate the energy spectrum of the elliptic systems. 

\subsection*{Trigonometric case}

Let us start by reviewing the known analytic solution to the 2-tRS system, 
along the same lines of Section \ref{tCMsection}. For generic half-period $\tilde{\omega}$ the 2-tRS Hamiltonian reads
\begin{equation}
\begin{split}
\widehat{H}^{(\text{T})} \psi^{(\text{T})}(x) &= \left[ \sqrt{\frac{\sin(\pi\frac{x + i g_x}{2\tilde{\omega}}) }{\sin(\pi\frac{x}{2\tilde{\omega}})}} e^{i \hbar_x \partial_x} \sqrt{\frac{\sin(\pi\frac{x - i g_x}{2\tilde{\omega}}) }{\sin(\pi\frac{x}{2\tilde{\omega}})}} 
+ \sqrt{\frac{\sin(\pi\frac{x - i g_x}{2\tilde{\omega}}) }{\sin(\pi\frac{x}{2\tilde{\omega}})}} e^{-i \hbar_x \partial_x} \sqrt{\frac{\sin(\pi\frac{x + i g_x}{2\tilde{\omega}}) }{\sin(\pi\frac{x}{2\tilde{\omega}})}} \right] \psi^{(\text{T})}(x) \\[10 pt]
& = \left( e^{\frac{\pi a}{2\tilde{\omega}}} + e^{-\frac{\pi a}{2\tilde{\omega}}} \right) \psi^{(\text{T})}(x) , \label{tRSbis}
\end{split}
\end{equation}
where for later convenience we reparametrized the energy $E^{(\text{T})}$ in terms of an auxiliary variable $a$ as
\begin{equation}
E^{(\text{T})}(a) = e^{\frac{\pi a}{2\tilde{\omega}}} + e^{-\frac{\pi a}{2\tilde{\omega}}}. \label{tRSenergygen}
\end{equation}
$L^2([0,2\tilde{\omega}])$ eigenfunctions for \eqref{tRSbis} can be constructed out of the Rogers polynomials, defined as (for $\vert q^{-1} \vert < 1$, $\vert \mu^{-1} \vert < 1$) \cite{gasper}
\begin{equation}
C_n(x;\mu^{-1}\vert q^{-1}) = \sum_{k = 0}^n \dfrac{(\mu^{-1};q^{-1})_{k}(\mu^{-1};q^{-1})_{n-k}}{(q^{-1};q^{-1})_{k}(q^{-1};q^{-1})_{n-k}} 
e^{-i(n-2k)\frac{x}{2}},
\end{equation}
also representable as the basic hypergeometric series
\begin{equation}
C_n(x;\mu^{-1}\vert q^{-1}) = \dfrac{(\mu^{-1}; q^{-1})_n}{(q^{-1};q^{-1})_n} e^{-\frac{i n x}{2}} 
{}_{2}\phi_{1}(q^{n}, \mu^{-1}; \mu q^{n-1};q^{-1}\mu \, e^{i x} \vert q^{-1});
\end{equation}
notice that we can move from $\vert q^{-1} \vert < 1$ to $\vert q^{-1} \vert >1$ thanks to
\begin{equation}
C_n(x;\mu^{-1}\vert q^{-1}) = (\mu^{-1} q)^n C_n(x;\mu\vert q).
\end{equation}
These orthogonal polynomials play the same role for 2-tRS of the Gegenbauer polynomials for 2-tCM. 
 The Rogers polynomials are known to satisfy the finite-difference equation \cite{2007arXiv0704.3123A} (for $q = e^{\frac{ \pi \hbar_x}{\tilde{\omega}}}$, $\mu = e^{\frac{\pi g_x}{\tilde{\omega}}}$) 
\begin{equation}
\begin{split}
& \left[ e^{\frac{\pi g_x}{2 \tilde{\omega}}} \dfrac{1 - e^{-\frac{\pi g_x}{\tilde{\omega}}} e^{\frac{i \pi x}{\tilde{\omega}}}}{1 - e^{\frac{i \pi x}{\tilde{\omega}}}} e^{i \hbar_x \partial_x} 
+ e^{\frac{\pi g_x}{2\tilde{\omega}}} \dfrac{1 - e^{-\frac{\pi g_x}{\tilde{\omega}}} e^{-\frac{i \pi x}{\tilde{\omega}}}}{1 - e^{-\frac{i \pi x}{\tilde{\omega}}}} e^{-i \hbar_x \partial_x} \right] C_n(\tfrac{\pi x}{\tilde{\omega}};e^{-\frac{\pi g_x}{\tilde{\omega}}}\vert e^{-\frac{\pi \hbar_x}{\tilde{\omega}}}) \\[10 pt]
& = \left[ \dfrac{\sin(\pi\frac{x + i g_x}{2 \tilde{\omega}})}{\sin(\frac{\pi x}{2 \tilde{\omega}})} e^{i \hbar_x \partial_x} 
+ \dfrac{\sin(\pi\frac{x - i g_x}{2\tilde{\omega}})}{\sin(\frac{\pi x}{2\tilde{\omega}})} e^{-i \hbar_x \partial_x} \right] 
C_n(\tfrac{\pi x}{\tilde{\omega}};e^{-\frac{\pi g_x}{\tilde{\omega}}}\vert e^{-\frac{\pi \hbar_x}{\tilde{\omega}}}) \\[10 pt]
& = \left( e^{\pi\frac{g_x + n \hbar_x}{2\tilde{\omega}}} + e^{- \pi\frac{g_x + n \hbar_x}{2\tilde{\omega}}} \right) 
C_n(\tfrac{\pi x}{\tilde{\omega}};e^{-\frac{\pi g_x}{\tilde{\omega}}}\vert e^{-\frac{\pi \hbar_x}{\tilde{\omega}}}).
\end{split}
\end{equation}
Therefore, if we multiply them by the square root $\sqrt{w(x)}$ of an appropriate weight function (which is actually the orthogonality measure for the Rogers polynomials)
\begin{equation}
w(x) = \dfrac{(e^{\frac{i \pi x}{\tilde{\omega}}};e^{-\frac{\pi \hbar_x}{\tilde{\omega}}})_{\infty}(e^{-\frac{i \pi x}{\tilde{\omega}}};e^{-\frac{\pi \hbar_x}{\tilde{\omega}}})_{\infty}}{(e^{\frac{i \pi x}{\tilde{\omega}}} e^{-\frac{\pi g_x}{\tilde{\omega}}};e^{-\frac{\pi \hbar_x}{\tilde{\omega}}})_{\infty}(e^{-\frac{i \pi x}{\tilde{\omega}}} e^{-\frac{\pi g_x}{\tilde{\omega}}};e^{-\frac{\pi \hbar_x}{\tilde{\omega}}})_{\infty}},
\end{equation}
which satisfies
\begin{equation}
\begin{split}
& w(x + i \hbar_x) = 
\dfrac{
(1- e^{-\frac{\pi g_x}{\tilde{\omega}}} e^{\frac{i \pi x}{\tilde{\omega}}})
(1-e^{\frac{\pi \hbar_x}{\tilde{\omega}}}e^{-\frac{i \pi x}{\tilde{\omega}}})
}{
(1-e^{\frac{i \pi x}{\tilde{\omega}}})
(1-e^{\frac{\pi \hbar_x}{\tilde{\omega}}} e^{-\frac{\pi g_x}{\tilde{\omega}}} e^{-\frac{i \pi x}{\tilde{\omega}}})
} w(x) 
= \dfrac{
\sin(\pi\frac{x + i g_x}{2\tilde{\omega}}) 
\sin(\pi\frac{x + i \hbar_x}{2\tilde{\omega}})
}{
\sin(\pi \frac{x}{2\tilde{\omega}})
\sin(\pi\frac{x - i g_x + i \hbar_x}{2\tilde{\omega}}) 
} w(x), \\[5 pt]
& w(x - i \hbar_x) = 
\dfrac{
(1- e^{-\frac{\pi g_x}{\tilde{\omega}}} e^{-\frac{i \pi x}{\tilde{\omega}}})
(1-e^{\frac{\pi \hbar_x}{\tilde{\omega}}} e^{\frac{i \pi x}{\tilde{\omega}}})
}{
(1-e^{-\frac{i \pi x}{\tilde{\omega}}})
(1-e^{\frac{\pi \hbar_x}{\tilde{\omega}}} e^{-\frac{\pi g_x}{\tilde{\omega}}} e^{\frac{i \pi x}{\tilde{\omega}}})
} w(x) 
= \dfrac{
\sin(\pi\frac{x - i g_x}{2\tilde{\omega}}) 
\sin(\pi\frac{x - i \hbar_x}{2\tilde{\omega}})
}{
\sin(\pi\frac{x}{2\tilde{\omega}})
\sin(\pi\frac{x + i g_x - i \hbar_x}{2\tilde{\omega}}) 
} w(x),
\end{split}
\end{equation}
it is easy to realize that
\begin{equation}
\psi_n^{(\text{T})}(x) \;\propto\; \sqrt{w(x)} C_n(\tfrac{\pi x}{\tilde{\omega}};e^{-\frac{\pi g_x}{\tilde{\omega}}}\vert e^{-\frac{\pi \hbar_x}{\tilde{\omega}}}) \label{unnorm}
\end{equation}
is an $L^2([0,2\tilde{\omega}])$ eigenfunction of the 2-tRS system with eigenvalue 
\begin{equation}
E^{(\text{T})}_n = e^{\pi\frac{g_x + n \hbar_x}{2\tilde{\omega}}} + e^{- \pi\frac{g_x + n \hbar_x}{2\tilde{\omega}}}; \label{tRSenergy}
\end{equation}
in fact
\begin{equation}
\begin{split}
\widehat{H}^{(\text{T})} \psi_n^{(\text{T})}(x)
 & = \sqrt{w(x)} \left[ \dfrac{\sin(\pi\frac{x + i g_x}{2 \tilde{\omega}})}{\sin(\frac{\pi x}{2 \tilde{\omega}})} e^{i \hbar_x \partial_x} 
+ \dfrac{\sin(\pi\frac{x - i g_x}{2 \tilde{\omega}})}{\sin(\frac{\pi x}{2 \tilde{\omega}})} 
e^{-i \hbar_x \partial_x}  \right] 
C_n(\tfrac{\pi x}{\tilde{\omega}};e^{-\frac{\pi g_x}{\tilde{\omega}}}\vert e^{-\frac{\pi \hbar_x}{\tilde{\omega}}}) \\[5 pt]
& = \left( e^{\pi\frac{g_x + n \hbar_x}{2\tilde{\omega}}} + e^{- \pi\frac{g_x + n \hbar_x}{2\tilde{\omega}}} \right) 
\psi_n^{(\text{T})}(x).
\end{split}
\end{equation}
Moreover, if we introduce the normalized Rogers polynomials
\begin{equation}
\begin{split}
& G_n(\tfrac{\pi x}{\tilde{\omega}};e^{-\frac{\pi g_x}{\tilde{\omega}}}\vert e^{-\frac{\pi \hbar_x}{\tilde{\omega}}}) = \\[5 pt]
& =\sqrt{\dfrac{1}{4\tilde{\omega}} \dfrac{1-e^{-\frac{\pi g_x}{\tilde{\omega}}}e^{-\frac{n \pi \hbar_x}{\tilde{\omega}}}}{1-e^{-\frac{\pi g_x}{\tilde{\omega}}}} 
\dfrac{(e^{-\frac{\pi \hbar_x}{\tilde{\omega}}}; e^{-\frac{\pi \hbar_x}{\tilde{\omega}}})_n}{(e^{-\frac{2\pi g_x}{\tilde{\omega}}}; e^{-\frac{\pi \hbar_x}{\tilde{\omega}}})_n}
\dfrac{(e^{-\frac{\pi \hbar_x}{\tilde{\omega}}}; e^{-\frac{\pi \hbar_x}{\tilde{\omega}}})_{\infty} (e^{-\frac{2\pi g_x}{\tilde{\omega}}}; e^{-\frac{\pi \hbar_x}{\tilde{\omega}}})_{\infty}}{(e^{-\frac{\pi g_x}{\tilde{\omega}}}e^{-\frac{\pi \hbar_x}{\tilde{\omega}}}; e^{-\frac{\pi \hbar_x}{\tilde{\omega}}})_{\infty} (e^{-\frac{\pi g_x}{\tilde{\omega}}}; e^{-\frac{\pi \hbar_x}{\tilde{\omega}}})_{\infty}}} 
C_n(\tfrac{\pi x}{\tilde{\omega}};e^{-\frac{\pi g_x}{\tilde{\omega}}}\vert e^{-\frac{\pi \hbar_x}{\tilde{\omega}}}), \label{Rogersnorm}
\end{split}
\end{equation}
then the normalized eigenfunctions
\begin{equation}
\psi_n^{(\text{T})}(x) = \sqrt{w(x)} G_n(\tfrac{\pi x}{\tilde{\omega}};e^{-\frac{\pi g_x}{\tilde{\omega}}}\vert e^{-\frac{\pi \hbar_x}{\tilde{\omega}}}) \label{tRSsolnorm}
\end{equation}
satisfy the orthonormality condition
\begin{equation}
\int_{0}^{2\tilde{\omega}} dx \, \psi_m^{(\text{T})}(x) \psi_n^{(\text{T})}(x) = 
\int_{0}^{2\tilde{\omega}} dx \,w(x) G_m(\tfrac{\pi x}{\tilde{\omega}};e^{-\frac{\pi g_x}{\tilde{\omega}}}\vert e^{-\frac{\pi \hbar_x}{\tilde{\omega}}}) G_n(\tfrac{\pi x}{\tilde{\omega}};e^{-\frac{\pi g_x}{\tilde{\omega}}}\vert e^{-\frac{\pi \hbar_x}{\tilde{\omega}}}) = \delta_{m,n}.
\end{equation}
To sum up, we have completely determined the normalized $L^2([0,2\tilde{\omega}])$ 2-tRS eigenfunctions $\psi_n^{(\text{T})}(x)$ \eqref{tRSsolnorm} and the corresponding discrete energies $E_n^{(\text{T})}$ \eqref{tRSenergy}. These energies can also be thought as $E_n^{(\text{T})} = E^{(\text{T})}(a_n)$, i.e. as coming from \eqref{tRSenergygen} for $a = a_n$ satisfying the quantization conditions
\begin{equation}
a_n = \pm ( g_x + n \hbar_x ) , \;\; n \in \mathbb{N}; \label{quanttRS}
\end{equation}
these are the same quantization conditions we previously found for 2-tCM \eqref{quanttCM}, modulo having redefined $a \rightarrow \frac{a}{2}$ and shifted $g_x$ by $\hbar_x/2$.

\subsection*{Hyperbolic case}

Let us now shortly comment on the 2-hRS system, whose Hamiltonian reads
\begin{equation}
\begin{split}
\widehat{H}^{(\text{H})} \psi^{(\text{H})}(y) = &
\Bigg[ \sqrt{\frac{\sinh(\pi\frac{y + i g_y}{2\tilde{\omega}}) }{\sinh(\frac{\pi y}{2 \tilde{\omega}})}} e^{i \hbar_y \partial_y} \sqrt{\frac{\sinh(\pi\frac{y - i g_y}{2\tilde{\omega}}) }{\sinh(\frac{\pi y}{2 \tilde{\omega}})}} \\[10 pt]
& + \sqrt{\frac{\sinh(\pi\frac{y - i g_y}{2\tilde{\omega}}) }{\sinh(\frac{\pi y}{2 \tilde{\omega}})}} e^{-i \hbar_y \partial_y} \sqrt{\frac{\sinh(\pi\frac{y + i g_y}{2\tilde{\omega}}) }{\sinh(\frac{\pi y}{2 \tilde{\omega}})}} \Bigg] \psi^{(\text{H})}(y) \\[10 pt]
= & \left( e^{\frac{\pi a}{2\tilde{\omega}}} + e^{-\frac{\pi a}{2\tilde{\omega}}} \right) \psi^{(\text{H})}(y), \label{hRSbis}
\end{split}
\end{equation}
where we again reparameterized the energy $E^{(\text{H})}$ in terms of an auxiliary variable $a$ as
\begin{equation}
E^{(\text{H})}(a) = e^{\frac{\pi a}{2\tilde{\omega}}} + e^{-\frac{\pi a}{2\tilde{\omega}}} . \label{hRSenergygen}
\end{equation}
This Hamiltonian can clearly be obtained from \eqref{tRSbis} by sending $x \rightarrow i y$, $g_x \rightarrow i g_y$, $\hbar_x \rightarrow i \hbar_y$. However, it is not possible to recover the 2-hRS solution from the ``off-shell'' (generic $a$) solution of 2-tRS (as we did instead in the Calogero-Moser case, Section \ref{hCMsection}). 
This is because the 2-hRS Hamiltonian is a finite-difference equation rather than a differential one, therefore any possible formal solution to \eqref{hRSbis} will be ambiguous: in fact if we multiply a formal solution by any $i\hbar_y$-periodic function (usually referred to as ``quasi-constant''), this will still be a formal solution. Of course, this would in principle be a problem also for the 2-tRS system; however, while such quasi-constants cannot be present for 2-tRS since they will spoil the polynomial nature of the eigenfunction, they may affect the 2-hRS solution which doesn't need to be polynomial \cite{rhilbert}.

From the works \cite{rhilbert,2012arXiv1206.3787H,2016arXiv160706672H}, it appears that the proper way to fix these quasi-constant ambiguities is to study the \textit{modular double} version of the problem, similarly to what happens for the relativistic open Toda chain \cite{Kharchev:2001rs}; that is, we should require $\psi^{(\text{H})}(y)$ to be a solution not only of the original 2-hRS Hamiltonian $\widehat{H}^{(\text{H})}$ \eqref{hRSbis}, but also of a ``modular dual'' 2-hRS Hamiltonian $\widehat{\widetilde{H}}^{(\text{H})}$, related to the original one by the exchange $2\tilde{\omega} \longleftrightarrow \hbar_y$:
\begin{equation}
\begin{split}
\widehat{\widetilde{H}}^{(\text{H})} \psi^{(\text{H})}(y) \;=\; &
\Bigg[ \sqrt{\frac{\sinh(\pi\frac{y + i g_y}{\hbar_y}) }{\sinh(\frac{\pi y}{\hbar_y})}} e^{i 2 \tilde{\omega} \partial_y} \sqrt{\frac{\sinh(\pi\frac{y - i g_y}{\hbar_y}) }{\sinh(\frac{\pi y}{\hbar_y})}} \\[10 pt]
& + \sqrt{\frac{\sinh(\pi\frac{y - i g_y}{\hbar_y}) }{\sinh(\frac{\pi y}{\hbar_y})}} e^{-i 2 \tilde{\omega} \partial_y} \sqrt{\frac{\sinh(\pi\frac{y + i g_y}{\hbar_y}) }{\sinh(\frac{\pi y}{\hbar_y})}} \Bigg] \psi^{(\text{H})}(y) \\[10 pt]
=\; & \left( e^{\frac{\pi a}{\hbar_y}} + e^{-\frac{\pi a}{\hbar_y}} \right) \psi^{(\text{H})}(y), \label{hRSbisdual}
\end{split}
\end{equation}
where the ``dual'' energy 
\begin{equation}
\widetilde{E}^{(\text{H})}(a) = e^{\frac{\pi a}{\hbar_y}} + e^{-\frac{\pi a}{\hbar_y}}
\end{equation}
is also related to the original one \eqref{hRSenergygen} by the exchange $2\tilde{\omega} \longleftrightarrow \hbar_y$. Since $\widehat{\widetilde{H}}^{(\text{H})}$ is a finite-difference equation in $2\tilde{\omega}$, requiring \eqref{hRSbisdual} in addition to \eqref{hRSbis} will completely fix the quasi-constant ambiguity of the solution $\psi^{(\text{H})}(y)$, which should then be $2\tilde{\omega} \longleftrightarrow \hbar_y$ symmetric \cite{2012arXiv1206.3787H,2016arXiv160706672H}. We will later see how modular duality plays an important role also for the 2-eRS$_{\text{A}}$ system.

\subsection{Elliptic case: B-model} \label{eRSBsection}

We are finally ready to study the 2-eRS$_{\text{B}}$ quantum mechanical problem, which in the form \eqref{ellRSB} reads
\begin{equation}
\begin{split}
\widehat{H}'_{\text{B}} \psi^{(\text{B})}(x,Q_{\text{5d}}) & = 
\Bigg[ \sqrt{\dfrac{\theta_1\big(\pi \frac{x + i g_x}{2\tilde{\omega}} , Q_{\text{5d}}^{1/2} \big)}{\theta_1(\pi \frac{x}{2 \tilde{\omega}}, Q_{\text{5d}}^{1/2})}} e^{i \hbar_x \partial_{x}} 
\sqrt{\dfrac{\theta_1\big(\pi \frac{x - i g_x}{2\tilde{\omega}} , Q_{\text{5d}}^{1/2} \big) }{\theta_1(\pi \frac{x}{2\tilde{\omega}} , Q_{\text{5d}}^{1/2})}}  \\[10 pt] 
& + \sqrt{\dfrac{\theta_1\big(\pi \frac{x - i g_x}{2\tilde{\omega}} , Q_{\text{5d}}^{1/2} \big)}{\theta_1(\pi \frac{x}{2 \tilde{\omega}}, Q_{\text{5d}}^{1/2})}} e^{-i \hbar_x \partial_{x}} 
\sqrt{\dfrac{\theta_1\big(\pi \frac{x + i g_x}{2\tilde{\omega}} , Q_{\text{5d}}^{1/2} \big) }{\theta_1(\pi \frac{x}{2\tilde{\omega}} , Q_{\text{5d}}^{1/2})}} \Bigg] \psi^{(\text{B})}(x,Q_{\text{5d}}) \\[10 pt]
& = E'^{(\text{B})} \psi^{(\text{B})}(x,Q_{\text{5d}}), \label{ellRSBbis}
\end{split}
\end{equation}
where we made explicit the dependence on $Q_{\text{5d}}$ of the eigenfunction $\psi^{(\text{B})}(x,Q_{\text{5d}})$ (which will depend also on the other parameters of the model). As for the 2-eCM$_{\text{B}}$ case, also this problem can be studied both analytically as a perturbation series in $Q_{\text{5d}}$ and numerically.

\subsubsection*{Analytical study}

To solve the 2-eRS$_{\text{B}}$ problem analytically on $L^2([0,2\tilde{\omega}])$ we may use perturbation theory around $Q_{\text{5d}} = 0$ (or $\tilde{\omega}' \rightarrow i \infty$), i.e. around the trigonometric limit. For this purpose it will be more convenient to rewrite the Hamiltonian operator $\widehat{H}'_{\text{B}}$ in a form better suited to study the small $Q_{\text{5d}}$ limit. To do this we factorize (in analogy with \eqref{tRSsolnorm})
\begin{equation}
\psi^{(\text{B})}(x,Q_{\text{5d}}) = \sqrt{w(x,Q_{\text{5d}})} \Phi^{(\text{B})}(x, Q_{\text{5d}});
\end{equation}
if now we introduce the elliptic Gamma function
\begin{equation}
\Gamma( z ; t, p) = \prod_{m \geqslant 0} \prod_{n \geqslant 0} \dfrac{1 - t^{m+1} p^{n+1}  z^{-1}}{1 - t^m p^n z} = \prod_{m \geqslant 0} \dfrac{(t^{m+1} p z^{-1}; p)_{\infty}}{(t^{m} z; p)_{\infty}} = \prod_{n \geqslant 0} \dfrac{(p^{n+1} t z^{-1};t)_{\infty}}{(p^{n} z;t)_{\infty}},
\end{equation}
then by choosing 
\begin{equation}
w(x,Q_{\text{5d}}) = \dfrac{\Gamma\big(e^{\frac{i \pi x}{\tilde{\omega}}}e^{-\frac{\pi g_x}{\tilde{\omega}}}; e^{-\frac{\pi \hbar_x}{\tilde{\omega}}},Q_{\text{5d}}\big) \Gamma\big(e^{-\frac{i \pi x}{\tilde{\omega}}}e^{-\frac{\pi g_x}{\tilde{\omega}}}; e^{-\frac{\pi \hbar_x}{\tilde{\omega}}},Q_{\text{5d}}\big)}{\Gamma\big(e^{\frac{i \pi x}{\tilde{\omega}}}; e^{-\frac{\pi \hbar_x}{\tilde{\omega}}},Q_{\text{5d}}\big) \Gamma\big(e^{-\frac{i \pi x}{\tilde{\omega}}}; e^{-\frac{\pi \hbar_x}{\tilde{\omega}}},Q_{\text{5d}}\big)},
\end{equation}
which is such that
\begin{equation}
\begin{split}
w(x + i \hbar_x, Q_{\text{5d}}) =  & 
\dfrac{
\theta_1(\pi \frac{x + i \hbar_x}{2\tilde{\omega}}, Q_{\text{5d}}^{1/2}) 
\theta_1(\pi \frac{x + i g_x}{2\tilde{\omega}}, Q_{\text{5d}}^{1/2})
}{
\theta_1(\pi \frac{x}{2 \tilde{\omega}}, Q_{\text{5d}}^{1/2}) 
\theta_1(\pi \frac{x + i \hbar_x - i g_x}{2\tilde{\omega}}, Q_{\text{5d}}^{1/2}) 
}  w(x, Q_{\text{5d}}), \\[5 pt]
w(x - i \hbar_x, Q_{\text{5d}}) =  &
\dfrac{
\theta_1(\pi \frac{x - i \hbar_x}{2\tilde{\omega}}, Q_{\text{5d}}^{1/2}) 
\theta_1(\pi \frac{x - i g_x}{2\tilde{\omega}}, Q_{\text{5d}}^{1/2})
}{
\theta_1(\pi \frac{x}{2 \tilde{\omega}}, Q_{\text{5d}}^{1/2}) 
\theta_1(\pi \frac{x - i \hbar_x + i g_x}{2\tilde{\omega}}, Q_{\text{5d}}^{1/2}) 
}  w(x, Q_{\text{5d}}),
\end{split}
\end{equation}
the problem \eqref{ellRSBbis} reduces to
\begin{equation}
\widehat{H}''_{\text{B}} \, \Phi(x, Q_{\text{5d}}) = E'^{(\text{B})} \Phi^{(\text{B})}(x,Q_{\text{5d}}),
\end{equation}
where the transformed Hamiltonian
\begin{equation}
\widehat{H}''_{\text{B}} = \dfrac{1}{\sqrt{w(x,Q_{\text{5d}})}} \widehat{H}'_{\text{B}} \sqrt{w(x,Q_{\text{5d}})} 
\end{equation}
reads explicitly
\begin{equation}
\widehat{H}''_{\text{B}} = \dfrac{\theta_1(\pi \frac{x + i g_x}{2 \tilde{\omega}}, Q_{\text{5d}}^{1/2})}{\theta_1(\frac{\pi x}{2 \tilde{\omega}}, Q_{\text{5d}}^{1/2})} e^{i \hbar_x \partial_x} + \dfrac{\theta_1(\pi \frac{x - i g_x}{2 \tilde{\omega}}, Q_{\text{5d}}^{1/2})}{\theta_1(\frac{\pi x}{2 \tilde{\omega}}, Q_{\text{5d}}^{1/2})} e^{-i \hbar_x \partial_x}.
\end{equation}
We can now consider the small $Q_{\text{5d}}$ expansion of the transformed Hamiltonian $\widehat{H}''_{\text{B}}$:
\begin{equation}
\widehat{H}''_{\text{B}} = \widehat{H}''_{\text{B}}{}^{\hspace{-0 cm}(0)} + Q_{\text{5d}} \widehat{H}''_{\text{B}}{}^{\hspace{-0 cm}(1)} + Q_{\text{5d}}^2 \widehat{H}''_{\text{B}}{}^{\hspace{-0 cm}(2)} + O(Q_{\text{5d}}^3),
\end{equation}
where
\begin{equation}
\begin{split}
\widehat{H}''_{\text{B}}{}^{\hspace{-0 cm}(0)} \,=\, & \dfrac{\sin(\pi \frac{x + i g_x}{2 \tilde{\omega}})}{\sin(\frac{\pi x}{2 \tilde{\omega}})} e^{i \hbar_x \partial_x} + \dfrac{\sin(\pi \frac{x - i g_x}{2 \tilde{\omega}})}{\sin(\frac{\pi x}{2 \tilde{\omega}})} e^{-i \hbar_x \partial_x}, \\
\widehat{H}''_{\text{B}}{}^{\hspace{-0 cm}(1)} \,=\, & 4 \dfrac{\sin(\frac{i \pi g_x}{2\tilde{\omega}})}{\sin(\frac{\pi x}{2 \tilde{\omega}})} \left[ \sin(\pi \tfrac{x + i g_x}{2 \tilde{\omega}}) \sin(\pi \tfrac{2 x + i g_x}{2 \tilde{\omega}}) e^{i \hbar_x \partial_x} - \sin(\pi \tfrac{x - i g_x}{2 \tilde{\omega}}) \sin(\pi \tfrac{2 x - i g_x}{2 \tilde{\omega}}) e^{-i \hbar_x \partial_x} \right], \\[5 pt]
\widehat{H}''_{\text{B}}{}^{\hspace{-0 cm}(2)} \,=\, & \ldots.
\end{split}
\end{equation}
The discrete energy levels and eigenfunctions will admit similar expansions:
\begin{equation}
\begin{split}
E'^{(\text{B})}_n & \,=\, E'^{(0)}_n + Q_{\text{5d}} E'^{(1)}_n + Q_{\text{5d}}^2 E'^{(2)}_n + O(Q_{\text{5d}}^3) , \\
\Phi_n^{(\text{B})}(x, Q_{\text{5d}}) & \,=\, \Phi_n^{(0)}(x) + Q_{\text{5d}} \Phi_n^{(1)}(x) + Q_{\text{5d}}^2 \Phi_n^{(2)}(x) + O(Q_{\text{5d}}^3).
\end{split}
\end{equation}
We already know from Section \ref{SectRS} that
\begin{equation}
E'^{(0)}_n = E^{(T)}_n = e^{\frac{\pi a_n}{2\tilde{\omega}}} + e^{-\frac{\pi a_n}{2\tilde{\omega}}}
\end{equation}
with $a_n$ as in \eqref{quanttRS}, while $\Phi_n^{(0)}(x)$ will be the normalized Rogers polynomial \eqref{Rogersnorm}:
\begin{equation}
\Phi_n^{(0)}(x) \equiv \vert \Phi_n^{(0)} \rangle = G_n(\tfrac{\pi x}{\tilde{\omega}}; e^{-\frac{\pi g_x}{\tilde{\omega}}} \vert e^{-\frac{\pi \hbar_x}{\tilde{\omega}}}).
\end{equation}
What we have to do then is to solve the transformed 2-eRS$_{\text{B}}$ problem
\begin{equation}
\widehat{H}''_{\text{B}} \, \Phi_n^{(\text{B})}(x, Q_{\text{5d}}) =  E'^{(\text{B})}_n \Phi_n^{(\text{B})}(x, Q_{\text{5d}})
\end{equation}
order by order in the $Q_{\text{5d}}$ expansion. This can be done if we further consider expanding all $\Phi^{(l)}_n(x) \equiv \vert \Phi^{(l)}_n \rangle$, $l \geqslant 1$ in terms of the 
basis $\Phi_n^{(0)}(x) \equiv  \vert  \Phi_n^{(0)} \rangle$, i.e.
\begin{equation}
\vert \Phi^{(l)}_n \rangle = \sum_{m = 0}^{\infty} c^{(l)}_{nm} \vert \Phi_m^{(0)} \rangle,
\end{equation}
where at the practical level the summation range can be truncated to $0 \leqslant m \leqslant n +2 l$ because of the properties of $\widehat{H}''_{\text{B}}{}^{\hspace{-0 cm}(l)}$. In this way we find for example\footnote{Computing $E'^{(l)}_n$ for generic $n$ as we did in Section \ref{seceCMB} would require the knowledge of a set of identities between $\Phi^{(0)}_n(x \pm i \hbar_x)$ and $\Phi^{(0)}_n(x)$  which we are not aware of.} 
\begin{equation}
\begin{split}
E'^{(\text{B})}_0 & = e^{\pi\frac{g_x}{2 \tilde{\omega}}} + e^{-\pi \frac{g_x}{2 \tilde{\omega}}} + Q_{\text{5d}} \dfrac{(e^{\pi \frac{g_x}{2 \tilde{\omega}}} - e^{-\pi \frac{g_x}{2 \tilde{\omega}}})(e^{\pi \frac{2 g_x}{2 \tilde{\omega}}} - e^{-\pi \frac{2 g_x}{2 \tilde{\omega}}})(e^{\pi \frac{g_x - \hbar_x}{2 \tilde{\omega}}} - e^{-\pi \frac{g_x - \hbar_x}{2 \tilde{\omega}}})}{e^{\pi \frac{g_x + \hbar_x}{2 \tilde{\omega}}} - e^{-\pi \frac{g_x + \hbar_x}{2 \tilde{\omega}}}} + O(Q^2_{\text{5d}}), \\
E'^{(\text{B})}_1 & = e^{\pi \frac{g_x + \hbar_x}{2 \tilde{\omega}}} + e^{-\pi \frac{g_x + \hbar_x}{2 \tilde{\omega}}} + Q_{\text{5d}} \dfrac{(e^{\pi \frac{g_x}{2 \tilde{\omega}}} - e^{-\pi \frac{g_x}{2 \tilde{\omega}}})(e^{\pi \frac{g_x + \hbar_x}{2 \tilde{\omega}}} + e^{-\pi \frac{g_x + \hbar_x}{2 \tilde{\omega}}})(e^{\pi \frac{g_x - \hbar_x}{2 \tilde{\omega}}} - e^{-\pi \frac{g_x - \hbar_x}{2 \tilde{\omega}}})^2}{e^{\pi \frac{g_x + 2\hbar_x}{2 \tilde{\omega}}} - e^{-\pi \frac{g_x + 2\hbar_x}{2 \tilde{\omega}}}} + O(Q^2_{\text{5d}}), \\
E'^{(\text{B})}_2 & = \ldots \label{list}
\end{split}
\end{equation}
Higher order terms in the $Q_{\text{5d}}$ expansion can easily be obtained with the help of some computer program, at least for the lowest energy states (lowest values of $n$).

\subsubsection*{Numerical study}

In those cases for which studying this problem analytically requires too much computational power, 
we can still compute the 2-eRS$_{\text{B}}$ spectrum numerically.
For numerical computations it may be easier to work with $\widehat{H}_{\text{B}}$ \eqref{ellRSBp} rather than $\widehat{H}'_{\text{B}}$ (or $\widehat{H}''_{\text{B}}$), due to the properties of the Weierstrass $\sigma$-function. In this case we need to diagonalize the matrix (truncated to some finite size) constructed out of the matrix elements
\begin{equation}
\langle m \vert \widehat{H}_{\text{B}} \vert n \rangle = 
\int_{0}^{2\tilde{\omega}} dx\, \psi_m(x) \widehat{H}_{\text{B}} \psi_n(x) ,
\end{equation}
at fixed $0 < g_x < \hbar_x + 2\tilde{\omega}$ and $Q_{\text{5d}}$. Here we are using the $L^2([0,2\tilde{\omega}])$ 2-tRS orthonormal basis $\psi_n^{(\text{T})}(x) = \vert n \rangle$ \eqref{tRSsolnorm} to perform diagonalization. The expectation is that by increasing the size of the matrix, the numerical eigenvalues will converge to the actual energies of the 2-eRS$_{\text{B}}$ system; in this way we can compute both the numerical spectrum and the numerical eigenfunctions of the 2-eRS$_{\text{B}}$ system.

\subsection{Elliptic case: A-model}

Lastly, let us comment on the 2-eRS$_{\text{A}}$ quantum mechanical problem in the form \eqref{ellRSA}, i.e.
\begin{equation}
\begin{split}
\widehat{H}'_{\text{A}} \psi^{(\text{A})}(y,Q_{\text{5d}}) & = 
\Bigg[ \sqrt{\dfrac{\theta_1\big(i \pi \frac{y + i g_y}{2\tilde{\omega}} , Q_{\text{5d}}^{1/2} \big)}{\theta_1(i \pi \frac{y}{2 \tilde{\omega}}, Q_{\text{5d}}^{1/2})}} e^{i \hbar_y \partial_{y}} 
\sqrt{\dfrac{\theta_1\big(i\pi \frac{y - i g_y}{2\tilde{\omega}} , Q_{\text{5d}}^{1/2} \big) }{\theta_1(i\pi \frac{y}{2\tilde{\omega}} , Q_{\text{5d}}^{1/2})}}  \\[10 pt] 
& + \sqrt{\dfrac{\theta_1\big(i\pi \frac{y - i g_y}{2\tilde{\omega}} , Q_{\text{5d}}^{1/2} \big)}{\theta_1(i\pi \frac{y}{2 \tilde{\omega}}, Q_{\text{5d}}^{1/2})}} e^{-i \hbar_y \partial_{y}} 
\sqrt{\dfrac{\theta_1\big(i\pi \frac{y + i g_y}{2\tilde{\omega}} , Q_{\text{5d}}^{1/2} \big) }{\theta_1(i\pi \frac{y}{2\tilde{\omega}} , Q_{\text{5d}}^{1/2})}} \Bigg] \psi^{(\text{A})}(y,Q_{\text{5d}}) \\[10 pt]
& = E'^{(\text{A})} \psi^{(\text{A})}(y,Q_{\text{5d}}). \label{ellRSAbis}
\end{split}
\end{equation}
For the same reasons of its 2-eCM$_{\text{A}}$ counterpart, it seems quite complicated to solve this problem analytically on $L^2([0,2\vert \tilde{\omega}' \vert])$ in terms of perturbation theory around $Q_{\text{5d}} = 0$ (or $\tilde{\omega}' \rightarrow i \infty$); in addition, already the solution to the 2-hRS system is not entirely understood. It may however be possible to construct eigenfunctions using the quantum Separation of Variables approach, which involves the study of entire solutions to the associated Baxter equation; we will sketch how gauge theory would suggest us to proceed in Section \ref{eRSeigenfunctions}. \\

Fortunately, for our purposes it will be sufficient to study the 2-eRS$_{\text{A}}$ problem numerically. In order to perform numerical diagonalization, it will again be more convenient to work with the Hamiltonian $\widehat{H}_{\text{A}}$ \eqref{ellRSAp} rather than $\widehat{H}'_{\text{A}}$. In fact, according to what we discussed around \eqref{relationeRS},
diagonalizing $\widehat{H}_{\text{A}}$ with periods $(\tilde{\omega}, \tilde{\omega}')$ is equivalent to diagonalizing $\widehat{H}_{\text{B}}$ with exchanged periods $(-i\tilde{\omega}', i \tilde{\omega})$ once we identify $\hbar_y \leftrightarrow \hbar_x$, $g_y \leftrightarrow g_x$, $y \leftrightarrow x$, and as a consequence
\begin{equation}
E^{(\text{A})}(\hbar_y, g_y, \tilde{\omega}, \tilde{\omega}') = 
E^{(\text{B})}(\hbar_y, g_y, -i\tilde{\omega}', i\tilde{\omega}).
\end{equation}
We can therefore apply the same procedure described in Section \ref{eRSBsection} to compute the spectrum $E^{(\text{A})}$ of the 2-eRS$_{\text{A}}$ model $\widehat{H}_{\text{A}}$ numerically, although we will need to use the $L^2([0,2\vert \tilde{\omega}' \vert])$ orthonormal basis $\vert n \rangle = \psi_n^{(\text{T})}(y)$ \eqref{tRSsolnorm} with half-period $\vert \tilde{\omega}' \vert = -\frac{1}{2} \frac{\pi}{\tilde{\omega}} \ln Q_{\text{5d}}$ rather than $\tilde{\omega}$ to perform diagonalization.

\subsection{Ruijsenaars-Schneider energy spectrum from gauge theory} \label{seceRSenergygauge}

\renewcommand{\arraystretch}{1.5}
\begin{table}
\begin{center}
\begin{tabular}{|c|c|c|c|}
\hline 
Gauge theory (NS) & 2-eRS model (complex) & 2-eRS$_{\text{B}}$ model & 2-eRS$_{\text{A}}$ model \\ 
\hline 
$\epsilon_1$ & $\hbar$ & $\hbar_x$ & $i \hbar_y$ \\ 
\hline 
$m$ & $g$ & $g_x$ & $i g_y$ \\ 
\hline 
$R^{-1}$ & $2\tilde{\omega}$ & $2\tilde{\omega}$ & $2\tilde{\omega}$ \\
\hline
$Q_{\text{5d}}$ & $e^{2\pi i \frac{\tilde{\omega}'}{\tilde{\omega}}}$ & 
$e^{2\pi i \frac{\tilde{\omega}'}{\tilde{\omega}}}$ & $e^{2\pi i \frac{\tilde{\omega}'}{\tilde{\omega}}}$ \\ 
\hline 
$ - \frac{i}{2} \frac{\tilde{\omega}}{\pi} \ln Q_{\text{5d}}$ &  $\tilde{\omega}'$ & $\tilde{\omega}'$ & $\tilde{\omega}'$ \\ 
\hline 
\end{tabular} \caption{Map of 5d $SU(2)$ $\mathcal{N} = 1^*$ gauge theory and 2-eRS$_{\text{B}}$, 2-eRS$_{\text{A}}$ parameters.} \label{mapRS}
\end{center}
\end{table}
\renewcommand{\arraystretch}{1}

We have seen in the previous Sections how the 2-eRS$_{\text{B}}$ and 2-eRS$_{\text{A}}$ problems, as well as their 2-tRS and 2-hRS limits, can be solved (analytically or numerically) with the standard tools of quantum mechanics. We now want to understand if it is possible to reproduce the same results from gauge theory arguments, at least at the level of energy spectrum. 

As we mentioned in the Introduction, it should be possible to do so by considering five-dimensional gauge theories or topological string theory quantities in the NS limit \citep{2010maph.conf..265N}, although for relativistic quantum integrable systems non-perturbative contributions in $\hbar$ are also expected \cite{2015arXiv151102860H,Franco:2015rnr}.
Based on the proposal by \citep{2010maph.conf..265N} and the analogy with the 2-eCM case, the 2-eRS$_{\text{B}}$ and 2-eRS$_{\text{A}}$ systems are expected to be related to the $\mathcal{N} = 1^*$ $SU(2)$ supersymmetric gauge theory on $\mathbb{R}^4_{\epsilon_1,\epsilon_2} \times S^1_R$, always in the NS limit $\epsilon_2 \rightarrow 0$. 
The dictionary between gauge theory and 2-eRS$_{\text{B}}$, 2-eRS$_{\text{A}}$ models is given in Table \ref{mapRS}, where in particular $R$ is the radius of the $S^1_R$ circle while $Q_{\text{5d}} = e^{2\pi i \tilde{\tau}}$ is the instanton counting parameter (with $\tilde{\tau}$ inverse gauge coupling); the real VEV $a$ of the scalar field in the $\mathcal{N} = 1$ vector multiplet will also play a role in what follows as a different way of parametrizing the energy.

As in Section \ref{seceCMenergygauge}, we should proceed in two steps:

\subsection*{Step I: general analysis}

First, we would need to construct two (generically) linearly independent formal solutions $\psi^{(i)}(z,a,\hbar,g,\tilde{\omega},\tilde{\omega}')$, $i = 1,2$ to the \textit{complex} 2-particle elliptic Ruijsenaars-Schneider system $\widehat{H}_{\text{eRS}}$ \eqref{ellRS2p}, such that
\begin{equation}
\begin{split}
\widehat{H}_{\text{eRS}} \psi^{(i)}(z,a,\hbar,g,\tilde{\omega},\tilde{\omega}') 
= E(a,\hbar,g,\tilde{\omega},\tilde{\omega}') \psi^{(i)}(z,a,\hbar,g,\tilde{\omega},\tilde{\omega}') , \label{eq}
\end{split}
\end{equation}
for any generic $z,a,\hbar,g \in \mathbb{C}$; clearly these will also solve 
\begin{equation}
\widehat{H}'_{\text{eRS}} \psi^{(i)}(z,a,\hbar,g,\tilde{\omega},\tilde{\omega}') 
= E'(a,\hbar,g,\tilde{\omega},\tilde{\omega}') \psi^{(i)}(z,a,\hbar,g,\tilde{\omega},\tilde{\omega}'), \label{eq2}
\end{equation}
with $\widehat{H}'_{\text{eRS}}$ as in \eqref{ellRS3} and
\begin{equation}
E(a,\hbar,g,\tilde{\omega},\tilde{\omega}')  = 
\exp \left( -\frac{ g(g-\hbar)}{24}
\frac{\pi^2}{\tilde{\omega}^2} E_2(Q_{\text{5d}}) \right)
E'(a,\hbar,g,\tilde{\omega},\tilde{\omega}') . \label{en}
\end{equation}
As for the elliptic Calogero-Moser case, five-dimensional gauge theory provides explicit, convergent $Q_{\text{5d}}$-series expressions for the energy and the formal solutions as functions of $a$; these again coincide with the NS limit of VEVs of codimension four and two defects with $\hbar = \epsilon_1$, $g = m$, $2\tilde{\omega} = R^{-1}$:
\begin{equation}
\begin{array}{ccc}
E'(a, \hbar, g, \tilde{\omega},\tilde{\omega}') &\;\;\;\; \Longleftrightarrow \;\;\;\; & \text{codim. 4 defect} \; \langle W^{SU(2)}_{\mathbf{2}}(a, \epsilon_1, m, R, Q_{\text{5d}}) \rangle_{\text{NS}}, \\[6 pt] 
\psi^{(i)}(z, a, \hbar, g, \tilde{\omega},\tilde{\omega}') & \;\;\;\; \Longleftrightarrow \;\;\;\; & \text{codim. 2 defect (monodromy, NS)}.
\end{array} 
\end{equation}
The relevant codimension two defects are still of monodromy (or Gukov-Witten) type: their explicit computation can be found for example in \cite{Bullimore:2014awa}, where it was also checked that such defects satisfy \eqref{eq2} at the first few orders in $Q_{\text{5d}}$. 
The codimension four defect associated to the energy is instead a line observable, and more precisely corresponds to the Wilson loop in the fundamental representation $\mathbf{2}$ of $SU(2)$ wrapping $S^1_R$, again as explained in \cite{Bullimore:2014awa}: its first few orders in the $Q_{\text{5d}}$ expansion are (from the formulae in Appendix \ref{appgauge})
\begin{equation}
\begin{split}
& E'(a, \hbar, g, \tilde{\omega},\tilde{\omega}') = \langle W_{\mathbf{2}}^{SU(2)}(a, \epsilon_1, m, R, Q_{\text{5d}}) \rangle_{\text{NS}}\Big\vert_{ \epsilon_1 = \hbar, m = g, R^{-1} = 2\tilde{\omega} } \\
&= \alpha^{1/2} + \alpha^{-1/2} - Q_{\text{5d}} \dfrac{(\alpha^{1/2} + \alpha^{-1/2})(1-\mu)^2(1-q_1 \mu^{-1})^2}{(1-q_1 \alpha)(1-q_1 \alpha^{-1})} + O(Q_{\text{5d}}^2), \label{wilson}
\end{split}
\end{equation}
where (for $a_1 = -a_2 = \frac{a}{2}$)
\begin{equation}
\alpha = e^{2\pi R a} = e^{\frac{2 \pi a}{2\tilde{\omega}}} \,,\;\;\;\;
q_1 = e^{2\pi R \epsilon_1} = e^{\frac{2 \pi \hbar}{2\tilde{\omega}}} \,,\;\;\;\;
\mu = e^{2\pi R m} = e^{\frac{2 \pi g}{2\tilde{\omega}}}. \label{par}
\end{equation}
This gauge theory expression for the energy 
 enjoys many symmetries, in particular it is invariant under the simultaneous exchange $(q_1, \mu) \leftrightarrow (q_1^{-1}, \mu^{-1})$ as well as under $\alpha \leftrightarrow \alpha^{-1}$.

\subsection*{Step II: Hilbert integrability}

Having constructed the formal solution to the generic complex 2-eRS system \eqref{eq2} from gauge theory, we can start discussing about Hilbert integrability. For generic values of $a$ (i.e. of the energy), our formal solution is not expected to belong to any Hilbert space of some quantum mechanical system, such as 2-eRS$_{\text{B}}$ or 2-eRS$_{\text{A}}$; however, as for the 2-eCM case, according to \citep{2010maph.conf..265N} Hilbert integrability should emerge in two special cases: 

\renewcommand{\arraystretch}{1.5}
\begin{table}
\begin{center}
\begin{tabular}{|c|c|c|}
\hline 
$\tilde{\omega} = \pi$, $\;\tilde{\omega}' = -\frac{i}{2} \ln(10^{-3})$ & $E^{(\text{B})}_{0}$ & $E^{(\text{B})}_{1}$ \\ 
\hline 
Gauge theory - $O(Q_{\text{5d}})$ & \underline{2.84}510702564\ldots & 
\underline{4.33}733466191\ldots \\ 
\hline 
Gauge theory - $O(Q_{\text{5d}}^2)$ & \underline{2.84895}354164\ldots & 
\underline{4.33886}116831\ldots \\ 
\hline 
Gauge theory - $O(Q_{\text{5d}}^3)$ & \underline{2.8489552}7508\ldots & 
\underline{4.338869}80397\ldots
\\ 
\hline 
Gauge theory - $O(Q_{\text{5d}}^4)$ & \underline{2.8489552886}5\ldots & 
\underline{4.338869797}49\ldots \\ 
\hline 
Numerical & \underline{2.84895528863}\ldots & \underline{4.33886979750}\ldots \\ 
\hline 
\end{tabular} \caption{Energies for the 2-eRS$_{\text{B}}$ model \eqref{ellRSBp} at $g_x = 2$, $\hbar_x = 1$.} \label{ex5}
\end{center}
\end{table}
\renewcommand{\arraystretch}{1}

\renewcommand{\arraystretch}{1.5}
\begin{table}
\begin{center}
\begin{tabular}{|c|c|c|}
\hline 
$\tilde{\omega} = \frac{\pi^2}{3}$, $\; \tilde{\omega}' = i \pi$ & $E^{(\text{B})}_{0}$ & $E^{(\text{B})}_{1}$ \\ 
\hline 
Gauge theory - $O(Q_{\text{5d}})$ & \underline{3}.39999656871\ldots & 
\underline{6.31}299288979\ldots \\ 
\hline 
Gauge theory - $O(Q_{\text{5d}}^2)$ & \underline{3.4163}2604117\ldots & 
\underline{6.318}80651657\ldots \\ 
\hline 
Gauge theory - $O(Q_{\text{5d}}^3)$ & \underline{3.416350}25554\ldots & 
\underline{6.31891}130483\ldots
\\ 
\hline 
Gauge theory - $O(Q_{\text{5d}}^4)$ & \underline{3.41635065}566\ldots & 
\underline{6.31891099}761\ldots \\ 
\hline 
Numerical & \underline{3.41635065432}\ldots & \underline{6.31891099879}\ldots \\ 
\hline 
\end{tabular} \caption{Energies for the 2-eRS$_{\text{B}}$ model \eqref{ellRSBp} at $g_x = \sqrt{7}$, $\hbar_x = \sqrt{2}$.} \label{ex6}
\end{center}
\end{table}
\renewcommand{\arraystretch}{1}

\paragraph{2-eRS$_{\text{B}}$ model:}
Fixing $\epsilon_1 = \hbar_x \in \mathbb{R}_+$ and $m = g_x \in \mathbb{R}_+$, if $a$ satisfies the B-type quantization conditions
\begin{equation}
a = g_x + n \hbar_x , \;\;\;\; n \in \mathbb{N}, \label{quantRSB}
\end{equation}
the gauge theory expression \eqref{wilson} is conjecturally expected to reproduce the discrete energy levels for the 2-eRS$_{\text{B}}$ model $\widehat{H}'_{\text{B}}$ \eqref{ellRSB} with Hilbert space $L^2([0,2\tilde{\omega}])$. 
This can indeed be checked analytically, for example by comparing \eqref{wilson} with \eqref{quantRSB} imposed against our previous computations \eqref{list}.
We can also perform some numerical test: remarkably enough, as we can see from Tables \ref{ex5} and \ref{ex6} our gauge theory results seem to converge to the numerical ones when adding more and more instanton corrections.\footnote{In the Tables we compute the numerical energy values $E$ for the model \eqref{ellRSBp} starting from $E'$ given by the gauge theory Wilson loop \eqref{wilson} and using the relation \eqref{en}.} It is also important to remark that the gauge theory proposal correctly reproduces the known 2-tRS energy spectrum \eqref{tRSenergy} in the trigonometric limit $Q_{\text{5d}} \rightarrow 0$, and that B-type quantization conditions for 2-eRS$_{\text{B}}$ \eqref{quantRSB} coincide with the 2-tCM quantization conditions \eqref{quanttRS}, i.e. they do not receive corrections in $Q_{\text{5d}}$. 

\paragraph{2-eRS$_{\text{A}}$ model:} Fixing $\epsilon_1 = i \hbar_y$ and $m = i g_y$ with $\hbar_y, g_y \in \mathbb{R}_+$, the conjecture by \citep{2010maph.conf..265N} states that if $a$ satisfies the A-type quantization conditions

\begin{equation}
\begin{split}
a_D^\text{na\"ive} & \; :=\; -\dfrac{4\pi i \tilde{\omega}' a}{2\tilde{\omega} \hbar_y} + \dfrac{2\pi}{2\tilde{\omega} \hbar_y}g_{y}(g_y - 2\tilde{\omega} - \hbar_y) \\
& + 2 \sum_{k \geqslant 1} \dfrac{1}{k} \dfrac{\cos(\frac{\pi k \hbar_y}{2\tilde{\omega}})}{\sin(\frac{\pi k \hbar_y}{2\tilde{\omega}})} e^{-\frac{2\pi k a}{2\tilde{\omega}}}
- 2 \sum_{k \geqslant 1} \dfrac{1}{k} \dfrac{\cos(\frac{\pi k (\hbar_y - 2 g_y)}{2\tilde{\omega}})}{\sin(\frac{\pi k \hbar_y}{2\tilde{\omega}})} e^{-\frac{2\pi k a}{2\tilde{\omega}}}
\\
& -2i \frac{\partial}{\partial a} \left[ 
\mathcal{W}_{\text{5d,inst}}^{SU(2)}(a, i\hbar_y, ig_y, (2\tilde{\omega})^{-1}, e^{\frac{2\pi i (2\tilde{\omega}')}{2\tilde{\omega}}}) \right] = 2\pi n \, , \;\;\;\; n \in \mathbb{N}, \label{quantRSAwrong}
\end{split}
\end{equation}
our gauge theory expression \eqref{wilson} should reproduce the discrete energy levels for the 2-eRS$_{\text{A}}$ model with Hilbert space $L^2([0,2\vert\tilde{\omega}'\vert])$. 
What enters in the A-type quantization conditions is the total (classical + 1-loop + instanton) twisted effective superpotential of the five-dimensional $\mathcal{N} = 1^*$ $SU(2)$ theory on flat space $\mathbb{R}^4_{\epsilon_1, \epsilon_2} \times S^1_R$, whose instanton part defined as (see Appendix \ref{appgauge})
\begin{equation}
\mathcal{W}^{SU(2)}_{\text{5d,\,inst}}(a,\epsilon_1,m,R,Q_{\text{5d}}) 
= \lim_{\epsilon_2 \to 0} \left[ - \epsilon_2 \ln Z^{SU(2)}_{\text{5d,\,inst}}(a,\epsilon_1,\epsilon_2,m,R,Q_{\text{5d}}) \right] 
\end{equation}
admits a $Q_{\text{5d}}$ expansion whose first orders are
\begin{equation}
\begin{split}
& \mathcal{W}^{SU(2)}_{\text{5d,\,inst}}(a,\epsilon_1,m,R,Q_{\text{5d}})  = \\
& - Q_{\text{5d}} \dfrac{(1-\mu)(1-q_1\mu^{-1})}{(1-q_1)(1-q_1\alpha)(1-q_1\alpha^{-1})}
\Big[ (1+\mu)(1+q_1^2 \mu^{-1}) +q_1(\mu+\mu^{-1}) \\
& \hspace*{6 cm} -2q_1(\alpha + \alpha^{-1}) -2q_1 \Big] + O(Q_{\text{5d}}^2) ,
\end{split}
\end{equation}
with parameters as in \eqref{par}. Despite not having analytic results for 2-eRS$_{\text{A}}$, we can still test whether the gauge theory conjecture by \citep{2010maph.conf..265N} makes sense by checking it against the energy spectrum computed numerically: however, even if it seems to work for the special case $g_y = 2 \hbar_y$\footnote{This is also the special case discussed in \cite{0305-4470-32-9-018}.} considered in Table \ref{exWKB1}, its validity seems to break down in more generic cases such as the one considered in Table \ref{exWKB2}, in which the gauge theory energies tend to converge to a value which does not correspond to the numerical one.

\renewcommand{\arraystretch}{1.5}
\begin{table}
\begin{center}
\begin{tabular}{|c|c|c|}
\hline 
$\tilde{\omega} = -\frac{1}{2} \ln(10^{-3})$, $\;\tilde{\omega}' = i \pi$ & $E^{(\text{A})}_{0}$ & $E^{(\text{A})}_{1}$ \\ 
\hline 
Gauge theory - $O(Q_{\text{5d}})$ & \underline{2.84}563654275\ldots & 
\underline{4.33}743438960\ldots \\ 
\hline 
Gauge theory - $O(Q_{\text{5d}}^2)$ & \underline{2.848}87535665\ldots & 
\underline{4.3388}3850492\ldots \\ 
\hline 
Gauge theory - $O(Q_{\text{5d}}^3)$ & \underline{2.84895}429911\ldots & 
\underline{4.338869}48412\ldots
\\ 
\hline 
Gauge theory - $O(Q_{\text{5d}}^4)$ & \underline{2.8489552}7734\ldots & 
\underline{4.33886979}429\ldots \\ 
\hline 
Numerical & \underline{2.84895528864}\ldots & \underline{4.33886979750}\ldots \\ 
\hline 
\end{tabular} \caption{Gauge theory energies for the 2-eRS$_{\text{A}}$ model \eqref{ellRSAp} at $g_y = 2$, $\hbar_y = 1$ \\ obtained from the na\"ive WKB quantization conditions \eqref{quantRSAwrong}.} \label{exWKB1}
\end{center}
\end{table}
\renewcommand{\arraystretch}{1}

\renewcommand{\arraystretch}{1.5}
\begin{table}
\begin{center}
\begin{tabular}{|c|c|c|}
\hline 
$\tilde{\omega} = \pi$, $\;\tilde{\omega}' = \frac{i \pi^2}{3}$ & $E^{(\text{A})}_{0}$ & $E^{(\text{A})}_{1}$ \\ 
\hline 
Gauge theory - $O(Q_{\text{5d}})$ & \underline{3.41}360804483\ldots & 
\underline{6.31}772996010\ldots \\ 
\hline 
Gauge theory - $O(Q_{\text{5d}}^2)$ & \underline{3.416}42587059\ldots & 
\underline{6.3189}0324086\ldots \\ 
\hline 
Gauge theory - $O(Q_{\text{5d}}^3)$ & \underline{3.416}44610533\ldots & 
\underline{6.31891}130715\ldots
\\ 
\hline 
Gauge theory - $O(Q_{\text{5d}}^4)$ & \underline{3.416}44619284\ldots & 
\underline{6.31891}133951\ldots \\ 
\hline 
Numerical & \underline{3.41635065432}\ldots & \underline{6.31891099879}\ldots \\ 
\hline  
\end{tabular} \caption{Gauge theory energies for the 2-eRS$_{\text{A}}$ model \eqref{ellRSAp} at $g_y = \sqrt{7}$, $\hbar_y = \sqrt{2}$ \\ obtained from the na\"ive WKB quantization conditions \eqref{quantRSAwrong}.} \label{exWKB2}
\end{center}
\end{table}
\renewcommand{\arraystretch}{1}

However, this is not completely unexpected. As we mentioned in the Introduction, we already know that for ``relativistic'' integrable systems such as relativistic Toda chains \cite{2015arXiv151102860H} and relativistic cluster integrable systems \cite{Franco:2015rnr}, the na\"ive all-order WKB quantization conditions obtained by the proposal of \citep{2010maph.conf..265N} are ill-defined and need to be corrected by non-perturbative terms in $\hbar_y$. Exactly the same situation arises in our case: indeed the na\"ive WKB quantization conditions \eqref{quantRSAwrong} are ill-defined for $\frac{\hbar_y}{2\tilde{\omega}} \in \mathbb{Q}$, as we can see by noticing that at those values they develop poles. In order to fix this problem we can proceed as in \cite{2015arXiv151102860H,Franco:2015rnr}, where it was pointed out that the poles at $\frac{\hbar_y}{2\tilde{\omega}} \in \mathbb{Q}$ disappear if we complete \eqref{quantRSAwrong} by adding non-perturbative contributions in $\hbar_y$; these can be completely determined by requiring the full (WKB + non-perturbative) exact quantization conditions to be invariant under ``modular duality'', i.e. under the exchange $\hbar_y \leftrightarrow 2\tilde{\omega}$. For the cases considered in \cite{2015arXiv151102860H,Franco:2015rnr} modular duality is encoded in the quantum group structure underlying the relativistic integrable system; in our 2-eRS case such modular duality also appears, and was already noticed in a number of papers \cite{0305-4470-32-9-018,Ruijsenaars1999,2015SIGMA..11..004R}.  

According to this prescription therefore the full A-type quantization conditions for 2-eRS$_{\text{A}}$ should be given by
\begin{equation}
\begin{split}
a_D :=
& -\dfrac{4\pi i \tilde{\omega}' a}{2\tilde{\omega} \hbar_y} + \dfrac{2\pi}{2\tilde{\omega} \hbar_y}g_{y}(g_y - 2\tilde{\omega} - \hbar_y) \\
& + 2 \sum_{k \geqslant 1} \dfrac{1}{k} \dfrac{\cos(\frac{\pi k \hbar_y}{2\tilde{\omega}})}{\sin(\frac{\pi k \hbar_y}{2\tilde{\omega}})} e^{-\frac{2\pi k a}{2\tilde{\omega}}}
- 2 \sum_{k \geqslant 1} \dfrac{1}{k} \dfrac{\cos(\frac{\pi k (\hbar_y - 2 g_y)}{2\tilde{\omega}})}{\sin(\frac{\pi k \hbar_y}{2\tilde{\omega}})} e^{-\frac{2\pi k a}{2\tilde{\omega}}}
\\
& + 2 \sum_{k \geqslant 1} \dfrac{1}{k} \dfrac{\cos(\frac{2 \pi k \tilde{\omega}}{\hbar_y})}{\sin(\frac{2 \pi k \tilde{\omega}}{\hbar_y})} e^{-\frac{2\pi k a}{\hbar_y}}
- 2 \sum_{k \geqslant 1} \dfrac{1}{k} \dfrac{\cos(\frac{\pi k (2\tilde{\omega} - 2 g_y)}{\hbar_y})}{\sin(\frac{2 \pi k \tilde{\omega}}{\hbar_y})} e^{-\frac{2\pi k a}{\hbar_y}}
\\
& -2i \dfrac{\partial}{\partial a} \Bigg[ 
\mathcal{W}_{\text{5d,inst}}^{SU(2)}(a, i\hbar_y, ig_y, (2\tilde{\omega})^{-1}, e^{\frac{2\pi i  (2\tilde{\omega}')}{2\tilde{\omega}}}) \\
& \hspace{1.3 cm} + 
\mathcal{W}_{\text{5d,inst}}^{SU(2)}(a, i 2\tilde{\omega}, ig_y, \hbar_y^{-1}, e^{\frac{2\pi i (2\tilde{\omega}')}{\hbar_y}}) \Bigg] 
= 2\pi n \, , \;\;\;\; n \in \mathbb{N}, \label{quantRSAtrue}
\end{split}
\end{equation}
where the symmetry under exchange $\hbar_y \leftrightarrow 2\tilde{\omega}$ is manifest. Poles at $\frac{\hbar_y}{2\tilde{\omega}} \in \mathbb{Q}$ disappear in these exact quantization conditions almost by construction, as one can show by following the same arguments used in \cite{2015arXiv151102860H,Franco:2015rnr}.\footnote{In \cite{2015arXiv151102860H,Franco:2015rnr} in order to ensure poles cancellation one sometimes needs to shift $a$, or the K\"ahler parameters, by a constant B-field. For the $N$-particle relativistic closed Toda chain, the B-field is only needed for $N$ odd; in our elliptic systems instead it seems that the B-field is not needed for any $N$.} Having such well-defined quantization conditions, we can now check if a ``non-perturbatively corrected'' version of the proposal by \citep{2010maph.conf..265N} can be valid: that is, if the values of $a$ satisfying \eqref{quantRSAtrue} reproduce the 2-eRS$_{\text{A}}$ numerical spectrum when inserted in the gauge theory expression for the energy. Quite remarkably, this indeed seems to be the case. This can be seen from the example in Table \ref{ex8}: differently from the previous version of this example (Table \ref{exWKB2}) in which we were using the na\"ive WKB quantization conditions, it is now evident that the gauge theory results approach the numerical ones by increasing the number of instanton corrections considered. Notice also that nothing changes for the example of Table \ref{exWKB1}, since we can easily see that non-perturbative corrections vanish when $g_y = k \hbar_y$ with $k \in \mathbb{N}$.

Finally, let us mention that we can also think of the exact A-type quantization conditions \eqref{quantRSAtrue} as an S-dual (electro-magnetic dual) version of the B-type conditions \eqref{quantRSB} where we quantize $a_D$ instead of $a$, with $a_D$, $a$ related as
\begin{equation}
a_D = -2i \partial_a \mathcal{W}^{SU(2)}_{\text{full}} (a)
; \label{rrr}
\end{equation}
here $\mathcal{W}^{SU(2)}_{\text{full}}$ is the full, non-perturbatively corrected twisted effective superpotential which contains also classical and 1-loop part. Equation \eqref{rrr} can be interpreted as an $\epsilon_1$-deformed version of the Seiberg-Witten relation between $a$ and $a_D$: however, differently from the analogue formula \eqref{finale} valid for four-dimensional theories, in the five-dimensional case non-perturbative corrections in $\epsilon_1$ are essential in order to restore S-duality when the Omega background is turned on. We will see in Section \ref{eRSeigenfunctions} how exact A-type quantization conditions may arise from a quantum Separation of Variables approach to the 2-eRS$_{\text{A}}$ problem.

\renewcommand{\arraystretch}{1.5}
\begin{table}
\begin{center}
\begin{tabular}{|c|c|c|}
\hline 
$\tilde{\omega} = \pi$, $\;\tilde{\omega}' = \frac{i \pi^2}{3}$ & $E^{(\text{A})}_{0}$ & $E^{(\text{A})}_{1}$ \\ 
\hline 
Gauge theory - $O(Q_{\text{5d}})$ & \underline{3.41}351261316\ldots & 
\underline{6.31}772961937\ldots \\ 
\hline 
Gauge theory - $O(Q_{\text{5d}}^2)$ & \underline{3.4163}3033157\ldots & 
\underline{6.3189}0290001\ldots \\ 
\hline 
Gauge theory - $O(Q_{\text{5d}}^3)$ & \underline{3.416350}56648\ldots & 
\underline{6.3189109}6630\ldots
\\ 
\hline 
Gauge theory - $O(Q_{\text{5d}}^4)$ & \underline{3.41635065}398\ldots & 
\underline{6.318910998}66\ldots \\ 
\hline 
Numerical & \underline{3.41635065432}\ldots & \underline{6.31891099879}\ldots \\ 
\hline  
\end{tabular} \caption{Gauge theory energies for the 2-eRS$_{\text{A}}$ model \eqref{ellRSAp} at $g_y = \sqrt{7}$, $\hbar_y = \sqrt{2}$, obtained from the exact quantization conditions \eqref{quantRSAtrue}.} \label{ex8}
\end{center}
\end{table}
\renewcommand{\arraystretch}{1}

To summarize, we have seen that although the conjecture by \citep{2010maph.conf..265N} seems to work nicely for the 2-eRS$_{\text{B}}$ system, it is not consistent in the 2-eRS$_{\text{A}}$ case: first of all the conjectural quantization conditions are ill-defined, that is have poles for $\frac{\hbar_y}{2\tilde{\omega}} \in \mathbb{Q}$, and even if we consider them for non-singular values of $\hbar_y$ the gauge theory results are in disagreement with the 2-eRS$_{\text{A}}$ numerical spectrum. This happens because the proposal by \citep{2010maph.conf..265N} does not take into account possible non-perturbative contributions in $\hbar_y$; in fact once these are properly taken into account, in a way dictated by modular duality \cite{2015arXiv151102860H,Franco:2015rnr}, we obtain pole-free quantization conditions which moreover seem to reproduce the 2-eRS$_{\text{A}}$ numerical spectrum. Therefore, similarly to what happened for the 2-eCM system, also in this case the discrete energy levels for 2-eRS$_{\text{B}}$ and 2-eRS$_{\text{A}}$ can be obtained from the \textit{same} gauge theory expression for the energy \eqref{wilson} 
 by quantizing $a$ in two different ways related by S-duality $\tilde{\tau} \leftrightarrow - \frac{1}{\tilde{\tau}}$, where however this time the S-duality relation involves non-perturbative corrections in $\epsilon_1$. 
 S-duality is also reflected in 
 \eqref{relationeRSbis}, which is indeed realized in gauge theory as evident from comparing Tables \ref{ex5}, \ref{exWKB1} or Tables \ref{ex6}, \ref{ex8} (rather than Tables \ref{ex6}, \ref{exWKB2}).

\subsection{Comments on Baxter equation and quantum Separation of Variables} \label{eRSeigenfunctions}

For the sake of completeness, in this Section we would like to comment on the possibility of constructing $L^2([0,2\vert \tilde{\omega}' \vert])$ eigenfunctions for the 2-eRS$_{\text{A}}$ system via the quantum Separation of Variables approach, appropriately reinterpreted in gauge theory language; we will follow the same line of reasoning as in Section \ref{eCMeigenfunctions}, again without any pretence of rigorousness.

The starting point in the quantum Separation of Variables approach is to consider the Baxter equation associated to our system: as we already mentioned, this is an auxiliary 1-dimensional problem which can be thought as a quantum version of the spectral/Seiberg-Witten curve. Following \cite{Nekrasov:2012xe,Nekrasov:2013xda} we can construct the Baxter equation for the 2-eRS$_{\text{A}}$ system from gauge theory by considering the NS limit of the VEV of the five-dimensional $\mathcal{N} = 1^*$ $SU(2)$ fundamental qq-character $\langle \chi_{\text{5d}}^{SU(2)}(\sigma, a, \epsilon_1, m, R, Q_{\text{5d}}) \rangle_{\text{NS}}$ \cite{Nekrasov:2015wsu,Tong:2014cha,Kim:2016qqs}. Introducing for later convenience the variables ($\sigma \in \mathbb{R}$)
\begin{equation}
X = e^{\frac{2\pi \sigma}{2\tilde{\omega}}}, \;\; \mu = e^{\frac{2 \pi i g_y}{2\tilde{\omega}}}, \;\; q_1 = e^{\frac{2\pi i \hbar_y}{2\tilde{\omega}}}, \;\;
\alpha = e^{\frac{2\pi a}{2\tilde{\omega}}}, \;\;
Q_{\text{5d}} = e^{2\pi i \frac{2\tilde{\omega}'}{2 \tilde{\omega}}},
\end{equation}
from the formulae in Appendix \ref{appgauge} we find
\begin{equation}
\big\langle \chi_{\text{5d}}^{SU(2)}(\sigma, a, \epsilon_1, m, R, Q_{\text{5d}}) \big\rangle_{\text{NS}} = X - E'(a,i \hbar_y, i g_y, \tilde{\omega}, \tilde{\omega}') + X^{-1}, \label{last}
\end{equation}
where $E' = E'(a,i \hbar_y, i g_y, \tilde{\omega}, \tilde{\omega}') $ is the VEV of the fundamental Wilson loop in the NS limit given in \eqref{wilson}.\footnote{The fundamental qq-character for five-dimensional theories can indeed be interpreted as the generating function of Wilson loops in all possible antisymmetric representations \cite{Tong:2014cha,Kim:2016qqs}.} The Baxter equation is then obtained as in \eqref{bax}; in a more compact form, using the $\overline{\theta}(X,Q)$ function defined in \eqref{thetabar} we get
\begin{equation}
X\, \overline{\theta}(\mu^{-1}q_1 e^{i \hbar_y \partial_{\sigma}},Q_{\text{5d}})
- E'\, \overline{\theta}(e^{i \hbar_y \partial_{\sigma}},Q_{\text{5d}}) 
+ X^{-1}\, \overline{\theta}(\mu q_1^{-1} e^{i \hbar_y \partial_{\sigma}},Q_{\text{5d}}) = 0, \label{eRSAbax}
\end{equation}
which again is a complicated infinite-order finite-difference operator on $\mathbb{R}$. However, we already know from previous examples \cite{Kharchev:2001rs,2012arXiv1206.3787H,2016arXiv160706672H,Sciarappa:2017hds,Kashaev:2017zmv,2018arXiv180306196B,2018arXiv180401749B} that when the integrable system is relativistic and has an underlying ``modular double'' structure, we should in addition consider a ``modular dual'' Baxter equation 
\begin{equation}
\tilde{X}\, \overline{\theta}(\tilde{\mu}^{-1}\tilde{q}_1 e^{i 2\tilde{\omega} \partial_{\sigma}},\tilde{Q}_{\text{5d}})
- \tilde{E}'\, \overline{\theta}(e^{i 2\tilde{\omega} \partial_{\sigma}},\tilde{Q}_{\text{5d}}) 
+ \tilde{X}^{-1}\, \overline{\theta}(\tilde{\mu} \tilde{q}_1^{-1} e^{i 2\tilde{\omega} \partial_{\sigma}},\tilde{Q}_{\text{5d}}) = 0, \label{eRSAbaxdual}
\end{equation}
related to the original one \eqref{eRSAbax} by $\hbar_y \leftrightarrow 2\tilde{\omega}$ exchange, where $\tilde{E}' = E'(a,i 2\tilde{\omega}, i g_y, \hbar_y/2, \tilde{\omega}') $ and
\begin{equation}
\tilde{X} = e^{\frac{2\pi \sigma}{\hbar_y}}, \;\; \tilde{\mu} = e^{\frac{2 \pi i g_y}{\hbar_y}}, \;\; \tilde{q}_1 = e^{\frac{2\pi i 2\tilde{\omega}}{\hbar_y}}, \;\;
\tilde{\alpha} = e^{\frac{2\pi a}{\hbar_y}}, \;\;
\tilde{Q}_{\text{5d}} = e^{2\pi i \frac{2\tilde{\omega}'}{\hbar_y}},
\end{equation}
and look for two generically linearly independent functions $\mathcal{Q}^{(f)}(\sigma, a, \hbar_y, g_y, \tilde{\omega}, \tilde{\omega}')$ and $\mathcal{Q}^{(af)}(\sigma, a, \hbar_y, g_y, \tilde{\omega}, \tilde{\omega}')$ formally satisfying both \eqref{eRSAbax} and \eqref{eRSAbaxdual} simultaneously.\footnote{This was also suggested in \cite{Kashani-Poor:2016edc,Sciarappa:2016ctj} in the context of open topological strings in the NS limit.} Similarly to Section \ref{eCMeigenfunctions}, such formal solutions can conjecturally be obtained by considering the NS limit of the VEVs of codimension two defects corresponding to coupling the five-dimensional theory to two free three-dimensional $\mathcal{N} = 2^*$ hypermultiplets, whose $SU(2)$ part of the flavour symmetry is gauged and identified with the five-dimensional gauge group
 \cite{Gaiotto:2014ina,Bullimore:2014awa,Gaiotto:2015una}.
Schematically, these two functions can be written as
\begin{equation}
\begin{split}
& \mathcal{Q}^{(f)}(\sigma, a, \hbar_y, g_y,\tilde{\omega}, \tilde{\omega}') \\
& = \mathcal{Q}^{(f)}_{0}(\sigma, a, \hbar_y, g_y,\tilde{\omega}) 
\big\langle \mathcal{Q}^{(f)}_{\text{5d,inst}}(\sigma, a, i\hbar_y, i g_y, (2\tilde{\omega})^{-1}, Q_{\text{5d}}) \big\rangle_{\text{NS}} \big\langle \mathcal{Q}^{(f)}_{\text{5d,inst}}(\sigma, a, i 2\tilde{\omega}, i g_y, \hbar_y^{-1}, \tilde{Q}_{\text{5d}}) \big\rangle_{\text{NS}} \;, \\[7 pt]
& \mathcal{Q}^{(af)}(\sigma, a, \hbar_y, g_y, \tilde{\omega}, \tilde{\omega}') \\
& = e^{\frac{4\pi \sigma \tilde{\omega}'}{2\tilde{\omega}\hbar_y }}\mathcal{Q}^{(af)}_{0}(\sigma, a, \hbar_y, g_y,\tilde{\omega}) 
\big\langle \mathcal{Q}^{(af)}_{\text{5d,inst}}(\sigma, a, i \hbar_y, i g_y, (2\tilde{\omega})^{-1}, Q_{\text{5d}}) \big\rangle_{\text{NS}} \big\langle \mathcal{Q}^{(af)}_{\text{5d,inst}}(\sigma, a, i 2\tilde{\omega}, i g_y, \hbar_y^{-1}, \tilde{Q}_{\text{5d}}) \big\rangle_{\text{NS}} \;; \label{kp}
\end{split}
\end{equation}
here
\begin{equation}
\begin{split}
\mathcal{Q}^{(f)}_{0}(\sigma, a, \hbar_y, g_y,\tilde{\omega}) & =
\dfrac{S_2^{-1}\left(-i(\sigma - \frac{a}{2}) \vert \hbar_y, 2\tilde{\omega}\right)
S_2^{-1}\left(-i(\sigma + \frac{a}{2}) \vert \hbar_y, 2\tilde{\omega}\right)}{S_2^{-1}\left(-i(\sigma - \frac{a}{2} -i g_y + i \hbar_y + i 2\tilde{\omega}) \vert \hbar_y, 2\tilde{\omega}\right)
S_2^{-1}\left(-i(\sigma + \frac{a}{2} - i g_y + i \hbar_y + i 2\tilde{\omega}) \vert \hbar_y, 2\tilde{\omega}\right)}, \\
\mathcal{Q}^{(af)}_{0}(\sigma, a, \hbar_y, g_y,\tilde{\omega}) & =
\dfrac{S_2^{-1}\left(i(\sigma - \frac{a}{2}) \vert \hbar_y, 2\tilde{\omega}\right)
S_2^{-1}\left(i(\sigma + \frac{a}{2}) \vert \hbar_y, 2\tilde{\omega}\right)}{S_2^{-1}\left(i(\sigma - \frac{a}{2} + i g_y - i \hbar_y - i 2\tilde{\omega}) \vert \hbar_y, 2\tilde{\omega}\right)
S_2^{-1}\left(i(\sigma + \frac{a}{2} + i g_y - i \hbar_y - i 2\tilde{\omega}) \vert \hbar_y, 2\tilde{\omega}\right)}, 
\end{split}
\end{equation}
where the double sine function is defined in Appendix \ref{appspecial}, while the instanton corrections contained in 
$\langle \mathcal{Q}^{(f)}_{\text{5d,inst}}(\sigma, a, \epsilon_1, m,R, Q_{\text{5d}}) \rangle_{\text{NS}}$, $\langle \mathcal{Q}^{(af)}_{\text{5d,inst}}(\sigma, a, \epsilon_1, m, R, Q_{\text{5d}}) \rangle_{\text{NS}}$
and their dual can be computed with the formulae in Appendix \ref{appgauge}. We will not try to report here their explicit expression since they are rather complicated already at one instanton, but we will just mention that, apart from their 
$a \leftrightarrow -a$ symmetry, they satisfy
\begin{equation}
\langle \mathcal{Q}^{(af)}_{\text{5d,inst}}(\sigma, a, \epsilon_1, m,R, Q_{\text{5d}}) \rangle_{\text{NS}}
= \langle \mathcal{Q}^{(f)}_{\text{5d,inst}}(\sigma, a, -\epsilon_1, -m,R, Q_{\text{5d}}) \rangle_{\text{NS}},
\end{equation}
as well as
\begin{equation}
\langle \mathcal{Q}^{(f)}_{\text{5d,inst}}(-\sigma, a, -\epsilon_1, -m,R, Q_{\text{5d}}) \rangle_{\text{NS}}
= \langle \mathcal{Q}^{(f)}_{\text{5d,inst}}(\sigma,a, \epsilon_1, m, R, Q_{\text{5d}}) \rangle_{\text{NS}}.
\end{equation}
Assuming that \eqref{kp} indeed satisfy both Baxter equations, then for any constant $\xi \in \mathbb{C}$ the linear combination
\begin{equation}
\mathcal{Q}(\sigma, a, \hbar_y, g_y,\tilde{\omega},\tilde{\omega}') = 
\mathcal{Q}^{(f)}(\sigma, a, \hbar_y, g_y,\tilde{\omega},\tilde{\omega}') 
- \xi^{-1} \mathcal{Q}^{(af)}(\sigma, a, \hbar_y, g_y,\tilde{\omega},\tilde{\omega}') \label{linearRS}
\end{equation}
formally satisfies
\begin{equation}
\begin{split}
&\left[X\, \overline{\theta}(\mu^{-1}q_1 e^{i \hbar_y \partial_{\sigma}},Q_{\text{5d}})
- E'\, \overline{\theta}(e^{i \hbar_y \partial_{\sigma}},Q_{\text{5d}}) 
+ X^{-1}\, \overline{\theta}(\mu q_1^{-1} e^{i \hbar_y \partial_{\sigma}},Q_{\text{5d}})\right]
\mathcal{Q}(\sigma, a, \hbar_y, g_y,\tilde{\omega},\tilde{\omega}') = 0, \\[6 pt]
& \left[ \tilde{X}\, \overline{\theta}(\tilde{\mu}^{-1}\tilde{q}_1 e^{i 2\tilde{\omega} \partial_{\sigma}},\tilde{Q}_{\text{5d}})
- \tilde{E}'\, \overline{\theta}(e^{i 2\tilde{\omega} \partial_{\sigma}},\tilde{Q}_{\text{5d}}) 
+ \tilde{X}^{-1}\, \overline{\theta}(\tilde{\mu} \tilde{q}_1^{-1} e^{i 2\tilde{\omega} \partial_{\sigma}},\tilde{Q}_{\text{5d}}) \right] \mathcal{Q}(\sigma, a, \hbar_y, g_y,\tilde{\omega},\tilde{\omega}') = 0.
\end{split}
\end{equation}
Exactly as in Section \ref{eCMeigenfunctions}, also in this case \eqref{linearRS} is not entire in $\sigma$ for generic values of $a$, $\xi$, since by studying the denominator structure of their first instanton corrections it appears that \eqref{kp} have simple poles at $\sigma = \pm \frac{a}{2} + i k \hbar_y + i n 2\tilde{\omega}$ and $\sigma = \pm \frac{a}{2} + i g_y + i k \hbar_y + i n 2\tilde{\omega}$ for $k, n \in \mathbb{Z}$. We notice however that requiring the poles at $\sigma = \pm \frac{a}{2}$ to vanish, i.e. requiring
\begin{equation}
\xi = \dfrac{
\underset{\sigma = a/2}{\text{Res}} \left[\mathcal{Q}^{(af)}(\sigma, a, \hbar_y, g_y,\tilde{\omega},\tilde{\omega}') \right]}{\underset{\sigma = a/2}{\text{Res}} 
\left[ \mathcal{Q}^{(f)}(\sigma, a, \hbar_y, g_y,\tilde{\omega},\tilde{\omega}') \right]} 
= \dfrac{\underset{\sigma = - a/2}{\text{Res}} 
\left[\mathcal{Q}^{(af)}(\sigma, a, \hbar_y, g_y,\tilde{\omega},\tilde{\omega}') \right]}{\underset{\sigma = - a/2}{\text{Res}} 
\left[ \mathcal{Q}^{(f)}(\sigma, a, \hbar_y, g_y,\tilde{\omega},\tilde{\omega}')\right]} , \label{quant5d}
\end{equation}
seems to be equivalent to imposing the full A-type quantization conditions \eqref{quantRSAtrue}; it would therefore be natural, although highly non-trivial, to expect that \textit{all} simple poles will cancel in the linear combination \eqref{linearRS} for the values of $a$, $\xi$ satisfying full A-type quantization conditions, in which case our solution \eqref{linearRS} to the two Baxter equations would be entire in $\sigma$.

Of course, all we said insofar is highly conjectural and we do not see any easy way to make it even slightly more rigorous. Let us however push this line of reasoning a bit further. Let us assume that \eqref{linearRS} is indeed an entire solution to the two 2-eRS$_{\text{A}}$ Baxter equations for the values of $a$, $\xi$ satisfying \eqref{quant5d}, and let us also assume it decays rapidly enough at infinity. We could then consider its Fourier transform
\begin{equation}
\Phi_a(y, \hbar_y, g_y,\tilde{\omega},\tilde{\omega}') = \int_{\mathbb{R}} d\sigma e^{-\frac{2 \pi i \sigma y}{2\tilde{\omega} \hbar_y}} \mathcal{Q}(\sigma,a,\hbar_y, g_y,\tilde{\omega},\tilde{\omega}');
\end{equation}
this will satisfy the modular double pair of equations
\begin{equation}
\begin{split}
& \left[ \dfrac{\overline{\theta}(\mu^{-1}e^{-\frac{2\pi y}{2\tilde{\omega}}},Q_{\text{5d}})}{\overline{\theta}(e^{-\frac{2\pi y}{2\tilde{\omega}}},Q_{\text{5d}}) } e^{i \hbar_y \partial_y}
+ \dfrac{\overline{\theta}(\mu e^{-\frac{2\pi y}{2\tilde{\omega}}},Q_{\text{5d}})}{\overline{\theta}(e^{-\frac{2\pi y}{2\tilde{\omega}}},Q_{\text{5d}}) } e^{-i \hbar_y \partial_y} - E' \right] \Phi_a(y, \hbar_y, g_y,\tilde{\omega},\tilde{\omega}') = 0 , \\[7 pt]
& \left[ \dfrac{\overline{\theta}(\mu^{-1}e^{-\frac{2\pi y}{\hbar_y}},Q_{\text{5d}})}{\overline{\theta}(e^{-\frac{2\pi y}{\hbar_y}},Q_{\text{5d}}) } e^{i 2\tilde{\omega} \partial_y}
+ \dfrac{\overline{\theta}(\mu e^{-\frac{2\pi y}{\hbar_y}},Q_{\text{5d}})}{\overline{\theta}(e^{-\frac{2\pi y}{\hbar_y}},Q_{\text{5d}}) } e^{-i 2\tilde{\omega} \partial_y} - \tilde{E}' \right] \Phi_a(y, \hbar_y, g_y,\tilde{\omega},\tilde{\omega}') = 0.
\end{split}
\end{equation}
If we now proceed as indicated in \cite{doi:10.1063/1.531809} and multiply $\Phi_a(y, \hbar_y, g_y,\tilde{\omega},\tilde{\omega}')$ by an appropriate weight function $\sqrt{w(y)}$ 
involving the elliptic Gamma function, the combination 
\begin{equation}
\psi^{(\text{A})}_n(y,Q_{\text{5d}}) \propto \sqrt{w(y)} \Phi_a(y, \hbar_y, g_y,\tilde{\omega},\tilde{\omega}')
\end{equation}
should provide the desired (unnormalized) eigenfunction for the 2-eRS$_{\text{A}}$ Quantum Mechanical problem \eqref{ellRSAbis} on the appropriate $L^2([0,2\tilde{\omega}'])$ Hilbert space, thanks to the analytic properties of the Baxter solution $\mathcal{Q}(\sigma,a,\hbar_y, g_y,\tilde{\omega},\tilde{\omega}')$. Eigenfunctions for the 2-hRS systems are then recovered in the limit $\tilde{\omega}' \rightarrow i \infty$.

\section{$N$-particle systems of Ruijsenaars-Schneider type} \label{NRSsection}

Up until now we restricted our attention to quantum elliptic integrable systems involving only two particles: these are the easiest ones one can study, since they are described by a single Hamiltonian which moreover reduces to a one-dimensional problem after decoupling the center of mass of the system. In the present Section we will consider instead $N$-particle systems; here we only focus on the elliptic Ruijsenaars-Schneider model, since the Calogero-Moser one and the various trigonometric/hyperbolic versions can be obtained as special limits of it.

The complex $N$-particle elliptic Ruijsenaars-Schneider system ($N$-eRS) is 
defined on a torus of half-periods $(\tilde{\omega}, \tilde{\omega}')$ and describes the motion of $N$ particles at positions $z_j$, dictated by the finite-difference equation
\begin{equation}
\widehat{H}_{\text{eRS}} \psi(\vec{z}, Q_{\text{5d}}) = E \psi(\vec{z}, Q_{\text{5d}}), \label{NRSeq}
\end{equation}
where the Hamiltonian is given by \cite{Ruijsenaars1999}
\begin{equation}
\widehat{H}_{\text{eRS}} = \sum_{j = 1}^N \mathrm{W}_j e^{i \hbar \partial_{z_j}} \mathrm{W}_j^*, \label{hRSgen}
\end{equation}
with
\begin{equation}
\mathrm{W}_j = \prod_{k \neq j}^N \sqrt{\dfrac{\sigma(z_j - z_k + i g \vert \tilde{\omega}, \tilde{\omega}')}{\sigma(z_j - z_k \vert \tilde{\omega}, \tilde{\omega}')}} 
\end{equation}
and all parameters $z_i$, $\hbar$, $g$, $E \in \mathbb{C}$. 
The Hamiltonian $\widehat{H}_{\text{eRS}} = \widehat{H}_1$ is actually part of a set of $N$ commuting linearly independent operators $\widehat{H}_{l}$,
\begin{equation}
[\widehat{H}_{l}, \widehat{H}_{m}] = 0 \;\;\;\;\;\; \forall \;\; l, m = 1, \ldots, N,
\end{equation}
which are such that (renaming $E = E_1$)
\begin{equation}
\widehat{H}_l \psi(\vec{z}, Q_{\text{5d}})  = E_l \psi(\vec{z}, Q_{\text{5d}}) \;\;\;\;\;\; , \;\; l = 1,\ldots, N; \label{other}
\end{equation}
we will however mostly focus only on the first Hamiltonian $\widehat{H}_{\text{eRS}}$ and the associated equation \eqref{NRSeq}. As in Section \ref{RSsection}, we can also equivalently study the problem
\begin{equation}
\widehat{H}'_{\text{eRS}} \psi(\vec{z}, Q_{\text{5d}}) = E' \psi(\vec{z}, Q_{\text{5d}}), \label{NRSeqp}
\end{equation}
where the new Hamiltonian 
\begin{equation}
\widehat{H}'_{\text{eRS}} = \sum_{j = 1}^N \mathrm{W'}_{\hspace{-0.1 cm} j} e^{i \hbar \partial_{z_j}} \mathrm{W'}_{\hspace{-0.1 cm} j}^* \label{hRSgenp}
\end{equation}
with
\begin{equation}
\mathrm{W'}_{\hspace{-0.1 cm} j} = \prod_{k \neq j}^N \sqrt{\dfrac{\theta_1(\pi \frac{z_j - z_k + i g}{2 \tilde{\omega}} ,Q_{\text{5d}}^{1/2})}{\theta_1(\pi \frac{z_j - z_k}{2 \tilde{\omega}} ,Q_{\text{5d}}^{1/2})}} 
\end{equation}
is related to the previous one by
\begin{equation}
\widehat{H}_{\text{eRS}} = \text{exp} \left( -(N-1)\frac{g(g - \hbar)}{24} \dfrac{\pi^2}{\tilde{\omega}^2} E_2(Q_{\text{5d}}) \right) \widehat{H}'_{\text{eRS}}. \label{relN}
\end{equation} 
As we did for the 2-particle case, by restricting the values of the parameters in \eqref{hRSgen}, \eqref{hRSgenp} to some real slice we can define the $N$-eRS$_{\text{B}}$ and $N$-eRS$_{\text{A}}$ quantum mechanical problems:

\paragraph{B-model:} Restricting to the real slice $z_j = x_j \in \mathbb{R}$ and requiring $\hbar = \hbar_x \in \mathbb{R}_+$, $g = g_x \in \mathbb{R}_+$, $E = E^{\text{(B)}} \in \mathbb{R}_+$ problems \eqref{NRSeq}, \eqref{NRSeqp} reduce to
\begin{equation}
\begin{split}
\widehat{H}_{\text{B}} \psi^{(\text{B})}(\vec{x},Q_{\text{5d}}) 
& = \Big[ \sum_{j = 1}^N \mathrm{W}_j e^{i \hbar_x \partial_{x_j}} \mathrm{W}_j^* \Big] \psi^{(\text{B})}(\vec{x},Q_{\text{5d}})
= E^{(\text{B})} \psi^{(\text{B})}(\vec{x},Q_{\text{5d}}), \\
\widehat{H}'_{\text{B}} \psi^{(\text{B})}(\vec{x},Q_{\text{5d}}) 
& = \Big[ \sum_{j = 1}^N \mathrm{W'}_{\hspace{-0.1 cm} j} e^{i \hbar_x \partial_{x_j}} \mathrm{W'}_{\hspace{-0.1 cm} j}^* \Big] \psi^{(\text{B})}(\vec{x},Q_{\text{5d}})
= E'^{(\text{B})} \psi^{(\text{B})}(\vec{x},Q_{\text{5d}}), \label{NeRSB}
\end{split}
\end{equation}
with
\begin{equation}
\begin{split}
\mathrm{W}_j = \prod_{k \neq j}^N \sqrt{\dfrac{\sigma(x_j - x_k + i g_x \vert \tilde{\omega}, \tilde{\omega}')}{\sigma(x_j - x_k \vert \tilde{\omega}, \tilde{\omega}')}} , \;\;\;\;\;\;
\mathrm{W'}_{\hspace{-0.1 cm} j} = \prod_{k \neq j}^N \sqrt{\dfrac{\theta_1(\pi \frac{x_j - x_k + i g_x}{2 \tilde{\omega}} ,Q_{\text{5d}}^{1/2})}{\theta_1(\pi \frac{x_j - x_k}{2 \tilde{\omega}} ,Q_{\text{5d}}^{1/2})}} .
\end{split}
\end{equation}
For $0 < g_x < \hbar_x + 2 \vert \tilde{\omega}' \vert$ and decoupling the center of mass of the system, \eqref{NeRSB} should define the $N$-eRS$_{\text{B}}$ quantum mechanical problem on $L^2_x([0,2\tilde{\omega}]^{N-1})$ whose energy spectrum (as well as the eigenvalues of the other commuting operators $\widehat{H}_2$, \ldots, $\widehat{H}_N$) is real and discrete \cite{0305-4470-32-9-018,Ruijsenaars1999,2015SIGMA..11..004R}. 

In the limit $\tilde{\omega}' \rightarrow i \infty$ (i.e. $Q_{\text{5d}} \rightarrow 0$) the second line of \eqref{NeRSB} reduces to the $N$-particle trigonometric Ruijsenaars-Schneider quantum integrable system ($N$-tRS)
\begin{equation}
\widehat{H}^{\text{(T)}}\psi^{\text{(T)}}(\vec{x}) = \Bigg[ \sum_{j=1}^N \prod_{k \neq j}^N \sqrt{\dfrac{\sin(\pi \frac{x_j - x_k + i g_x}{2 \tilde{\omega}})}{\sin(\pi \frac{x_j - x_k}{2 \tilde{\omega}})}} e^{i \hbar_x \partial_{x_j}} \sqrt{\dfrac{\sin(\pi \frac{x_j - x_k - i g_x}{2 \tilde{\omega}})}{\sin(\pi \frac{x_j - x_k}{2 \tilde{\omega}})}} \Bigg] \psi^{\text{(T)}}(\vec{x}) = E^{\text{(T)}} \psi^{\text{(T)}}(\vec{x}),
\end{equation}
which is again a quantum mechanical problem on $L^2_x([0,2\tilde{\omega}]^{N-1})$ with discrete energy levels and known analytic solution.

\paragraph{A-model:} Restricting instead on the slice $z_j = i y_j \in i \mathbb{R}$ and requiring 
$\hbar = i \hbar_y \in i\mathbb{R}_+$, $g = i g_y \in i\mathbb{R}_+$, $E = E^{(\text{A})} \in \mathbb{R}_+$ problems \eqref{NRSeq}, \eqref{NRSeqp} reduce to
\begin{equation}
\begin{split}
\widehat{H}_{\text{A}} \psi^{(\text{A})}(\vec{y},Q_{\text{5d}}) 
& = \Big[ \sum_{j = 1}^N \mathrm{W}_j e^{i \hbar_y \partial_{y_j}} \mathrm{W}_j^* \Big] \psi^{(\text{A})}(\vec{y},Q_{\text{5d}})
= E^{(\text{A})} \psi^{(\text{A})}(\vec{y},Q_{\text{5d}}), \\
\widehat{H}'_{\text{A}} \psi^{(\text{A})}(\vec{y},Q_{\text{5d}}) 
& = \Big[ \sum_{j = 1}^N \mathrm{W'}_{\hspace{-0.1 cm} j} e^{i \hbar_y \partial_{y_j}} \mathrm{W'}_{\hspace{-0.1 cm} j}^* \Big] \psi^{(\text{A})}(\vec{y},Q_{\text{5d}})
= E'^{(\text{A})} \psi^{(\text{A})}(\vec{y},Q_{\text{5d}}), \label{NeRSA}
\end{split}
\end{equation}
with
\begin{equation}
\begin{split}
\mathrm{W}_j = \prod_{k \neq j}^N \sqrt{\dfrac{\sigma\big( i(y_j - y_k + i g_y) \vert \tilde{\omega}, \tilde{\omega}'\big)}{\sigma\big(i(y_j - y_k) \vert \tilde{\omega}, \tilde{\omega}'\big)}} , \;\;\;\;\;\;
\mathrm{W'}_{\hspace{-0.1 cm} j} = \prod_{k \neq j}^N \sqrt{\dfrac{\theta_1(i\pi \frac{y_j - y_k + i g_y}{2 \tilde{\omega}} ,Q_{\text{5d}}^{1/2})}{\theta_1(i\pi \frac{y_j - y_k}{2 \tilde{\omega}} ,Q_{\text{5d}}^{1/2})}} .
\end{split}
\end{equation}
For $0 \leqslant g_y \leqslant \hbar_y + 2\tilde{\omega}$ and decoupling the center of mass of the system, \eqref{NeRSA} should define the $N$-eRS$_{\text{A}}$ quantum mechanical problem on $L^2_y([0,2\vert \tilde{\omega}'\vert]^{N-1})$ whose energy spectrum is real and discrete. 

In the limit $\tilde{\omega}' \rightarrow i \infty$ (i.e. $Q_{\text{5d}} \rightarrow 0$) the second line of \eqref{NeRSA} reduces to the $N$-particle hyperbolic Ruijsenaars-Schneider quantum integrable system ($N$-hRS)
\begin{equation}
\widehat{H}^{\text{(H)}}\psi^{\text{(H)}}(\vec{y}) = \Bigg[ \sum_{j=1}^N \prod_{k \neq j}^N \sqrt{\dfrac{\sinh(\pi \frac{y_j - y_k + i g_y}{2 \tilde{\omega}})}{\sinh(\pi \frac{y_j - y_k}{2 \tilde{\omega}})}} e^{i \hbar_y \partial_{y_j}} \sqrt{\dfrac{\sinh(\pi \frac{y_j - y_k - i g_y}{2 \tilde{\omega}})}{\sinh(\pi \frac{y_j - y_k}{2 \tilde{\omega}})}} \Bigg] \psi^{\text{(H)}}(\vec{y}) = E^{\text{(H)}} \psi^{\text{(H)}}(\vec{y});
\end{equation}
this is still a quantum mechanical problem, although with continuous spectrum, whose analytic solution was partially studied in \cite{2012arXiv1206.3787H,2016arXiv160706672H} for the lowest values of $N$.

\noindent Exacly as in the 2-particle case, the two problems $N$-eRS$_{\text{B}}$ and $N$-eRS$_{\text{A}}$ are related as
\begin{equation}
\widehat{H}_{\text{A}}(y, \hbar_y, g_y, \tilde{\omega}, \tilde{\omega}') 
\;
=
\; 
\widehat{H}_{\text{B}}(y, \hbar_y, g_y, -i\tilde{\omega}', i\tilde{\omega}) ,
\end{equation}
which implies
\begin{equation}
E^{(\text{A})}(\hbar_y, g_y, \tilde{\omega}, \tilde{\omega}')  = 
E^{(\text{B})}(\hbar_y, g_y, -i \tilde{\omega}', i\tilde{\omega}).  \label{speciale}
\end{equation}
The gauge theory approach will again treat problems B and A differently, however the different treatments should be such that \eqref{speciale} is respected. 

\subsection{Elliptic case: B- and A-models} \label{SecNeRS}

As for the 2-particle case discussed in Section \ref{eRSBsection}, we can in principle study the $N$-eRS$_{\text{B}}$ quantum mechanical problem both numerically and analytically in terms of a convergent perturbation series in small $Q_{\text{5d}}$; the energy levels for the $N$-eRS$_{\text{A}}$ model will then be obtained by using the relation \eqref{speciale}. In order to do this it is however necessary to know the analytic solution to the $N$-particle trigonometric Ruijsenaars-Schneider model, which we review next.

\subsubsection*{Trigonometric case}

Let us then consider the $N$-tRS problem 
\begin{equation}
\widehat{H}^{\text{(T)}}\psi^{\text{(T)}}(\vec{x}) = \Bigg[ \sum_{j=1}^N \prod_{k \neq j}^N \sqrt{\dfrac{\sin(\pi \frac{x_j - x_k + i g_x}{2 \tilde{\omega}})}{\sin(\pi \frac{x_j - x_k}{2 \tilde{\omega}})}} e^{i \hbar_x \partial_{x_j}} \sqrt{\dfrac{\sin(\pi \frac{x_j - x_k - i g_x}{2 \tilde{\omega}})}{\sin(\pi \frac{x_j - x_k}{2 \tilde{\omega}})}} \Bigg] \psi^{\text{(T)}}(\vec{x}) = E^{\text{(T)}} \psi^{\text{(T)}}(\vec{x}). \label{tRSNproblem}
\end{equation}
It is well known that eigenfunctions for this system can be constructed out of the Macdonald polynomials\footnote{In our convention, they are polynomials in the exponentiated variables $e^{\frac{i  \pi}{\omega}x_i}$.} $P_{\lambda}(\vec{x}; q, \mu)$, which depend on two parameters 
\begin{equation}
q = e^{\frac{2 \pi \hbar_x}{2\tilde{\omega}}}\,, \;\;\;\; \mu = e^{\frac{2 \pi g_x}{2\tilde{\omega}}}. \label{parameters}
\end{equation}
The definition of Macdonald polynomials is not immediate and requires several steps \cite{macdonald2015symmetric}.
We should first introduce the degree $k$ monomial symmetric functions $m_{\lambda}(\vec{x})$ in $N$ variables $\vec{x} = \{ x_1, \ldots, x_N \}$:
\begin{equation}
m_{\lambda}(\vec{x}) = \sum_{\sigma \in \mathcal{S}_N}\prod_{j = 1}^N \left (e^{\frac{i \pi}{\omega} x_{\sigma(j)}} \right )^{\lambda_j},
\end{equation}
where $\mathcal{S}_N$ is the symmetric group of degree $N$ and $\lambda = (\lambda_1, \ldots, \lambda_N)$ is a partition of $k$ in $N$ parts, that is $\lambda_1 \geqslant \ldots \geqslant \lambda_N \geqslant 0$ and $\vert \lambda \vert = \sum_{j = 1}^N \lambda_j = k$. Next we should introduce the degree $k = \vert \lambda \vert$ power-sum symmetric function $p_{\lambda}(\vec{x})$ for the same partition $\lambda$:
\begin{equation}
p_{\lambda}(\vec{x}) = \prod_{j = 1}^N p_{\lambda_j}(\vec{x}) \,, \;\;\;\;\;\;
p_{\lambda_j}(\vec{x}) = \sum_{n = 1}^N \left ( e^{\frac{i \pi}{\omega} x_{n}} \right )^{\lambda_j},
\end{equation}
as well as the inner product $\langle p_{\lambda}(\vec{x}), p_{\lambda'}(\vec{x}) \rangle_{q,\mu}$ between two such functions, which depends on the parameters $q,\mu$:
\begin{equation}
\langle p_{\lambda}(\vec{x}), p_{\lambda'}(\vec{x}) \rangle_{q,\mu} = \delta_{\lambda, \lambda'} z_{\lambda} \prod_{j = 1}^{l(\lambda)} \dfrac{1 - q^{\lambda_j}}{1 - \mu^{\lambda_j}} \,, \;\;\;\;\;\; z_{\lambda} = \prod_{a \geqslant 1} a^{n_{\lambda}(a)}n_{\lambda}(a)! \,,
\end{equation}
where $l(\lambda) = \# \{j$ such that $\lambda_j \neq 0 \}$ and $n_{\lambda}(a) = \# \{j$ such that $\lambda_j = a \}$. The Macdonald symmetric polynomial $P_{\lambda}(\vec{x}; q, \mu)$ associated to a partition $\lambda$ is then uniquely determined by the following two conditions \cite{macdonald2015symmetric}:
\begin{itemize}
\item[(1)] $P_{\lambda}(\vec{x}; q, \mu) = m_{\lambda}(\vec{x}) + \sum_{\lambda' < \lambda} c_{\lambda \lambda'}(q,\mu) m_{\lambda'}(\vec{x})\,$ with $\,c_{\lambda \lambda'}(q,\mu) \in \mathbb{Q}(q,t)\,$ and $\,\vert \lambda' \vert = k$;
\item[(2)] $\langle P_{\lambda}(\vec{x};q,\mu), P_{\lambda'}(\vec{x};q,\mu) \rangle_{q,\mu} = 0 \;$ for $\; \lambda \neq \lambda'$.
\end{itemize}
Here we are using the natural ordering for partitions, i.e. we say that $\lambda \geqslant \lambda'$ if $\lambda_1 + \ldots + \lambda_j \geqslant \lambda_1' + \ldots + \lambda_j'$ for all $1 \leqslant j \leqslant N$. For our purposes it is actually more convenient to consider the functions $P'_{\lambda}(\vec{x}; q, \mu)$ defined as\footnote{The difference between $P_{\lambda}(\vec{x}; q, \mu)$ and $P'_{\lambda}(\vec{x}; q, \mu)$ is related to decoupling of the center of mass: more precisely while $P_{\lambda}(\vec{x}; q, \mu)$ has (exponentiated) total momentum $q^{\vert \lambda \vert}$, that is $\prod_{j = 1}^N e^{i \hbar_x \partial_{x_j}} P_{\lambda}(\vec{x}; q, \mu) = q^{\vert \lambda \vert} P_{\lambda}(\vec{x}; q, \mu) $, $P'_{\lambda}(\vec{x}; q, \mu)$ has instead (exponentiated) total momentum 1.}
\begin{equation}
P'_{\lambda}(\vec{x}; q, \mu) = \left(m_{(1,\ldots,1)}(\vec{x})\right)^{-\frac{\vert \lambda \vert}{N}} P_{\lambda}(\vec{x}; q, \mu); \label{Macnorm}
\end{equation}
these can be shown to be eigenfunctions of the quantum mechanical problem \cite{macdonald2015symmetric}
\begin{equation}
\left[ \sum_{j=1}^N \prod_{k \neq j}^N \dfrac{\sin(\pi \frac{x_j - x_k + i g_x}{2 \tilde{\omega}})}{\sin(\pi \frac{x_j - x_k}{2 \tilde{\omega}})} e^{i \hbar_x \partial_{x_j}} \right] P'_{\lambda}(\vec{x}; q, \mu) = E^{(\text{T})}_{\lambda} P'_{\lambda}(\vec{x}; q, \mu) \label{tRSNmod}
\end{equation}
with eigenvalue
\begin{equation}
E^{(\text{T})}_{\lambda} = q^{- \frac{\vert \lambda \vert}{N}} \sum_{j = 1}^N q^{\lambda_j} \mu^{\frac{N + 1}{2} - j}, \label{tRSNenergy}
\end{equation}
which implies that the functions
\begin{equation}
\psi^{\text{(T)}}_{\lambda}(\vec{x}) \propto \sqrt{w(\vec{x})} P'_{\lambda}(\vec{x}; q, \mu)
\end{equation}
satisfy the $N$-tRS problem \eqref{tRSNproblem} with eigenvalue \eqref{tRSNenergy} for 
\begin{equation}
w(\vec{x}) = 
\prod_{j\neq k}^N \dfrac{
\big(e^{\frac{i \pi}{\tilde{\omega}}(x_j - x_k)};e^{-\frac{\pi \hbar_x}{\tilde{\omega}}}\big)_{\infty}
}{
\big(e^{\frac{i \pi}{\tilde{\omega}}(x_j - x_k)} e^{-\frac{\pi g_x}{\tilde{\omega}}};e^{-\frac{\pi \hbar_x}{\tilde{\omega}}}\big)_{\infty}
}.
\end{equation}
Let us also point out that because of the ``normalization'' performed in \eqref{Macnorm}, sometimes different partitions are associated to the same eigenfunction: more precisely
\begin{equation}
P'_{\lambda}(\vec{x}; q, \mu) \equiv P'_{\lambda'}(\vec{x}; q, \mu) \;\;\;\; \text{if} \;\;\;\; \lambda' = \lambda + n (1, \ldots, 1) , \;\; n \in \mathbb{Z}.
\end{equation}
Among other things, this means that we can limit ourselves to consider partitions with $\lambda_N = 0$. In the $N = 2$ case we are therefore left with partitions $\lambda = (n,0)$ with $n \geqslant 0$, and the associated functions \eqref{Macnorm} coincide with the Rogers polynomials already introduced in Section \ref{SectRS}. More in general, we can use this fact to alternatively label eigenfunctions and eigenvalues by a set of $N-1$ positive integers $\vec{n} = (n_1, \ldots, n_{N-1})$ rather than by partitions $(\lambda_1, \ldots, \lambda_{N-1}, 0)$, where $n_i = \lambda_i - \lambda_{i+1}$ for $i = 1, \ldots, N-1$:
\begin{equation}
\lambda = (\lambda_1, \ldots, \lambda_{N-1}, 0) \;\;\;\; \Longleftrightarrow \;\;\;\; \vec{n} = (n_1, \ldots, n_{N-1}) \;\;\;\; \text{for} \;\;\;\; n_i = \lambda_i - \lambda_{i+1}. \label{relabelling}
\end{equation}
This relabelling will be helpful in making contact with the gauge theory approach to the problem.

\subsubsection*{Elliptic case}

We can now consider the $N$-eRS$_{\text{B}}$ quantum mechanical problem, for example in the form
\begin{equation}
\begin{split}
\widehat{H}'_{\text{B}} \psi^{(\text{B})}(\vec{x},Q_{\text{5d}}) 
& = \Big[ \sum_{j = 1}^N \mathrm{W'}_{\hspace{-0.1 cm} j} e^{i \hbar_x \partial_{x_j}} \mathrm{W'}_{\hspace{-0.1 cm} j}^* \Big] \psi^{(\text{B})}(\vec{x},Q_{\text{5d}})
= E'^{(\text{B})} \psi^{(\text{B})}(\vec{x},Q_{\text{5d}}), \label{NeRSBbis}
\end{split}
\end{equation}
with
\begin{equation}
\begin{split}
\mathrm{W'}_{\hspace{-0.1 cm} j} = \prod_{k \neq j}^N \sqrt{\dfrac{\theta_1(\pi \frac{x_j - x_k + i g_x}{2 \tilde{\omega}} ,Q_{\text{5d}}^{1/2})}{\theta_1(\pi \frac{x_j - x_k}{2 \tilde{\omega}} ,Q_{\text{5d}}^{1/2})}} .
\end{split}
\end{equation}
The spectrum of this operator can always be evaluated numerically, by diagonalizing it in terms of the $L^{2}([0, 2\tilde{\omega}]^{N-1})$ $N$-tRS basis constructed out of Macdonald polynomials, similarly to what we did in previous cases. This procedure however appears to be quite time consuming for generic $N$; trying to solve this problem analytically as a series in $Q_{\text{5d}}$ instead appears to be a more efficient way to proceed. To this purpose it is more convenient to consider 
the transformed operator
\begin{equation}
\widehat{H}''_{\text{B}} \,=\, \dfrac{1}{\sqrt{w(\vec{x}, Q_{\text{5d}})}} \widehat{H}'_{\text{B}} \sqrt{w(\vec{x}, Q_{\text{5d}})} \,=\, \sum_{j = 1}^N \prod_{k \neq j}^N \dfrac{\theta_1(\pi \frac{x_j - x_k + i g_x}{2 \tilde{\omega}} ,Q_{\text{5d}}^{1/2})}{\theta_1(\pi \frac{x_j - x_k}{2 \tilde{\omega}} ,Q_{\text{5d}}^{1/2})} e^{i \hbar_x \partial_{x_j}},
\end{equation}
where
\begin{equation}
w(\vec{x}, Q_{\text{5d}}) = \prod_{j \neq k}^N \dfrac{\Gamma \Big( e^{\frac{i \pi}{\tilde{\omega}}(x_j - x_k)} e^{-\frac{\pi g_x}{\tilde{\omega}}}; e^{-\frac{\pi \hbar_x}{\tilde{\omega}}}, Q_{\text{5d}} \Big)}{\Gamma \Big( e^{\frac{i \pi}{\tilde{\omega}}(x_j - x_k)}; e^{-\frac{\pi \hbar_x}{\tilde{\omega}}}, Q_{\text{5d}} \Big)},
\end{equation}
and study the problem
\begin{equation}
\widehat{H}''_{\text{B}} \Phi^{(\text{B})}(\vec{x}, Q_{\text{5d}}) = E'^{(\text{B})} \Phi^{(\text{B})}(\vec{x}, Q_{\text{5d}}) \label{lastproblem}
\end{equation}
order by order in $Q_{\text{5d}}$. We can now proceed exactly as in Section \ref{eRSBsection}, the only difference being that the discrete set of eigenvalues and eigenfunctions will be labelled by a
partition $\lambda = (\lambda_1, \ldots, \lambda_{N-1},\lambda_N)$ (or, equivalently, by a set of $N-1$ integers $\vec{n} = (n_1, \ldots, n_{N-1})$ \eqref{relabelling} if we fix $\lambda_N = 0$) rather than a single integer $n$ (or its associated partition $(n,0)$):
 we will then have 
\begin{equation}
\begin{split}
\widehat{H}''_{\text{B}} & \,=\, \widehat{H}''_{\text{B}}{}^{\hspace{-0 cm}(0)} + Q_{\text{5d}} \widehat{H}''_{\text{B}}{}^{\hspace{-0 cm}(1)} + Q_{\text{5d}}^2 \widehat{H}''_{\text{B}}{}^{\hspace{-0 cm}(2)} + O(Q_{\text{5d}}^3), \\[5 pt]
E'^{(\text{B})}_{\lambda} & \,=\, E'^{(0)}_{\lambda} + Q_{\text{5d}} E'^{(1)}_{\lambda} + Q_{\text{5d}}^2 E'^{(2)}_{\lambda} + O(Q_{\text{5d}}^3) , \\[5 pt]
\Phi_{\lambda}^{(\text{B})}(\vec{x}, Q_{\text{5d}}) & \,=\, \Phi_{\lambda}^{(0)}(\vec{x}) + Q_{\text{5d}} \Phi_{\lambda}^{(1)}(\vec{x}) + Q_{\text{5d}}^2 \Phi_{\lambda}^{(2)}(\vec{x}) + O(Q_{\text{5d}}^3),
\end{split}
\end{equation}
where since
\begin{equation}
\widehat{H}''_{\text{B}}{}^{\hspace{-0 cm}(0)} = \sum_{j=1}^N \prod_{k \neq j}^N \dfrac{\sin(\pi \frac{x_j - x_k + i g_x}{2 \tilde{\omega}})}{\sin(\pi \frac{x_j - x_k}{2 \tilde{\omega}})} e^{i \hbar_x \partial_{x_j}} 
\end{equation}
as in \eqref{tRSNmod} we can immediately conclude that 
\begin{equation}
E'^{(0)}_{\lambda} = E^{(\text{T})}_{\lambda} = q^{- \frac{\vert \lambda \vert}{N}} \sum_{j = 1}^N q^{\lambda_j} \mu^{\frac{N + 1}{2} - j} \,,\;\;\;\;\;\; \Phi_{\lambda}^{(0)}(\vec{x}) = P'_{\lambda}(\vec{x}; q, \mu).
\end{equation}
By further expanding all $\Phi^{(l)}_{\lambda}(\vec{x}) = \vert \Phi^{(l)}_{\lambda} \rangle$, $l \geqslant 1$ in terms of the basis $\Phi^{(0)}_{\lambda}(\vec{x}) = \vert \Phi^{(0)}_{\lambda} \rangle$, i.e. 
\begin{equation}
\vert \Phi^{(l)}_{\lambda} \rangle = \sum_{\lambda'} c_{\lambda \lambda'}^{(l)} \vert \Phi^{(0)}_{\lambda'} \rangle,
\end{equation}
and noticing that at the practical level the summation can be truncated to those $\lambda'$ such that\footnote{Here we are having in mind to consider partitions $\lambda'$ with $\lambda'_N$ not restricted to zero.} $\vert \lambda' \vert = \vert \lambda \vert + N l$ because of the properties of $\widehat{H}''_{\text{B}}{}^{\hspace{-0 cm}(l)}$, we can completely solve the problem \eqref{lastproblem} at each order in the $Q_{\text{5d}}$ expansion. In this way we find, for example for $N = 3$:
\begin{equation}
\begin{split}
E'^{(\text{B})}_{(0,0,0)} &= \mu + 1 + \mu^{-1} + Q_{\text{5d}} \dfrac{(\mu - \mu^{-1})(\mu^{\frac{3}{2}} - \mu^{-\frac{3}{2}})(\mu^{\frac{1}{2}}q^{-\frac{1}{2}} - \mu^{-\frac{1}{2}}q^{\frac{1}{2}})}{\mu q^{\frac{1}{2}} - \mu^{-1} q^{-\frac{1}{2}}} + O(Q_{\text{5d}}^2), \\
E'^{(\text{B})}_{(1,0,0)} &= \mu \,q^{\frac{2}{3}} + q^{-\frac{1}{3}} + \mu^{-1} q^{-\frac{1}{3}} \\
& + Q_{\text{5d}} \dfrac{(\mu - \mu^{-1})(\mu^{\frac{1}{2}} q^{-\frac{1}{2}} - \mu^{-\frac{1}{2}} q^{\frac{1}{2}})(\mu^2 q - \mu q^2 + \mu q - \mu + 1 + \mu^{-1} - \mu^{-1} q^{-1} - \mu^{-1} q + \mu^{-2} q^{-1} - q^{-1})}{q^{\frac{1}{3}}(\mu^{\frac{1}{2}} q^{\frac{1}{2}} - \mu^{-\frac{1}{2}} q^{-\frac{1}{2}})(\mu q - \mu^{-1} q^{-1})} \\
& + O(Q_{\text{5d}}^2), \\[5 pt]
E'^{(\text{B})}_{(1,1,0)} & = \mu \,q^{\frac{1}{3}} + q^{\frac{1}{3}} + \mu^{-1} q^{-\frac{2}{3}} \\
& + Q_{\text{5d}} \dfrac{(\mu - \mu^{-1})(\mu^{\frac{1}{2}} q^{-\frac{1}{2}} - \mu^{-\frac{1}{2}} q^{\frac{1}{2}})(\mu^2 q^2 - \mu q^2 + \mu q - \mu + q - q^2 + \mu^{-1} - \mu^{-1} q^{-1} - \mu^{-1} q + \mu^{-2})}{q^{\frac{2}{3}}(\mu^{\frac{1}{2}} q^{\frac{1}{2}} - \mu^{-\frac{1}{2}} q^{-\frac{1}{2}})(\mu q - \mu^{-1} q^{-1})} \\
& + O(Q_{\text{5d}}^2), \\[5 pt]
E'^{(\text{B})}_{(2,1,0)} & = \ldots,\label{3energy}
\end{split}
\end{equation}
where $q, \mu$ are as defined in \eqref{parameters}. Higher order corrections are too lengthy to be reported here, but can be computed straightforwardly with the help of some computer program; the same procedure also applies for higher $N$.

\subsection{Energy spectrum from gauge theory}
\label{sec4.2}

Having discussed how to compute, both numerically and analytically, the energy levels of the
 $N$-eRS$_{\text{B}}$ and $N$-eRS$_{\text{A}}$ problems, let us try to reproduce the same results from $\mathcal{N} = 1^*$ $SU(N)$ gauge theory quantities, following what we did in Section \ref{seceRSenergygauge} for the 2-particle case and what was done in \cite{2015arXiv151102860H} for the similar case of the $N$-particle relativistic Toda chain. \\

The first thing we should do is to look for a basis of linearly independent formal solutions $\psi^{(i)}(\vec{z},\vec{a},\hbar,g,\tilde{\omega},\tilde{\omega}')$ to the complex $N$-particle elliptic Ruijsenaars-Schneider problem \eqref{NRSeq} or \eqref{NRSeqp} for generic $\vec{z}, \vec{a}, \hbar, g \in \mathbb{C}$, such that
\begin{equation}
\begin{split}
\widehat{H}_{\text{eRS}} \psi^{(i)}(\vec{z}, \vec{a}, \hbar, g, \tilde{\omega}, \tilde{\omega}') 
& = E(\vec{a},\hbar, g, \tilde{\omega}, \tilde{\omega}') \psi^{(i)}(\vec{z}, \vec{a}, \hbar, g, \tilde{\omega}, \tilde{\omega}'), \\[4 pt]
\widehat{H}'_{\text{eRS}} \psi^{(i)}(\vec{z}, \vec{a}, \hbar, g, \tilde{\omega}, \tilde{\omega}') 
& = E'(\vec{a},\hbar, g, \tilde{\omega}, \tilde{\omega}') \psi^{(i)}(\vec{z}, \vec{a}, \hbar, g, \tilde{\omega}, \tilde{\omega}'),
\end{split}
\end{equation}
where  
\begin{equation}
E(\vec{a},\hbar, g, \tilde{\omega}, \tilde{\omega}') = 
\text{exp} \left( -(N-1)\frac{g(g - \hbar)}{24} \dfrac{\pi^2}{\tilde{\omega}^2} E_2(Q_{\text{5d}}) \right)
E'(\vec{a},\hbar, g, \tilde{\omega}, \tilde{\omega}') \label{eccoci}
\end{equation}
as follows from \eqref{relN}. Both the energy and the formal solutions can be computed from gauge theory as convergent power series in $Q_{\text{5d}}$ with coefficients depending on the set of auxiliary variables $\vec{a} = (a_1, \ldots, a_N)$ satisfying $\sum_{l = 1}^N a_l = 0$ (which parametrize the VEV of the scalar field in the vector supermultiplet). As in the 2-particle case previously discussed, these correspond to the NS limit of VEVs of certain codimension four and two defects; more precisely the energy $E'$ will be given by the Wilson loop in the fundamental representation $\mathbf{N}$ of $SU(N)$, while the formal solutions are associated to full monodromy defects \cite{Bullimore:2014awa}:
\begin{equation}
\begin{array}{ccc}
E'(\vec{a}, \hbar, g, \tilde{\omega},\tilde{\omega}') &\;\;\;\; \Longleftrightarrow \;\;\;\; & \text{codim. 4 defect} \; \langle W^{SU(N)}_{\mathbf{N}}(\vec{a}, \epsilon_1, m, R, Q_{\text{5d}}) \rangle_{\text{NS}}, \\[6 pt] 
\psi^{(i)}(\vec{z}, \vec{a}, \hbar, g, \tilde{\omega},\tilde{\omega}') & \;\;\;\; \Longleftrightarrow \;\;\;\; & \text{codim. 2 defect (full monodromy, NS)}. \label{questa}
\end{array} 
\end{equation}
Although we will not be interested in them, let us mention that these formal solutions actually also satisfy the finite-difference equations \eqref{other} associated to the whole set of $N$ commuting operators $\widehat{H}_l$, with the various energies $E_l$ corresponding to Wilson loops in the rank $l$ antisymmetric representation of $SU(N)$; both the fundamental and the various antisymmetric Wilson loops can be computed via the formulae collected in Appendix \ref{appgauge}. \\

Once we have a formal solution to the complex $N$-eRS system, we can restrict ourselves to  consider only those values of parameters associated to quantum mechanical problems with discrete energy spectrum, such as $N$-eRS$_{\text{B}}$ or $N$-eRS$_{\text{A}}$. Let us discuss these two cases separately.

\paragraph{$N$-eRS$_{\text{B}}$ model:} After fixing $\epsilon_1 = \hbar_x \in \mathbb{R}_+$ and $m = g_x \in \mathbb{R}_+$, conjecturally when the parameters $\vec{a}$ satisfy the B-type quantization conditions
\begin{equation}
a_l - a_{l + 1} = g_x + n_l \hbar_x , \;\;\;\; n_l \in \mathbb{N}, \;\;\;\; l = 1, \ldots, N-1, \label{quantNRSB}
\end{equation}
the $\mathcal{N} = 1^*$ $SU(N)$ fundamental Wilson loop \eqref{questa} should reproduce the discrete energy levels $E'^{(\text{B})}$ for the $N$-eRS$_{\text{B}}$ model $\widehat{H}'_{\text{B}}$ \eqref{NeRSB} with Hilbert space $L^2([0,2\tilde{\omega}]^{N-1})$. 

\paragraph{$N$-eRS$_{\text{A}}$ model:} After fixing $\epsilon_1 = i \hbar_y$ and $m = i g_y$ (for $\hbar_y$, $g_y \in \mathbb{R}_+$), based on what we understood from the 2-particle case studied in Section \ref{seceRSenergygauge} and from the relativistic Toda chain case analysed in \cite{2015arXiv151102860H}, we conjecturally expect that if the parameters $\vec{a}$ satisfy the A-type quantization conditions
\begin{equation}
- i \sum_{l = 1}^{N-1} C_{kl} \frac{\partial \mathcal{W}^{SU(N)}_{\text{full}}(\vec{a}, \hbar_y, g_y, \tilde{\omega}, \tilde{\omega}')}{\partial a_{l\,l+1}} = 2 \pi n_k, \;\;\;\;\;\; k = 1, \ldots, N-1, \label{quantNRSA}
\end{equation}
then the fundamental Wilson loop should reproduce the discrete energy levels $E'^{(\text{A})}$ for the $N$-eRS$_{\text{A}}$ model $\widehat{H}'_{\text{A}}$ \eqref{NeRSA} with Hilbert space $L^2([0,2 \vert \tilde{\omega}' \vert ]^{N-1})$. The matrix $C_{kl}$ entering \eqref{quantNRSA} is the Cartan matrix of $SU(N)$, while the variables $a_{l\,l+1}$ are defined as $a_{l\,l+1} = a_l - a_{l+1}$. The function $\mathcal{W}^{SU(N)}_{\text{full}}(\vec{a}, \hbar_y, g_y, \tilde{\omega}, \tilde{\omega}')$ instead is the full, non-perturbatively corrected twisted effective superpotential for the $\mathcal{N} = 1^*$ $SU(N)$ theory, free of poles at $\frac{\hbar_y}{2\tilde{\omega}} \in \mathbb{Q}$\footnote{As we already mentioned earlier, differently from the $N$-particle relativistic closed Toda chain, it appears that for the $N$-eRS$_{\text{A}}$ model no shift by the B-field is required for poles cancellation at any $N$.} and symmetric under the exchange $\hbar_y \leftrightarrow 2\tilde{\omega}$. In order to compute this function we will follow the approach suggested in \cite{2015arXiv150704799H,Sciarappa:2017hds}: 
that is, we start by considering the integrand $Z_{S^5}^{\text{int}}$ of the partition function on the squashed 5-sphere $S^5_{\omega_1, \omega_2, \omega_3}$ for the $\mathcal{N} = 1^*$ $SU(N)$ theory, which factorizes into a perturbative (classical + 1-loop) and instanton part \cite{Kim:2012ava,Kim:2012qf,Kallen:2012cs,Hosomichi:2012ek,Kallen:2012va,Imamura:2012xg,Imamura:2012bm,2012arXiv1210.5909L}
\begin{equation}
Z_{S^5}^{\text{int}} = Z_{S^5}^{\text{pert}} Z_{S^5}^{\text{inst}} ,
\end{equation}
and we define the exact twisted effective superpotential by taking the ``NS limit'' $\omega_3 \rightarrow 0$ on $S^5_{\omega_1, \omega_2, \omega_3}$,\footnote{This procedure may be interpreted as computing the twisted effective superpotential for a theory on $S^3_{\omega_1, \omega_2} \times \mathbb{R}^2$, whose extrema (i.e. solutions to the quantization conditions \eqref{quantNRSA}, aka Bethe Ansatz equations) enter into the computation of the partition function for five-dimensional theories on $S^3_{\omega_1, \omega_2} \times \Sigma_g$ with partial topological A-twist along the Riemann surface $\Sigma_g$ \cite{Crichigno:2018adf} (see also \cite{Hosseini:2018uzp} for closely related set-ups).}
\begin{equation}
\mathcal{W}^{SU(N)}_{\text{full}} =
\lim_{\omega_3 \to 0} \left[ -i \omega_3 \ln Z_{S^5}^{\text{int}} \right].
\end{equation}
The perturbative part of $Z_{S^5}^{\text{int}}$ admits an expression in terms of triple sine functions (see Appendix \ref{appspecial}) as
\begin{equation}
Z_{S^5}^{\text{pert}} = 
\prod_{l = 1}^N e^{ \frac{2 i \tilde{\omega}' \pi a_l^2}{\omega_1 \omega_2 \omega_3}} 
\prod_{\alpha > 0} \dfrac{S_3\left(i \alpha(\vec{a}) \,\vert\, \omega_1, \omega_2, \omega_3 \right)S_3\left(i \alpha(\vec{a}) + \vert \omega \vert  \,\vert\, \omega_1, \omega_2, \omega_3 \right)}{S_3\left(i \alpha(\vec{a}) + g_y \,\vert\, \omega_1, \omega_2, \omega_3 \right)S_3\left(i \alpha(\vec{a}) - g_y + \vert \omega \vert \,\vert\, \omega_1, \omega_2, \omega_3 \right)},
\end{equation}
where we identify $\omega_1 = \hbar_y$, $\omega_2 = 2\tilde{\omega}$
and define $\vert \omega \vert = \omega_1 + \omega_2 + \omega_3$. The instanton part instead is often assumed to factorize into three copies of the instanton partition function on $\mathbb{R}^4_{\epsilon_1, \epsilon_2} \times S^1_R$, where the identification between $\omega_1$, $\omega_2$, $\omega_3$ with $\epsilon_1$, $\epsilon_2$, $R$ is different for each of the copies; in the ``NS limit'' $\omega_3 \rightarrow 0$ this contributes to the full twisted effective superpotential as
\begin{equation}
\begin{split}
\lim_{\omega_3 \to 0} \left[ -i \omega_3 \ln Z_{S^5}^{\text{inst}} \right] =
& - i \mathcal{W}^{SU(N)}_{\text{5d,inst}}(\vec{a}, i \hbar_y, i g_y, (2\tilde{\omega})^{-1},e^{\frac{4\pi i \tilde{\omega}'}{2\tilde{\omega}}}) \\
& - i \mathcal{W}^{SU(N)}_{\text{5d,inst}}(\vec{a}, i 2\tilde{\omega}, i g_y, \hbar_y^{-1},e^{\frac{4\pi i \tilde{\omega}'}{\hbar_y}}) ,
 \end{split}
\end{equation}
with $\mathcal{W}^{SU(N)}_{\text{5d,inst}}$ the instanton part of the twisted effective superpotential for the theory on $\mathbb{R}^4_{\epsilon_1, \epsilon_2} \times S^1_R$ defined in \eqref{ultimo}.

\subsection{An example: the 3-particle case}

Although it may be hard to prove the conjectures stated in Section \ref{sec4.2}, we can still test them against numerical or analytical results (obtained as discussed in Section \ref{SecNeRS}): let us do this explicitly for the 3-particle case. 

The main quantity we should consider is the NS limit of the Wilson loop in the fundamental representation $\mathbf{3}$ of $SU(3)$, which should correspond to the energy of the complex 3-eRS system for generic complex values of the parameters. From the formulae in Appendix \ref{appgauge} we find
\begin{equation}
\begin{split}
& E'(a_{12}, a_{23}, \hbar, g, \tilde{\omega}, \tilde{\omega}') = \langle W^{SU(3)}_{\mathbf{3}}(a_{12}, a_{23}, \epsilon_1, m, R, Q_{\text{5d}}) \rangle_{\text{NS}}\Big\vert_{\epsilon_1 = \hbar, m = g, R^{-1} = 2\tilde{\omega}} \\[5 pt]
& = \alpha_{12}^{\frac{2}{3}} \alpha_{23}^{\frac{1}{3}} + \alpha_{12}^{-\frac{1}{3}} \alpha_{23}^{\frac{1}{3}} + \alpha_{12}^{-\frac{1}{3}} \alpha_{23}^{-\frac{2}{3}} \\[5 pt]
& \;\; + Q_{\text{5d}} \dfrac{\alpha_{12}^{-\frac{1}{3}} \alpha_{23}^{\frac{1}{3}}}{\mu^3 q_1^3} \dfrac{(1 - \mu)^2(q_1 - \mu)^2 \Big( \alpha_{12}(1 + \alpha_{23} + \alpha_{12} \alpha_{23})(1 + q_1 + q_1^2) - q_1 (1 + \alpha_{12} + \alpha_{12} \alpha_{23})^2 \Big)}{(1 - q_1 \alpha_{12})(1 - q_1^{-1} \alpha_{12})(1 - q_1 \alpha_{23})(1 - q_1^{-1} \alpha_{23})(1 - q_1 \alpha_{12}\alpha_{23})(1 - q_1^{-1} \alpha_{12}\alpha_{23})} \\[5 pt]
& \;\; \times \Big( (1 + \mu)(1 + q_1)(\mu + q_1) \alpha_{12} \alpha_{23} - \mu q_1 (1 + \alpha_{12})(1 + \alpha_{23})(1 + \alpha_{12}\alpha_{23}) \Big) + O(Q_{\text{5d}}^2), \label{W3}
\end{split}
\end{equation}
where similarly to \eqref{par} we defined (in terms of the variables $a_{12} = a_1 - a_2$, $a_{23} = a_2 - a_3$)\footnote{When needed we use $a_1 = \frac{2a_{12}}{3} + \frac{a_{23}}{3}$, $a_2 = -\frac{a_{12}}{3} + \frac{a_{23}}{3}$, $a_3 = -\frac{a_{12}}{3} - \frac{2a_{23}}{3}$.}
\begin{equation}
\alpha_{12} = e^{2\pi R a_{12}} = e^{\frac{2 \pi a_{12}}{2\tilde{\omega}}} \,,\;\;\;\;
\alpha_{23} = e^{2\pi R a_{23}} = e^{\frac{2 \pi a_{23}}{2\tilde{\omega}}} \,,\;\;\;\;
q_1 = e^{2\pi R \epsilon_1} = e^{\frac{2 \pi \hbar}{2 \tilde{\omega}}} \,,\;\;\;\;
\mu = e^{2\pi R m} = e^{\frac{2 \pi g}{2\tilde{\omega}}}.
\end{equation}
According to the general discussion in Section \ref{sec4.2}, the gauge theory Wilson loop \eqref{W3} should reproduce the $3$-eRS$_{\text{B}}$ energy levels once we restrict to $\epsilon_1 = \hbar_x \in \mathbb{R}_+$,  $m = g_x \in \mathbb{R}_+$ and impose the B-type quantization conditions \eqref{quantNRSB}, which in our case simply reduce to
\begin{equation}
a_{12} = g_x + n_1 \hbar_x \,, \;\;\;\; a_{23} = g_x + n_2 \hbar_x \,, \;\;\;\; n_1, n_2 \in \mathbb{N};
\end{equation}
more precisely, a state labelled by $(n_1,n_2)$ should correspond to a state labelled by the partition $(\lambda_1, \lambda_2, 0)$ via $n_1 = \lambda_1 - \lambda_2$, $n_2 = \lambda_2$, as dictated by \eqref{relabelling}. This can be easily checked analytically, for example by noticing that our Wilson loop indeed reproduces the energy levels we previously computed in \eqref{3energy} when the B-type quantization conditions are imposed. Tests against numerical results can also be performed (see for example Table \ref{ex3B}), which further confirm the validity of the gauge theory prescription.

\renewcommand{\arraystretch}{1.5}
\begin{table}
\begin{center}
\begin{tabular}{|c|c|c|c|}
\hline 
$\tilde{\omega} = \frac{\pi^2}{3}$, $\;\tilde{\omega}' = i \pi$ & $E^{(\text{B})}_{0,0}$ & $E^{(\text{B})}_{0,1}$ & $E^{(\text{B})}_{1,0}$ \\ 
\hline 
Gauge theory - $O(Q_{\text{5d}})$ & \underline{10}.7676364819\ldots & \underline{16.8}160803048\ldots & 
\underline{24.9}327146426\ldots \\ 
\hline 
Gauge theory - $O(Q_{\text{5d}}^2)$ & \underline{10.822}4190473\ldots & \underline{16.897}6767493\ldots & 
\underline{24.9}594982573\ldots \\ 
\hline 
Gauge theory - $O(Q_{\text{5d}}^3)$ & \underline{10.822975}2647\ldots & \underline{16.8979}569599\ldots & 
\underline{24.9600}693306\ldots \\ 
\hline 
Gauge theory - $O(Q_{\text{5d}}^4)$ & \underline{10.8229758}745\ldots & \underline{16.89796018}49\ldots & 
\underline{24.9600712}318\ldots \\ 
\hline 
Numerical & \underline{10.8229758847}\ldots & \underline{16.8979601851}\ldots & \underline{24.9600712270}\ldots \\ 
\hline 
\end{tabular} \caption{Energies $E^{(\text{B})}_{n_1,n_2}$ for the 3-eRS$_{\text{B}}$ model at $g_x = \sqrt{7}$, $\hbar_x = \sqrt{2}$.} \label{ex3B}
\end{center}
\end{table}
\renewcommand{\arraystretch}{1}

\renewcommand{\arraystretch}{1.5}
\begin{table}
\begin{center}
\begin{tabular}{|c|c|c|c|}
\hline 
$\tilde{\omega} = \pi$, $\;\tilde{\omega}' = \frac{i \pi^2}{3}$ & $E^{(\text{A})}_{0,0}$ & $E^{(\text{A})}_{0,1}$ & $E^{(\text{A})}_{1,0}$ \\ 
\hline 
Gauge theory - $O(Q_{\text{5d}})$ &  \underline{10.8}124160070\ldots & 
\underline{16.8}833616281\ldots & \underline{24.9}545309653\ldots \\ 
\hline 
Gauge theory - $O(Q_{\text{5d}}^2)$ & \underline{10.822}8828498\ldots & 
\underline{16.897}8499870\ldots & \underline{24.9600}268738\ldots \\ 
\hline 
Gauge theory - $O(Q_{\text{5d}}^3)$ & \underline{10.822975}2954\ldots & 
\underline{16.8979}596523\ldots & \underline{24.96007}09871\ldots
\\ 
\hline 
Gauge theory - $O(Q_{\text{5d}}^4)$ & \underline{10.82297588}11\ldots & 
\underline{16.89796018}21\ldots & \underline{24.96007122}58\ldots \\ 
\hline 
Numerical & \underline{10.8229758847}\ldots & \underline{16.8979601851}\ldots & \underline{24.9600712270}\ldots \\ 
\hline 
\end{tabular} \caption{Energies $E^{(\text{A})}_{n_1,n_2}$ for the 3-eRS$_{\text{A}}$ model at $g_y = \sqrt{7}$, $\hbar_y = \sqrt{2}$.} \label{ex3A}
\end{center}
\end{table}
\renewcommand{\arraystretch}{1}

For what the $3$-eRS$_{\text{A}}$ model is concerned instead, in addition to the energy \eqref{W3} and the requirement $\epsilon_1 = i \hbar_y \in i \mathbb{R}_+$, $m = i g_y \in i \mathbb{R}_+$ we also need the explicit expression for the A-type quantization conditions \eqref{quantNRSA}, which for the case at hand are given by
\begin{equation}
\left\{
\begin{array}{ccc}
-i (2 \partial_{a_{12}} - \partial_{a_{23}}) \mathcal{W}^{SU(3)}_{\text{exact}}(a_{12}, a_{23}, \hbar_y, g_y, \tilde{\omega}, \tilde{\omega}') &=& 2 \pi n_{1},\\[5 pt]
-i (-\partial_{a_{12}} + 2 \partial_{a_{23}}) \mathcal{W}^{SU(3)}_{\text{exact}}(a_{12}, a_{23}, \hbar_y, g_y, \tilde{\omega}, \tilde{\omega}') &=& 2 \pi n_{2},
\end{array} \right. \label{ex3}
\end{equation}
where
\begin{equation}
\begin{split}
& -i \mathcal{W}^{SU(3)}_{\text{exact}}(a_{12}, a_{23}, \hbar_y, g_y, \tilde{\omega}, \tilde{\omega}') = 
-\frac{4 \pi i \tilde{\omega}'}{6 \tilde{\omega} \hbar_y} (a_{12}^2 + a_{12} a_{23} + a_{23}^2)
+ \dfrac{2\pi}{2\tilde{\omega} \hbar_y} g_y (g_y - 2\tilde{\omega} - \hbar_y)  \\
& + \sum_{k \geqslant 1} \left( - \dfrac{2 \tilde{\omega}}{2 \pi k} \right) \dfrac{1}{k} \dfrac{\cos(\frac{\pi k \hbar_y}{2\tilde{\omega}})}{\sin(\frac{\pi k \hbar_y}{2\tilde{\omega}})} \left[ e^{-\frac{2\pi k a_{12}}{2\tilde{\omega}}} + e^{-\frac{2\pi k a_{23}}{2\tilde{\omega}}} + e^{-\frac{2\pi k (a_{12} + a_{23})}{2\tilde{\omega}}} \right] \\
& - \sum_{k \geqslant 1} \left( - \dfrac{2 \tilde{\omega}}{2 \pi k} \right) \dfrac{1}{k} \dfrac{\cos(\frac{\pi k (\hbar_y - 2 g_y)}{2\tilde{\omega}})}{\sin(\frac{\pi k \hbar_y}{2\tilde{\omega}})} \left[ e^{-\frac{2\pi k a_{12}}{2\tilde{\omega}}} + e^{-\frac{2\pi k a_{23}}{2\tilde{\omega}}} + e^{-\frac{2\pi k (a_{12} + a_{23})}{2\tilde{\omega}}} \right] 
\\
& + \sum_{k \geqslant 1} \left( - \dfrac{\hbar_y}{2 \pi k} \right) \dfrac{1}{k} \dfrac{\cos(\frac{2\pi k\tilde{\omega}}{\hbar_y})}{\sin(\frac{2\pi k\tilde{\omega}}{\hbar_y})} \left[ e^{-\frac{2\pi k a_{12}}{\hbar_y}} + e^{-\frac{2\pi k a_{23}}{\hbar_y}} + e^{-\frac{2\pi k (a_{12} + a_{23})}{\hbar_y}} \right] \\
& - \sum_{k \geqslant 1} \left( - \dfrac{\hbar_y}{2 \pi k} \right) \dfrac{1}{k} \dfrac{\cos(\frac{\pi k (2\tilde{\omega} - 2 g_y)}{\hbar_y})}{\sin(\frac{2\pi k\tilde{\omega}}{\hbar_y})} \left[ e^{-\frac{2\pi k a_{12}}{\hbar_y}} + e^{-\frac{2\pi k a_{23}}{\hbar_y}} + e^{-\frac{2\pi k (a_{12} + a_{23})}{\hbar_y}} \right]
\\
& - i \mathcal{W}^{SU(3)}_{\text{5d,inst}}(a_{12}, a_{23}, i \hbar_y, i g_y, (2\tilde{\omega})^{-1},e^{\frac{2\pi i(2\tilde{\omega}')}{2\tilde{\omega}}}) 
 - i \mathcal{W}^{SU(3)}_{\text{5d,inst}}(a_{12}, a_{23}, i 2\tilde{\omega}, i g_y, \hbar_y^{-1},e^{\frac{2\pi i(2\tilde{\omega}')}{\hbar_y}}). \label{tweff3}
\end{split}
\end{equation}
Conjecturally, at fixed $n_1$, $n_2$ the values of $a_{12}$, $a_{23}$ satisfying \eqref{ex3}  should reproduce the $3$-eRS$_{\text{A}}$ energy levels when inserted in \eqref{W3}. This is indeed what seems to happen, as can be seen for example from Table \ref{ex3A}; as for the 2-particle case, in order to achieve agreement with numerical results and to satisfy the S-duality relation \eqref{speciale} it was crucial to take into account non-perturbative corrections to the A-type quantization conditions, in a way dictated by the modular double symmetry $\hbar_y \leftrightarrow 2\tilde{\omega}$.

\section{Conclusions}

In this paper we proposed and tested full, non-perturbatively corrected quantization conditions for the $N$-particle elliptic Ruijsenaars-Schneider system, inspired by previous results on relativistic quantum integrable systems of cluster type \cite{2015arXiv151102860H,Franco:2015rnr}, and we also checked the validity of the elliptic Calogero-Moser quantization conditions proposed by \cite{2010maph.conf..265N}. Furthermore, we gave a few comments on how gauge theory could help us constructing  entire solutions to the Baxter equation of these quantum integrable systems, which after taking an appropriate integral transformation produce the normalizable eigenfunctions for the systems in the spirit of quantum Separation of Variables. 

Among the many open problems, the main one would be to provide a rigorous proof of our conjectural full quantization conditions, which at the moment can only be tested against numerical computations. 
Another interesting question concerns the problem of constructing eigenfunctions for the elliptic Calogero-Moser or Ruijsenaars-Schneider systems: in particular it would be nice to be able to make more rigorous our sketchy derivation based on the quantum Separation of Variables approach.

Finally, it may be interesting to look at the $N$-particle elliptic Ruijsenaars-Schneider quantum spectral curve from the point of view of \cite{Grassi:2014zfa,Codesido:2015dia} (concering the energy spectrum) and \cite{Marino:2016rsq,Marino:2017gyg,Zakany:2017txl} (concerning the eigenfunctions). From this point of view the quantum spectral curve is not interpreted as the Baxter equation associated to a quantum integrable system, but is regarded instead as a quantum mechanical operator on $L^2(\mathbb{R})$; as such, boundary conditions for its eigenfunctions will in general be different from the ones required for the solution to the Baxter equation (unless $N = 2$). Clearly this quantum mechanical problem is more general than the one associated to the underlying integrable system, but should be equivalent to the latter at those special values of energies corresponding to quantum integrability: this is indeed what seems to happen for relativistic cluster integrable systems, and can also be seen in the enhanced decay properties of the eigenfunction \cite{Marino:2017gyg}. We expect similar structures will appear when studying the $N$-particle elliptic Ruijsenaars-Schneider quantum spectral curve as a quantum mechanical operator.

\section*{Acknowledgements}

We would like to thank Marcos Mari\~no for his contributions in an earlier version of this project, as well as his many useful comments in the latter stages, and Simon Ruijsenaars for useful correspondence and clarifications. 
The work of SZ is partly supported by the NCCR 51NF40-141869 ``The Mathematics of Physics'' (SwissMAP). The work of YH is supported by JSPS KAKENHI Grant Number JP18K03657.


\appendix

\section{Special functions}
\label{appA}

\subsection{Elliptic functions} \label{appell}

Let us introduce, in the standard notation of \cite{abramowitz70a}, the parameter
\begin{equation}
q = e^{i \pi \tau}\,, \;\;\; \tau = \dfrac{\omega'}{\omega} = \dfrac{ i K'}{K}\,, \label{tau}
\end{equation}
where $\omega \in \mathbb{R}_+$, $\omega' \in i \mathbb{R}_+$ are respectively the real and imaginary half-periods of the torus of modulus $\tau$. 
In the main text we also often work with the variable
\begin{equation}
Q = q^2 = e^{2\pi i \tau} , 
\end{equation}
($Q = Q_{\text{4d}}$ in Section \ref{CMsection} and $Q = Q_{\text{5d}}$ in Sections \ref{RSsection}, \ref{NRSsection}), since this is more natural from the gauge theory point of view; in terms of this variable the imaginary half-period reads
\begin{equation}
\omega' = -i \dfrac{\omega}{2 \pi} \ln Q.
\end{equation}

\noindent Always according to \cite{abramowitz70a}, the first Jacobi theta function $\theta_1(x,q)$ is defined as ($x \in \mathbb{C}$)
\begin{equation}
\begin{split}
\theta_1(x,q) & = \sum_{n \in \mathbb{Z}} (-1)^{n-1/2}q^{(n+1/2)^2}e^{(2n+1)ix} \\ 
& = -i q^{1/4} e^{i x} \sum_{n \in \mathbb{Z}} (-1)^n q^{n(n-1)} e^{-2inx} \\
& = 2 q^{1/4} \sum_{n \geqslant 0} (-1)^n q^{n(n+1)} \sin[(2n+1)x].
\end{split}
\end{equation}
From the Jacobi theta we can construct the Weierstrass sigma function 
\begin{equation}
\sigma(x \vert \omega, \omega') = \dfrac{2 \omega}{\pi \theta_1'(0,q)} \exp\left(-\dfrac{\pi^2 \theta_1'''(0,q)}{24 \omega^2 \theta_1'(0,q)}x^2 \right) \theta_1\left(\dfrac{\pi x}{2 \omega},q\right),
\end{equation}
where $q$ is related to $(\omega, \omega')$ as in \eqref{tau} and
\begin{equation}
\dfrac{\theta_1'''(0,q)}{\theta_1'(0,q)} = - \dfrac{\sum_{n \geqslant 0} (-1)^n (2n+1)^3 q^{n(n+1)}}{\sum_{n \geqslant 0} (-1)^n (2n+1) q^{n(n+1)}} = -1 + 24 \sum_{n \geqslant 1} \dfrac{q^{2n}}{(1-q^{2n})^2};
\end{equation}
for notational convenience we will sometimes write this last factor in terms of the quasi-modular form $E_2(Q)$ as
\begin{equation}
E_2(Q) = E_2(q^2) = - \dfrac{\theta_1'''(0,q)}{\theta_1'(0,q)} 
= 1 - 24 \sum_{n \geqslant 1} \dfrac{Q^n}{(1-Q^n)^2}.
\label{eq:E2}
\end{equation}
The logarithm of the sigma function therefore is
\begin{equation}
\ln \sigma(x \vert \omega, \omega') = \ln \left( \dfrac{2 \omega}{\pi \theta_1'(0,q)} \right) -\dfrac{\pi^2 \theta_1'''(0,q)}{24 \omega^2 \theta_1'(0,q)}x^2 +\ln(-i q^{1/4}) + i \dfrac{\pi x}{2 \omega}
+ \ln \left[ \sum_{n \in \mathbb{Z}} (-1)^n q^{n(n-1)} e^{-\frac{i\pi n x}{\omega}} \right],
\end{equation}
and its second $x$ derivative is related to the Weierstrass elliptic function $\wp(x \vert \omega, \omega')$ of half-periods $(\omega, \omega')$ as
\begin{equation}
\begin{split}
\wp(x\vert \omega, \omega') &= - \dfrac{d^2}{dx^2} \ln \sigma(x\vert \omega, \omega') 
= - \dfrac{d^2}{dx^2} \ln \left[ \sum_{n \in \mathbb{Z}} (-1)^n q^{n(n-1)} e^{-n x} \right] + \dfrac{\pi^2}{12 \omega^2} \dfrac{\theta_1'''(0,q)}{\theta_1'(0,q)} \\
& = -\dfrac{d^2}{dx^2} \ln \theta_1\left(\frac{ \pi x}{2 \omega},q\right) + \dfrac{\pi^2}{12 \omega^2} \dfrac{\theta_1'''(0,q)}{\theta_1'(0,q)} ,
\end{split}
\end{equation}
which in the limit $q\rightarrow 0$ (i.e. $\omega' \rightarrow i \infty$) behaves as
\begin{equation}
\wp(x \vert \omega, i \infty) = -\dfrac{\pi^2}{12 \omega^2} + \dfrac{\pi^2}{4 \omega^2 \sin^2 \frac{\pi x}{2\omega}} = \dfrac{1}{x^2} + \dfrac{\pi^4 x^2}{240 \omega^4} + \dfrac{\pi^6 x^4}{6048 \omega^6} + \ldots .
\end{equation}
It may also be useful to consider the Weierstrass invariants
\begin{equation}
\begin{split}
g_2(\omega, \omega') &= \dfrac{\pi^4}{12 \,\omega^4}\left( 1+ \dfrac{2}{\zeta(-3)} \sum_{n \geqslant 1} \dfrac{n^3 q^{2n}}{1-q^{2n}} \right), \\
g_3(\omega, \omega') &= \dfrac{\pi^6}{216 \,\omega^6}\left( 1+ \dfrac{2}{\zeta(-5)} \sum_{n \geqslant 1} \dfrac{n^5 q^{2n}}{1-q^{2n}} \right),
\end{split}
\end{equation}
which are such that
\begin{equation}
\wp(x \vert \omega, \omega') = \dfrac{1}{x^2} + \dfrac{g_2(\omega, \omega')}{20}x^2 +  \dfrac{g_3(\omega, \omega')}{28}x^4 + O(x^6).
\end{equation}
Finally, we sometimes use in the main text the functions $\theta(x,Q)$ and $\overline{\theta}(X,Q)$ defined as
\begin{eqnarray}
\theta(x,Q) & = & i Q^{-1/8} e^{-i x} \theta_1(x,Q^{1/2}) \nonumber  \\
& = & (Q;Q)_{\infty} (e^{-2ix};Q)_{\infty} (Q e^{2ix};Q)_{\infty} 
= \sum_{n \in \mathbb{Z}} (-1)^n Q^{\frac{n(n-1)}{2}} e^{-2inx},  \label{theta} \\
\overline{\theta}(X,Q) & = & \theta(\frac{i}{2} \ln X,Q) \nonumber \\ 
& = & (Q;Q)_{\infty} (X;Q)_{\infty} (Q X^{-1};Q)_{\infty} 
= \sum_{n \in \mathbb{X}} (-1)^n Q^{\frac{n(n-1)}{2}} X^n, \label{thetabar}
\end{eqnarray}
where the infinite q-Pochhammer symbol is 
\begin{equation}
(X;q)_{\infty} = \prod_{n \geqslant 0} (1-q^n X).
\end{equation}

\subsection{Sine functions} \label{appspecial}

\subsubsection*{Double sine function}
Let us consider the function $\mathcal{S}(x \vert \omega_1, \omega_2)$ defined as
\begin{equation}
\mathcal{S}(x \vert \omega_1, \omega_2)  = \text{exp} \left( \int_{\mathbb{R} + i 0} \dfrac{e^{xz}}{(e^{\omega_1 z} - 1)(e^{\omega_2 z} - 1)} \dfrac{dz}{z} \right),
\end{equation}
in the strip $0 < \text{Re}\,z  < \text{Re} (\omega_1 + \omega_2)$ when all $\text{Re} (\omega_j) > 0$.  This function is related to the quantum dilogarithm $\Phi(x \vert \omega_1, \omega_2)$ introduced in \cite{1995LMaPh..34..249F}
\begin{equation}
\Phi(x \vert \omega_1, \omega_2) = \text{exp} \left( \int_{\mathbb{R} + i 0} \dfrac{e^{-2ixz}}{2\sinh(\omega_1 z)\cdot 2\sinh(\omega_2 z)} \dfrac{dz}{z} \right),
\end{equation}
according to
\begin{equation}
\mathcal{S}\left(-ix + \frac{\omega_1 + \omega_2}{2} \Big\vert \omega_1, \omega_2 \right) 
= \Phi(x \vert \omega_1, \omega_2).
\end{equation}
When Im$\,(\omega_1/\omega_2) > 0$ or $\omega_1/\omega_2 \notin \mathbb{Q}$ it admits the infinite product representation
\begin{equation}
\mathcal{S}\left(ix + \omega_1 + \omega_2  \vert \omega_1, \omega_2 \right) = 
\Phi\left(-x + i \frac{\omega_1 + \omega_2}{2}\Big\vert  \omega_1, \omega_2 \right) = 
\dfrac{(q e^{-2\pi x/\omega_2};q)_{\infty}}{(e^{-2\pi x/\omega_1};\widetilde{q}^{-1})_{\infty}}, \label{pizza}
\end{equation}
where we introduced the parameters
\begin{equation}
q = e^{2\pi i \omega_1 /\omega_2} \;\;\;,\;\;\; \widetilde{q} = e^{2\pi i \omega_2 /\omega_1},
\end{equation}
and the $q$-Pochhammer symbol
\begin{equation}
(w;q)_{\infty} = \prod_{k=0}^{\infty} (1-q^k w).
\end{equation}
A useful formula related to the $q$-Pochhammer symbol is
\begin{equation}
\text{exp} \left( - \sum_{k \geqslant 1} \dfrac{q^k w^{k}}{k(1 - q^{k})} \right) =
\begin{cases}
\prod_{k \geqslant 0} (1-q^{k+1}w) \;\;\;, \;\;\; \vert q \vert < 1; \\
\prod_{k \geqslant 0} \dfrac{1}{(1-q^{-k}w)} \;\;\;, \;\;\; \vert q \vert > 1. 
\end{cases} 
\label{continuation}
\end{equation}
The $\mathcal{S}(x \vert \omega_1, \omega_2)$ function satisfies the identities
\begin{equation}
\begin{split}
& \mathcal{S}\left(ix + \omega_1 + \omega_2 - \omega_1  \vert \omega_1, \omega_2 \right) 
= \left( 1 - e^{-2\pi x/\omega_2} \right) 
\mathcal{S}\left(ix + \omega_1 + \omega_2  \vert \omega_1, \omega_2 \right), \\
& \mathcal{S}\left(ix + \omega_1 + \omega_2 - \omega_2  \vert \omega_1, \omega_2 \right) 
= \left( 1 - e^{-2\pi x/\omega_1} \right) 
\mathcal{S}\left(ix + \omega_1 + \omega_2  \vert \omega_1, \omega_2 \right), 
\end{split}
\end{equation}
and
\begin{equation}
\begin{split}
& \mathcal{S}\left(ix + \omega_1 + \omega_2 + \omega_1 \vert \omega_1, \omega_2 \right) 
= \dfrac{1}{\left( 1 - q e^{-2\pi x/\omega_2} \right)}
\mathcal{S}\left(ix + \omega_1 + \omega_2 \vert \omega_1, \omega_2 \right), \\
& \mathcal{S}\left(ix + \omega_1 + \omega_2 +\omega_2 \vert \omega_1, \omega_2 \right) 
= \dfrac{1}{\left( 1 - \widetilde{q}e^{-2\pi x/\omega_1} \right)}
\mathcal{S}\left(ix + \omega_1 + \omega_2 \vert \omega_1, \omega_2 \right), 
\end{split}
\end{equation}
as well as
\begin{equation}
\begin{split}
& \mathcal{S}\left(-ix + \frac{\omega_1 + \omega_2}{2} \Big\vert \omega_1, \omega_2 \right)\mathcal{S}\left(ix + \frac{\omega_1 + \omega_2}{2} \Big\vert \omega_1, \omega_2 \right) = \\
& = \; \Phi(x \vert \omega_1, \omega_2) \Phi(-x \vert \omega_1, \omega_2)
\; = \; e^{i \pi \frac{x^2}{\omega_1 \omega_2} + i \pi \frac{\omega_1^2 + \omega_2^2}{12\, \omega_1 \omega_2}}.
\end{split}
\end{equation}
Closely related to $\mathcal{S}(x \vert \omega_1, \omega_2)$ is the double sine function $S_2(x \vert \omega_1, \omega_2)$ defined as
\begin{equation}
S_2(x \vert \omega_1, \omega_2) = e^{\frac{i \pi}{2}B_{2,2}(x\vert \omega_1, \omega_2)} \mathcal{S}(x \vert \omega_1, \omega_2),
\end{equation}
where
\begin{equation}
B_{2,2}(x\vert \omega_1, \omega_2) = \dfrac{x^2}{\omega_1 \omega_2} - \dfrac{\omega_1 + \omega_2}{\omega_1 \omega_2}x + \dfrac{\omega_1^2 + 3 \omega_1 \omega_2 + \omega_2^2}{6\omega_1 \omega_2}. 
\end{equation}
The double sine function can be thought as the regularization of the infinite product
\begin{equation}
S_2(x \vert \omega_1, \omega_2)  = \prod_{m,n \geqslant 0}\dfrac{m \omega_1 + n \omega_2 + x}{m \omega_1 + n \omega_2 + \omega_1 + \omega_2 - x}.
\end{equation}
This function satisfies 
\begin{equation}
S_2(x \vert \omega_1, \omega_2) S_2(-x +\omega_1 + \omega_2 \vert \omega_1, \omega_2) = 1,
\end{equation}
as well as
\begin{equation}
\overline{S_2(x \vert \omega_1, \omega_2)} = S_2(\overline{x} \vert \overline{\omega}_1, \overline{\omega}_2) \label{complconj}
\end{equation}
and
\begin{equation}
\begin{split}
& S_2^{-1}(-i x + \omega_1 \vert \omega_1, \omega_2)
= - i \cdot 2 \sinh\left[\frac{\pi x}{\omega_2}\right] 
S_2^{-1}(-i x \vert \omega_1, \omega_2), \\
& S_2^{-1}(-i x + \omega_2 \vert \omega_1, \omega_2)
= - i \cdot 2 \sinh\left[\frac{\pi x}{\omega_1}\right] 
S_2^{-1}(-i x \vert \omega_1, \omega_2), \\
& S_2^{-1}(-i x - \omega_1 \vert \omega_1, \omega_2)
= \dfrac{1}{- i \cdot 2 \sinh\left[\frac{\pi x}{\omega_2} - i \pi \frac{\omega_1}{\omega_2}\right]} S_2^{-1}(-i x \vert \omega_1, \omega_2),  \\
& S_2^{-1}(-i x - \omega_2 \vert \omega_1, \omega_2)
= \dfrac{1}{- i \cdot 2 \sinh\left[\frac{\pi x}{\omega_1} - i \pi \frac{\omega_2}{\omega_1}\right]} S_2^{-1}(-i x \vert \omega_1, \omega_2). \label{property}
\end{split}
\end{equation}

\subsubsection*{Triple sine function}

We can also introduce the function $\mathcal{S}(x \vert \omega_1, \omega_2, \omega_3)$ defined as
\begin{equation}
\mathcal{S}(x \vert \omega_1, \omega_2, \omega_3)  = 
\text{exp} \left( - \int_{\mathbb{R} + i 0} \dfrac{e^{xz}}{(e^{\omega_1 z} - 1)(e^{\omega_2 z} - 1)(e^{\omega_3 z} - 1)} \dfrac{dz}{z} \right)
\end{equation}
in the strip $0 < \text{Re}\,z  < \text{Re} (\omega_1 + \omega_2 + \omega_3)$ when all $\text{Re} (\omega_j) > 0$. When Im$\,(\omega_1/\omega_2) > 0$, Im$\,(\omega_1/\omega_3) > 0$, Im$\,(\omega_3/\omega_2) > 0$ or when $\omega_1/\omega_2, \omega_1/\omega_3, \omega_3/\omega_2 \notin \mathbb{Q}$ this admits the infinite product representation
\begin{equation}
\begin{split}
& \mathcal{S}(x \vert \omega_1, \omega_2, \omega_3) = \\
& = \dfrac{(e^{2\pi i x/\omega_2};e^{2\pi i \omega_1/\omega_2}; e^{2\pi i \omega_3/\omega_2})_{\infty} 
(e^{-2\pi i \omega_3/\omega_1} e^{-2\pi i \omega_2/\omega_1}e^{2\pi i x/\omega_1};e^{-2\pi i \omega_3/\omega_1}; e^{-2\pi i \omega_2/\omega_1})_{\infty}}{(e^{-2\pi i \omega_2/\omega_3} e^{2\pi i x/\omega_3};e^{2\pi i \omega_1/\omega_3}; e^{-2\pi i \omega_2/\omega_3})_{\infty}}, 
\end{split}
\end{equation}
where we introduced the notation
\begin{equation}
\begin{split}
& \left(e^{2\pi i x/\omega_2};e^{2\pi i \omega_1/\omega_2}; e^{2\pi i \omega_3/\omega_2} \right)_{\infty} = \prod_{j,k\geqslant 0}\left(1 - e^{2\pi i x/\omega_2}e^{2\pi i \omega_1 j/\omega_2} e^{2\pi i \omega_3 k/\omega_2} \right) = \\
& = \text{exp}\left( - \sum_{n \geqslant 1} \dfrac{e^{2\pi n i x/\omega_2}}{n(1 - e^{2\pi i \omega_1 n /\omega_2})(1 - e^{2\pi i \omega_3 n/\omega_2})} \right).
\end{split}
\end{equation}
Closely related to $\mathcal{S}(x \vert \omega_1, \omega_2, \omega_3)$ is the triple sine function defined as
\begin{equation}
S_3(x \vert \omega_1, \omega_2, \omega_3) = e^{-\frac{i \pi}{6}B_{3,3}(x\vert \omega_1, \omega_2, \omega_3)} \mathcal{S}(x \vert \omega_1, \omega_2, \omega_3) \label{pppp}
\end{equation}
satisfying
\begin{equation}
S_3(x \vert \omega_1, \omega_2, \omega_3) = S_3(- x + \omega_1 + \omega_2 + \omega_3 \vert \omega_1, \omega_2, \omega_3),
\end{equation}
where
\begin{equation}
\begin{split}
B_{3,3}(x\vert \omega_1, \omega_2, \omega_3) & \;=\;  
\dfrac{x^3}{\omega_1 \omega_2 \omega_3} - \dfrac{3}{2}\dfrac{\omega_1 + \omega_2 + \omega_3}{\omega_1, \omega_2, \omega_3} x^2 \\
& \;+\; \dfrac{\omega_1^2 + \omega_2^2 + \omega_3^2 + 3 (\omega_1 \omega_2 + \omega_1 \omega_3 + \omega_2 \omega_3)}{2\omega_1 \omega_2 \omega_3} x \\
& \; - \; \dfrac{(\omega_1 + \omega_2 + \omega_3)(\omega_1 \omega_2 + \omega_1 \omega_3 + \omega_2 \omega_3)}{4\omega_1 \omega_2 \omega_3}.
\end{split}
\end{equation}

\section{Gauge theory formulae} \label{appgauge}


\subsection{Four dimensional case}

The instanton part of the partition function for an $\mathcal{N} = 2^*$ $U(N)$ theory on $\mathbb{R}^4_{\epsilon_1, \epsilon_2}$ is given by the $Q_{\text{4d}}$-series 
\begin{equation}
Z_{\text{4d,inst}}^{U(N)}(\vec{a}, \epsilon_1, \epsilon_2, m, Q_{\text{4d}}) 
= \sum_{k \geqslant 0} Q_{\text{4d}}^k Z_{\text{4d,inst}}^{U(N),(k)}(\vec{a}, \epsilon_1, \epsilon_2, m),
\end{equation}
where $\vec{a} = (a_1, \ldots, a_N)$ are the VEVs of the complex scalar field in the $\mathcal{N}=2$ vector multiplet. The coefficients of the $Q_{\text{4d}}$-series can be computed from the contour integral formula
\begin{equation}
Z_{\text{4d,inst}}^{U(N),(k)}(\vec{a}, \epsilon_1, \epsilon_2, m)
 = \dfrac{1}{k!} \oint \prod_{s=1}^k \frac{d\phi_s}{2\pi i}
Z_{\text{4d,ADHM}}^{U(N),(k)}(\vec{\phi}, \vec{a}, \epsilon_1, \epsilon_2, m),
\end{equation}
with $\vec{\phi} = (\phi_1, \ldots, \phi_k)$ integration variables and
\begin{equation}
\begin{split}
& Z_{\text{4d,ADHM}}^{U(N),(k)}(\vec{\phi}, \vec{a}, \epsilon_1, \epsilon_2, m) = \\
& \left( -\frac{(2\epsilon_+)}{(\epsilon_1 \epsilon_2)} 
\frac{(m + \epsilon_-)(m - \epsilon_-)}{(m + \epsilon_+)(m - \epsilon_+)} \right)^k 
 \prod_{s=1}^k \prod_{l=1}^N \dfrac{(\phi_s - a_l + m)(\phi_s - a_l - m)}{(\phi_s - a_l + \epsilon_+)(\phi_s - a_l - \epsilon_+)} \\
& \prod_{s\neq t}^k \dfrac{\phi_{st}^2 (\phi_{st}^2 - 4\epsilon_+^2)}{(\phi_{st}^2 - \epsilon_1^2 )(\phi_{st}^2 - \epsilon_2^2 )}
\dfrac{\left(\phi_{st}^2 - (m + \epsilon_-)^2 \right) \left(\phi_{st}^2 - (m - \epsilon_-)^2 \right)}{\left(\phi_{st}^2 - (m + \epsilon_+)^2 \right) \left(\phi_{st}^2 - (m - \epsilon_+)^2 \right)} .
\end{split}
\end{equation}
Here we are using the short-hand notation $\epsilon_+ = \frac{\epsilon_1 + \epsilon_2}{2}$, $\epsilon_- = \frac{\epsilon_1 - \epsilon_2}{2}$ and $\phi_{st} = \phi_s - \phi_t$. The integral is evaluated by residues, and the relevant poles are labelled by $N$-tuples of Young tableaux $\vec{Y} = (Y_1, \ldots, Y_N)$, $\sum_{l=1}^N \vert Y_l \vert = k$ corresponding to
\begin{equation}
\phi_s = a_l + \epsilon_+ + (i-1)\epsilon_1 + (j-1)\epsilon_2, \;\;\;\;\;\; l = 1, \ldots, N,
\end{equation}
with $(i,j)$ position of a box in the $l$-th Young tableau $Y_l$. When in the main text we refer to $SU(N)$ quantities, we simply mean that we impose $\sum_{l=1}^N a_l = 0$: for example, we define $SU(2)$ quantities as the $U(2)$ ones with $a_1 = -a_2 = a$. In the NS limit, the instanton partition function reduces to the instanton part of the twisted effective superpotential defined as
\begin{equation}
\mathcal{W}^{U(N)}_{\text{4d,\,inst}}(\vec{a},\epsilon_1,m,Q_{\text{4d}}) 
= \lim_{\epsilon_2 \to 0} \left[ -\epsilon_1 \epsilon_2 \ln Z^{U(N)}_{\text{4d,\,inst}}(\vec{a},\epsilon_1,\epsilon_2,m,Q_{\text{4d}})  \right];
\end{equation}
this is the quantity entering in the A-type quantization conditions for the $N$-particle elliptic Calogero-Moser system of type A. \\

The instanton part of the fundamental qq-character, 
which enters in the construction of the Baxter equation for the $N$-particle elliptic Calogero-Moser system of type A,
is instead given by
\begin{equation}
\chi_{\text{4d,inst}}^{U(N)}(\sigma, \vec{a}, \epsilon_1, \epsilon_2, m, Q_{\text{4d}}) 
= \sum_{k \geqslant 0} Q_{\text{4d}}^k \chi_{\text{4d,inst}}^{U(N),(k)}(\sigma, \vec{a}, \epsilon_1, \epsilon_2, m),
\end{equation}
where the coefficients of the $Q_{\text{4d}}$-series are obtained from
\begin{equation}
\chi_{\text{4d,inst}}^{U(N),(k)}(\sigma,\vec{a}, \epsilon_1, \epsilon_2, m)
 = \dfrac{1}{k!} \oint \prod_{s=1}^k \frac{d\phi_s}{2\pi i}
Z_{\text{4d,ADHM}}^{U(N),(k)}(\vec{\phi}, \vec{a}, \epsilon_1, \epsilon_2, m)
\chi_{\text{4d,ADHM}}^{U(N),(k)}(\vec{\phi},\sigma, \vec{a}, \epsilon_1, \epsilon_2, m),
\end{equation}
and the extra part of the integrand reads
\begin{equation}
\chi_{\text{4d,ADHM}}^{U(N),(k)}(\vec{\phi},\sigma, \vec{a}, \epsilon_1, \epsilon_2, m) 
= \prod_{l=1}^N (\sigma - a_l) \prod_{s=1}^k \dfrac{(\phi_s - \sigma + \epsilon_-)(\phi_s - \sigma - \epsilon_-)}{(\phi_s - \sigma + \epsilon_+)(\phi_s - \sigma - \epsilon_+)}.
\end{equation}
The poles to be considered in this case are labelled by ($N$+$1$)-tuples of Young tableaux $\vec{Y} = (Y_1, \ldots, Y_N, Y_{N+1})$, $\sum_{n=1}^{N+1} \vert Y_n \vert = k$ corresponding to
\begin{equation}
\begin{split}
\phi_s & = a_n + \epsilon_+ + (i-1)\epsilon_1 + (j-1)\epsilon_2, \;\;\;\; n = 1, \ldots, N \;\;\;\; \text{or} \\
\phi_s & = \sigma - \epsilon_+ - (i-1)(m + \epsilon_+) - (j-1)(-m+\epsilon_+), \;\;\;\; n = N + 1.
\end{split}
\end{equation}
Being a string theory observable defined as a system of two stacks of branes intersecting on a point, its natural normalization involves the partition function of the theory living on both stacks, that is \cite{Agarwal:2018tso}
\begin{equation}
\Big\langle \chi_{\text{4d}}^{U(N)}(\sigma, \vec{a}, \epsilon_1, \epsilon_2, m, Q_{\text{4d}})  \Big\rangle = \dfrac{\chi_{\text{4d,inst}}^{U(N)}(\sigma, \vec{a}, \epsilon_1, \epsilon_2, m, Q_{\text{4d}}) }{Z_{\text{4d,inst}}^{U(N)}(\vec{a}, \epsilon_1, \epsilon_2, m, Q_{\text{4d}})  Z_{\text{4d,inst}}^{U(1)}(\sigma, -\epsilon_+ + m, -\epsilon_+ - m, \epsilon_-, Q_{\text{4d}}) };
\end{equation}
its NS limit is then defined as
\begin{equation}
\Big\langle \chi_{\text{4d}}^{U(N)}(\sigma, \vec{a}, \epsilon_1, m, Q_{\text{4d}})  \Big\rangle_{\text{NS}}  = 
\Big\langle \chi_{\text{4d}}^{U(N)}(\sigma, \vec{a}, \epsilon_1, 0, m, Q_{\text{4d}})  \Big\rangle .
\end{equation}

Finally, the instanton part of the partition function for the theory coupled to an $N$-plet of two-dimensional free hypermultiplets is
\begin{equation}
\mathcal{Q}_{\text{4d,inst}}^{(f)}(\sigma, \vec{a}, \epsilon_1, \epsilon_2, m, Q_{\text{4d}}) 
= \sum_{k \geqslant 0} Q_{\text{4d}}^k \mathcal{Q}_{\text{4d,inst}}^{(f),(k)}(\sigma, \vec{a}, \epsilon_1, \epsilon_2, m),
\end{equation}
where the coefficients are given by
\begin{equation}
\mathcal{Q}_{\text{4d,inst}}^{(f),(k)}(\sigma,\vec{a}, \epsilon_1, \epsilon_2, m)
 = \dfrac{1}{k!} \oint \prod_{s=1}^k \frac{d\phi_s}{2\pi i}
Z_{\text{4d,ADHM}}^{U(N),(k)}(\vec{\phi}, \vec{a}, \epsilon_1, \epsilon_2, m)
\mathcal{Q}_{\text{4d,ADHM}}^{(f),(k)}(\vec{\phi},\sigma, \vec{a}, \epsilon_1, \epsilon_2, m),
\end{equation}
with
\begin{equation}
\mathcal{Q}_{\text{4d,ADHM}}^{(f),(k)}(\vec{\phi},\sigma, \vec{a}, \epsilon_1, \epsilon_2, m) 
= \prod_{s=1}^k \dfrac{(\phi_s - \sigma + \epsilon_2 + \epsilon_-)(\phi_s - \sigma - m)}{(\phi_s - \sigma + \epsilon_-)(\phi_s - \sigma + \epsilon_2 - m)}.
\end{equation}
Relevant poles are again labelled by ($N$+$1$)-tuples of Young tableaux $\vec{Y} = (Y_1, \ldots, Y_N, Y_{N+1})$, $\sum_{n=1}^{N+1} \vert Y_n \vert = k$, where this time
\begin{equation}
\begin{split}
\phi_s & = a_n + \epsilon_+ + (i-1)\epsilon_1 + (j-1)\epsilon_2, \;\;\;\; n = 1, \ldots, N \;\;\;\; \text{or} \\
\phi_s & = \sigma - m - \epsilon_2+ (i-1)\epsilon_1 - (j-1)(m + \epsilon_+) , \;\;\;\; n = N + 1.
\end{split}
\end{equation}
From the normalized observable
\begin{equation}
\langle \mathcal{Q}_{\text{4d,inst}}^{(f)}(\sigma, \vec{a}, \epsilon_1, \epsilon_2, m, Q_{\text{4d}})  \rangle = \dfrac{\mathcal{Q}_{\text{4d,inst}}^{(f)}(\sigma, \vec{a}, \epsilon_1, \epsilon_2, m, Q_{\text{4d}}) }{Z_{\text{4d,inst}}^{U(N)}(\vec{a}, \epsilon_1, \epsilon_2, m, Q_{\text{4d}}) }
\end{equation}
we obtain its NS limit as
\begin{equation}
\langle \mathcal{Q}_{\text{4d,inst}}^{(f)}(\sigma, \vec{a}, \epsilon_1, m, Q_{\text{4d}})  \rangle_{\text{NS}} = \langle \mathcal{Q}_{\text{4d,inst}}^{(f)}(\sigma, \vec{a}, \epsilon_1, 0, m, Q_{\text{4d}})  \rangle  ,
\end{equation}
which is part of the solution to the Baxter equation of the $N$-particle elliptic Calogero-Moser system of type A.

\subsection{Five dimensional case}

The instanton part of the partition function for an $\mathcal{N} = 1^*$ $U(N)$ theory on $\mathbb{R}^4_{\epsilon_1, \epsilon_2} \times S^1_R$ is given by the $Q_{\text{5d}}$-series
\begin{equation}
Z_{\text{5d,inst}}^{U(N)}(\vec{a},\epsilon_1,\epsilon_2,m,R,Q_{\text{5d}}) = \sum_{k \geqslant 0} Q_{\text{5d}}^k Z_{\text{5d,inst}}^{U(N),(k)}(\vec{a},\epsilon_1,\epsilon_2,m,R) ,
\end{equation}
whose coefficients are determined by the contour integral formula
\begin{equation}
Z_{\text{5d,inst}}^{U(N),(k)}(\vec{a},\epsilon_1,\epsilon_2,m,R) 
= \dfrac{1}{k!} \oint \prod_{s=1}^k \dfrac{d \varphi_s}{\varphi_s} 
Z_{\text{5d,ADHM}}^{U(N),(k)}(\vec{\varphi},\vec{a},\epsilon_1,\epsilon_2,m,R),
\end{equation}
with $\vec{\varphi} = (\varphi_1, \ldots, \varphi_k)$ integration variables (related to the 4d ones by $\varphi_s = e^{2\pi R \phi_s}$) and
\begin{equation}
\begin{split}
Z_{\text{5d,ADHM}}^{U(N),(k)}(\vec{\varphi},\vec{a},\epsilon_1,\epsilon_2,m,R) = & \left(\dfrac{(1-q_1 q_2)}{(1-q_1)(1-q_2)} \dfrac{(1 - \mu q_1^{-1})(1 - \mu q_2^{-1})}{(1-\mu)(1-\mu q_1^{-1}q_2^{-1}))} \right)^k  \\
& \prod_{s \neq t}^k \dfrac{(1-\varphi_s \varphi_t^{-1})(1-q_1 q_2 \varphi_s \varphi_t^{-1})}{(1-q_1 \varphi_s \varphi_t^{-1})(1-q_2\varphi_s \varphi_t^{-1})}
\dfrac{(1- \mu q_1^{-1}\varphi_s \varphi_t^{-1})(1- \mu q_2^{-1}\varphi_s \varphi_t^{-1})}{(1- \mu \varphi_s \varphi_t^{-1})(1- \mu q_1^{-1}q_2^{-1}\varphi_s \varphi_t^{-1})} \\
& \prod_{s=1}^k \prod_{l=1}^N q_1 q_2 \mu^{-1} 
\dfrac{(1-\varphi_s \alpha_l^{-1} \mu q_1^{-\frac{1}{2}} q_2^{-\frac{1}{2}})(1-\varphi_s^{-1} \alpha_l \mu q_1^{-\frac{1}{2}} q_2^{-\frac{1}{2}})}{(1-\varphi_s \alpha_l^{-1} q_1^{\frac{1}{2}} q_2^{\frac{1}{2}})(1-\varphi_s^{-1} \alpha_l q_1^{\frac{1}{2}} q_2^{\frac{1}{2}})},
\end{split}
\end{equation}
where for convenience we used the exponentiated variables
\begin{equation}
\alpha_l = e^{2\pi R a_l} \,, \;\;\; q_1 = e^{2\pi R \epsilon_1} \,, \;\;\; 
q_2 = e^{2\pi R \epsilon_2} \,,\;\;\; \mu = e^{2\pi R m}.
\end{equation}
This integral is still evaluated by residues, and receives contributions only from poles labelled by $N$-tuples of Young tableaux $\vec{Y} = (Y_1, \ldots, Y_N)$, $\sum_{l=1}^N \vert Y_l \vert = k$ corresponding to
\begin{equation}
\varphi_s = \alpha_l q_1^{\frac{1}{2}} q_2^{\frac{1}{2}} q_1^{i-1} q_2^{j-1}, \;\;\;\;\;\; l = 1, \ldots, N,
\end{equation}
with $(i,j)$ position of a box in the $l$-th Young tableau $Y_l$. As for the four-dimensional case, when in the main text we refer to $SU(N)$ quantities we simply mean to impose $\sum_{l=1}^N a_l = 0$: in particular, we define $SU(2)$ quantities as the $U(2)$ ones with $a_1 = -a_2 = \frac{a}{2}$. From the NS limit of the instanton partition function we obtain the instanton part of the twisted effective superpotential defined as
\begin{equation}
\mathcal{W}^{U(N)}_{\text{5d,\,inst}}(\vec{a},\epsilon_1,m,R,Q_{\text{5d}}) 
= \lim_{\epsilon_2 \to 0} \left[ - \epsilon_2 \ln Z^{U(N)}_{\text{5d,\,inst}}(\vec{a},\epsilon_1,\epsilon_2,m,R,Q_{\text{5d}})  \right]; \label{ultimo}
\end{equation}
this is the quantity entering in the A-type quantization conditions for the $N$-particle elliptic Ruijsenaars-Schneider system of type A. \\

The instanton corrections to fundamental qq-character, whose NS limit contains the energies of the $N$-particle elliptic Ruijsenaars-Schneider system, are similarly given by
\begin{equation}
\chi_{\text{5d,inst}}^{U(N)}(\sigma, \vec{a}, \epsilon_1, \epsilon_2, m, R, Q_{\text{5d}}) = \sum_{k \geqslant 0} Q_{\text{5d}}^k \chi_{\text{5d,inst}}^{U(N),(k)}(\sigma, \vec{a}, \epsilon_1, \epsilon_2, m, R),
\end{equation}
where the coefficients read
\begin{equation}
\begin{split}
& \chi_{\text{5d,inst}}^{U(N),(k)}(\sigma, \vec{a}, \epsilon_1, \epsilon_2, m, R) \\
& = \dfrac{1}{k!} \oint \prod_{s=1}^k \dfrac{d \varphi_s}{\varphi_s} 
Z_{\text{5d,ADHM}}^{U(N),(k)}(\vec{\varphi}, \vec{a}, \epsilon_1, \epsilon_2, m, R) \chi_{\text{5d, ADHM}}^{U(N),(k)}(\vec{\varphi}, \sigma, \vec{a}, \epsilon_1, \epsilon_2, m, R),
\end{split}
\end{equation}
with (for $X = e^{2\pi R \sigma}$)
\begin{equation}
\begin{split}
&\chi_{\text{5d, ADHM}}^{U(N),(k)}(\vec{\varphi}, \sigma, \vec{a}, \epsilon_1, \epsilon_2, m, R) \\
& = \prod_{l=1}^N (X^{\frac{1}{2}}\alpha_l^{-\frac{1}{2}} - X^{-\frac{1}{2}}\alpha_l^{\frac{1}{2}})
\prod_{s=1}^k \dfrac{(1-\varphi_s X^{-1} q_1^{\frac{1}{2}} q_2^{-\frac{1}{2}})(1-\varphi_s X^{-1} q_1^{-\frac{1}{2}} q_2^{\frac{1}{2}})}{(1-\varphi_s X^{-1} q_1^{\frac{1}{2}} q_2^{\frac{1}{2}})(1-\varphi_s X^{-1} q_1^{-\frac{1}{2}} q_2^{-\frac{1}{2}})}.
\end{split}
\end{equation}
Poles are labelled by ($N$+$1$)-tuples of Young tableaux $\vec{Y} = (Y_1, \ldots, Y_N, Y_{N+1})$, $\sum_{n=1}^{N+1} \vert Y_n \vert = k$ corresponding to
\begin{equation}
\begin{split}
\varphi_s &= \alpha_n q_1^{\frac{1}{2}} q_2^{\frac{1}{2}} q_1^{i-1} q_2^{j-1} , \;\;\;\; n = 1, \ldots, N \;\;\;\; \text{or} \\
\varphi_s &= X q_1^{-\frac{1}{2}} q_2^{-\frac{1}{2}} \mu^{-(i-1)} (\mu q_1^{-1} q_2^{-1} )^{j-1}, \;\;\;\; n = N+1.
\end{split}
\end{equation}
Since the fundamental qq-character is the partition function of a system of $N$ D4 branes intersecting a single different D4' brane along a circle, its natural normalization involves the $U(N)$ partition function of the D4 theory as well as the $U(1)$ partition function of the D4' theory, that is \cite{Agarwal:2018tso}
\begin{equation}
\Big\langle \chi_{\text{5d}}^{U(N)}(\sigma, \vec{a}, \epsilon_1, \epsilon_2, m, R, Q_{\text{5d}})  \Big\rangle = \dfrac{\chi_{\text{5d,inst}}^{U(N)}(\sigma, \vec{a}, \epsilon_1, \epsilon_2, m,R, Q_{\text{5d}}) }{Z_{\text{5d,inst}}^{U(N)}(\vec{a}, \epsilon_1, \epsilon_2, m, R, Q_{\text{5d}})  Z_{\text{5d,inst}}^{U(1)}(\sigma, -2\epsilon_+ + m, - m, -\epsilon_2,R, Q_{\text{5d}}) }.
\end{equation}
Its NS limit defined as
\begin{equation}
\Big\langle \chi_{\text{5d}}^{U(N)}(\sigma, \vec{a}, \epsilon_1, m, R, Q_{\text{5d}})  \Big\rangle_{\text{NS}}  = 
\Big\langle \chi_{\text{5d}}^{U(N)}(\sigma, \vec{a}, \epsilon_1, 0, m, R, Q_{\text{5d}})  \Big\rangle 
\end{equation}
can be used both to construct the Baxter equation for the $N$-particle elliptic Ruijsenaars-Schneider system of type A and to compute its energies: for example in the $SU(2)$ case we have \eqref{last} where the energy is nothing but the NS limit of the fundamental Wilson loop \eqref{wilson}, while for general $SU(N)$ we find
\begin{equation}
\Big\langle \chi_{\text{5d,inst}}^{U(N)}(\sigma, \vec{a}, \epsilon_1, m, R, Q_{\text{5d}})  \Big\rangle_{\text{NS}} = \sum_{l = 0}^N (-1)^n X^{\frac{N}{2} - l} \langle W^{SU(N)}_{\bigwedge^l} \rangle_{\text{NS}} = \sum_{l = 0}^N (-1)^l X^{\frac{N}{2} - l} E'_{l},
\end{equation}
with $\langle W^{SU(N)}_{\bigwedge^l} \rangle_{\text{NS}}$ the NS limit of the $SU(N)$ Wilson loop in the $l$-th antisymmetric representation and $E'_{l}$ energy of the $l$-th eRS Hamiltonian $\widehat{H}'_{l}$ (here $\widehat{H}'_{1} = \widehat{H}'_{\text{eRS}}$ in \eqref{hRSgenp}). \\

To conclude, the instanton corrections to the partition function coupled to an $N$-plet of three-dimensional free hypermultiplets is
\begin{equation}
\mathcal{Q}_{\text{5d,inst}}^{(f)}(\sigma, \vec{a}, \epsilon_1, \epsilon_2, m, R, Q_{\text{5d}}) = \sum_{k \geqslant 0} Q_{\text{5d}}^k \mathcal{Q}_{\text{5d,inst}}^{(f),(k)}(\sigma, \vec{a}, \epsilon_1, \epsilon_2, m, R),
\end{equation}
where the coefficients are 
\begin{equation}
\begin{split}
& \mathcal{Q}_{\text{5d,inst}}^{(f),(k)}(\sigma, \vec{a}, \epsilon_1, \epsilon_2, m, R) \\
& = \dfrac{1}{k!} \oint \prod_{s=1}^k \dfrac{d \varphi_s}{\varphi_s} 
Z_{\text{5d,ADHM}}^{U(N),(k)}(\vec{\varphi}, \vec{a}, \epsilon_1, \epsilon_2, m, R) 
\mathcal{Q}_{\text{5d,ADHM}}^{(f),(k)}(\vec{\varphi}, \sigma, \vec{a}, \epsilon_1, \epsilon_2, m, R),
\end{split}
\end{equation}
with
\begin{equation}
\mathcal{Q}_{\text{5d,ADHM}}^{(f),(k)}(\vec{\varphi}, \sigma, \vec{a}, \epsilon_1, \epsilon_2, m, R) = 
\prod_{s=1}^k \dfrac{(1-\varphi_s X^{-1} q_1^{\frac{1}{2}} q_2^{\frac{1}{2}})(1-\varphi_s X^{-1} \mu q_1^{-\frac{1}{2}} q_2^{-\frac{1}{2}})}{(1-\varphi_s X^{-1} q_1^{\frac{1}{2}} q_2^{-\frac{1}{2}})(1-\varphi_s X^{-1} \mu q_1^{-\frac{1}{2}} q_2^{\frac{1}{2}})}.
\end{equation}
Relevant poles are once more labelled by ($N$+$1$)-tuples of Young tableaux $\vec{Y} = (Y_1, \ldots, Y_N, Y_{N+1})$, $\sum_{n=1}^{N+1} \vert Y_n \vert = k$ corresponding to
\begin{equation}
\begin{split}
\varphi_s &= \alpha_n q_1^{\frac{1}{2}} q_2^{\frac{1}{2}} q_1^{i-1} q_2^{j-1} \,, \;\;\;\; n = 1, \ldots, N \;\;\;\; \text{or} \\
\varphi_s &= X \mu^{-1} q_1^{\frac{1}{2}} q_2^{-\frac{1}{2}} q_1^{i-1} \mu^{-(j-1)}  \,, \;\;\;\; n = N+1.
\end{split}
\end{equation}
From the normalized observable
\begin{equation}
\langle \mathcal{Q}_{\text{5d,inst}}^{(f)}(\sigma, \vec{a}, \epsilon_1, \epsilon_2, m, R, Q_{\text{5d}})  \rangle = \dfrac{\mathcal{Q}_{\text{5d,inst}}^{(f)}(\sigma, \vec{a}, \epsilon_1, \epsilon_2, m, R, Q_{\text{5d}}) }{Z_{\text{5d,inst}}^{U(N)}(\vec{a}, \epsilon_1, \epsilon_2, m, R, Q_{\text{5d}}) }
\end{equation}
we obtain its NS limit as
\begin{equation}
\langle \mathcal{Q}_{\text{5d,inst}}^{(f)}(\sigma, \vec{a}, \epsilon_1, m, R, Q_{\text{5d}})  \rangle_{\text{NS}} = \langle \mathcal{Q}_{\text{5d,inst}}^{(f)}(\sigma, \vec{a}, \epsilon_1, 0, m, R, Q_{\text{5d}})  \rangle  ,
\end{equation}
which is part of the solution to the Baxter equation of the $N$-particle elliptic Ruijsenaars-Schneider system of type A.

\bibliography{bibl}
\bibliographystyle{JHEP}

\end{document}